\documentclass[11pt,oneside,letterpaper]{article}
\usepackage{amssymb}
\usepackage{amsmath}
\usepackage[dvips]{graphicx}
\usepackage{setspace}
\usepackage{amsfonts}
\usepackage{fancyhdr}
\usepackage{xcolor}
\usepackage{graphicx}
\usepackage{rotating}
\usepackage{comment}
\usepackage{color}
\usepackage{cite}
\usepackage{braket}

\usepackage{tikz}

\usepackage{subfigure}
\usetikzlibrary{decorations.pathmorphing}

\definecolor{darkgreen}{rgb}{0,0.5,0}
\definecolor{darkblue}{rgb}{0,0,0.6}
\definecolor{purple}{rgb}{0.4,.2,0.7}

\newcommand{\be}{\begin{equation}}
\newcommand{\ee}{\end{equation}}

\usepackage[colorlinks=true,citecolor=darkgreen,linkcolor=black,urlcolor=purple]{hyperref}

\usepackage{pdfsync}

\makeatletter
\newcommand*{\defeq}{\mathrel{\rlap{%
                     \raisebox{0.3ex}{$\m@th\cdot$}}%
                     \raisebox{-0.3ex}{$\m@th\cdot$}}%
                     =} 
\makeatother

\def\be{\begin{eqnarray}}
\def\ee{\end{eqnarray}}

\newcommand{\la}{\langle}

\newcommand{\bea}{\begin{eqnarray}}
\newcommand{\eea}{\end{eqnarray}}
\def\ben{\begin{equation}}
\def\een{\end{equation}}

     \let\r=v

\let\la=\label

\def\be{\begin{equation}}
\def\ee{\end{equation}}
\def\ba{\begin{array}}
\def\ea{\end{array}}

\def\ba#1\ea{\begin{align}#1\end{align}}
\def\bs#1\es{\begin{split}#1\end{split}}

\allowdisplaybreaks  

\numberwithin{equation}{section}

\def\nref#1{(\ref{#1})}
\def \la {\label}   
\def \be {\begin{equation}}
\def \ee {\end{equation}}
	
\def \JM#1 {{\color{blue}  JM: #1 }}
\def \VG#1 {{\color{red}  VG: #1 }}
\def \YC#1 {{\color{darkgreen}  YC: #1 }}

\begin{document}
\onehalfspacing

\begin{center}

~
\vskip5mm

{\LARGE  {   Bra-ket wormholes in gravitationally prepared states \\
}}

\vskip10mm

Yiming Chen$^1$, \ Victor Gorbenko$^{2,3}$,\ \ Juan Maldacena$^{3}$ 

\vskip15mm

{\it $^{1}$ Jadwin Hall, Princeton University, Princeton, New Jersey, USA } \\
\vskip5mm

{\it $^{2}$ SITP, Stanford University, Palo Alto, California, USA } \\
\vskip5mm

{\it $^{3}$ Institute for Advanced Study, Princeton, New Jersey, USA } \\
\vskip5mm

\vskip5mm

\end{center}

\vspace{4mm}

\begin{abstract}
\noindent
We consider two dimensional CFT states that are produced by a gravitational path integral. 

As a first case, we consider a state produced by Euclidean $AdS_2$ evolution followed by flat space evolution. We use the fine grained entropy formula to explore the nature of the state. We find that the naive hyperbolic space geometry leads to a paradox. This is solved if we include a geometry that connects the bra with the ket, a bra-ket wormhole. The semiclassical Lorentzian interpretation leads to CFT state entangled with an expanding and collapsing Friedmann cosmology.  

As a second case, we consider a state produced by Lorentzian $dS_2$ evolution, again followed by flat space evolution. The most naive geometry also leads to a similar paradox. We explore several possible   bra-ket wormholes.  The most  obvious  one leads to a badly divergent temperature. The most promising one also leads to a divergent temperature but by making a projection onto low energy states we find that it has  features that look similar to the previous Euclidean case. In particular, the maximum entropy of an interval in the future is set by the de Sitter entropy. 
 
 \end{abstract}

\pagebreak
\pagestyle{plain}

\setcounter{tocdepth}{2}
{}
\vfill
\tableofcontents

\newpage


\section{Introduction}

\subsection{General motivation} 

Recent work has highlighted the useful properties of the fine grained entropy formula, the Ryu-Takayanagi formula and its generalizations \cite{Ryu:2006bv,Hubeny:2007xt,Faulkner:2013ana,Engelhardt:2014gca}, for understanding properties of states in quantum gravity. 
This formula is believed to give an accurate estimate for the fine grained entropy of the exact state in terms of a computation that can be done knowing only the semiclassical solution, a background geometry plus propagating quantum fields. 
This formula was very useful for computing the entropy of Hawking radiation coming from black holes
\cite{Penington:2019npb,Almheiri:2019psf,Almheiri:2019qdq,Penington:2019kki}.  

 \begin{figure}[h]
\begin{center}
\includegraphics[scale=.35]{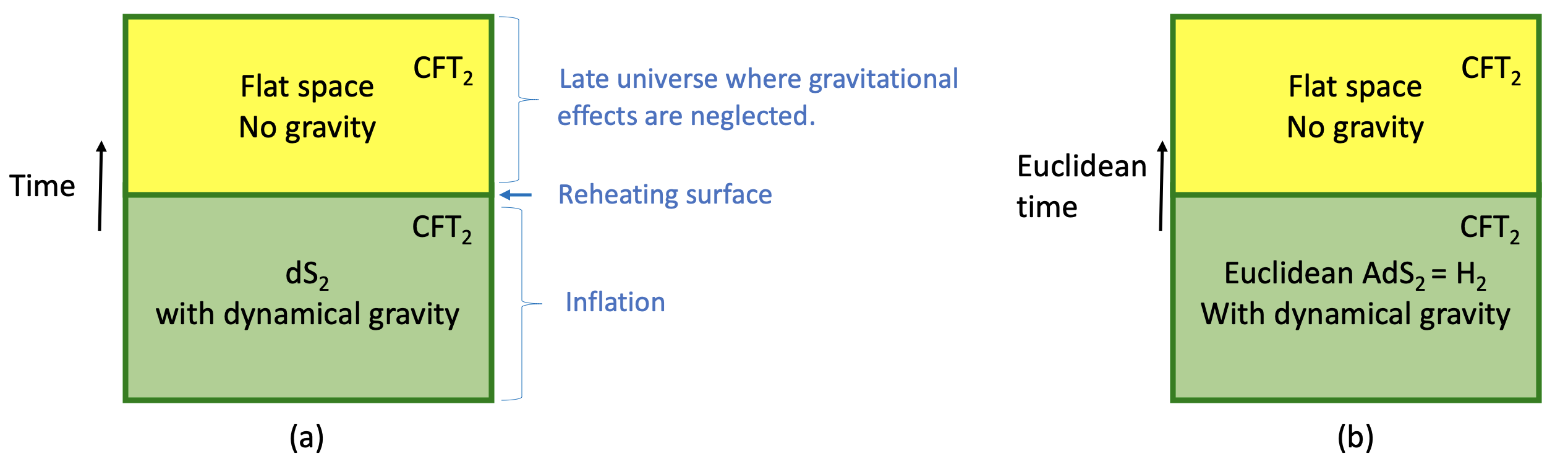}
\caption{The two main cases we study. (a) We model an inflationary spacetime as follows. We consider a two dimensional de Sitter JT gravity theory coupled to a matter CFT$_2$. Inflation ends when the dilaton reaches a large value. We then transition to a flat space region, where we neglect the effects of gravity. This procedure produces a certain quantum state for the CFT$_2$ in the flat space region. (b) Another way to produce a state in a flat space CFT. We do Euclidean evolution in Euclidean AdS$_2$ JT gravity followed by Euclidean evolution in flat space. We impose transparent boundary conditions for the CFT$_2$ at the $AdS_2$ boundary. 
}
\label{dSIntro}
\end{center}
\end{figure} 

It is then natural to ask whether we can use the gravitational fine grained entropy formula  for cosmological problems in order to gain some insight about properties of the exact state in terms of the semiclassical state. The formula is best understood in situations  where there is a  region where we can neglect gravity.  This is important for defining the density matrix (and the associated Hilbert space) whose entropy we are computing. At first sight, we do not seem to have such a region in cosmology. However,   we can think of the late universe as classical, neglecting quantum gravity effects at late times. Then we can define entropies of subregions and study the effects of quantum gravity on the early universe. This is routinely done in the study of small primordial fluctuations. Here the idea is to do something similar to learn about some non-perturbative effects that can lead to major corrections to the entropy of subregions. We are imagining that the state is produced by some version of the no boundary proposal \cite{Hartle:1983ai}.
 With this motivation in mind,  we set out to study a very simple two dimensional toy model for inflation. We consider de Sitter JT gravity \cite{Jackiw:1984je,Teitelboim:1983ux} coupled to matter consisting of a conformal field theory (see also \cite{Cotler:2019nbi,Maldacena:2019cbz}.) To the future of the de Sitter region, we imagine a flat space region with no gravity, see figure \ref{dSIntro}(a).  We can think of the de Sitter evolution as producing a particular state for the subsequent flat space evolution.

 We considered the entropy of intervals. When the interval is very long, we found that there is a kind of island which seems to imply that the entropy saturates at the de Sitter entropy. While this is an initially pleasing answer, further study revealed that there were inconsistencies with this result.  

\subsection{Boundary states generated by Euclidean AdS evolution } 

 This prompted us to study a similar problem in the more familiar AdS setup. In this case, we consider $AdS_2$ JT gravity, again coupled to a CFT. We use Euclidean AdS evolution to produce a certain state in the CFT, which we then further evolve in flat space, with no gravity, see figure \ref{dSIntro}(b). For this case, we expect a holographic interpretation as a boundary state for the CFT. Also in this case, when we apply the gravitational fine grained entropy formula we find islands for long enough intervals. As in the de Sitter case,  these islands are inconsistent with the entropy strong subadditivity formula, see section \ref{sec:paradox}.
 
The fact that the bra and the ket get connected when we compute traces of the density matrix motivates the introduction of bra-ket wormholes \cite{Page:1986vw}. We find   bra-ket wormholes that are a simple Euclidean rotation of the wormholes considered in \cite{Maldacena:2018lmt}.  This leads to a new semiclassical configuration, discussed in section \ref{sec:BraKet}. When we apply the fine grained entropy formula for this new configuration we find also that the entropy of the interval saturates at a finite value for large intervals. In addition, the inconsistencies that we discussed above disappear, see section \ref{ParadoxLost}. 
 
  This new bra-ket wormhole semiclassical geometry is such that it appears to prepare a mixed thermal state in the flat space region. The reason is that the wormhole connection has a finite Euclidean time circle which leads to a thermal looking state.  Nevertheless, the fine grained entropy formula tells us that the entropy of the whole universe is zero for the following reason. The Lorentzian continuation of the state produces a semiclassical state that consists of a thermal state in flat space with no gravity plus a  thermal state in an FLRW geometry with gravity which collapses both in the past and the future to a singularity,  where the dilaton goes to minus infinity, making the gravitational description very strongly coupled. 
 In a situation like this, the fine grained entropy formula applied to the whole flat space region, includes an island consisting of the whole FLRW region. The  entropy is zero, since we have a  pure state in the union of the two universes. 
 Since the entanglement wedge of the non-gravitational region includes the FLRW region, we can view this setup as a holographic description of such an expanding and collapsing FLRW universe, as explained in section \ref{sec:ClosedUniverse}. 
 
 These bra-ket wormholes are closely related to the Euclidean wormholes that are puzzling from the point of view of factorization of amplitudes. Namely, a small variant of the states we considered, where we project onto pure states give examples of Euclidean wormholes solutions that pose a factorization problem interpretation, in a manner very similar in spirit to the one discussed in \cite{Stanford:2020wkf}, which inspired section \ref{sec:Typical}.

 \subsection{Bra-ket wormholes and de Sitter }

 Motivated by the success of bra-ket wormholes in the Euclidean AdS problem, we 
 searched for similar bra-ket wormholes in de Sitter in section \ref{sec:dSbraket}. To produce such a geometry, we explored analytic continuations of the familiar de Sitter solutions. As in the no-boundary proposal, we allowed the contour to deviate into the Euclicean time direction in the past. We considered three choices of contour which give three different candidate states. The simplest contour does not include any Euclidean time evolution and produces a rather singular state. The next contour produces a state very similar to the Hartle-Hawking state, but it does not appear to solve the problem. Finally, there is an intermediate contour that produces a semiclassically thermal  state, but with a temperature that is driven to very large values. Except for this detail, the state solves the paradox. This large value of the temperatures is the same as saying that the inflationary period becomes very short. This is also a problem that arises when we apply the Hartle Hawking proposal in four dimensional inflationary models, a problem for which there is no generally accepted answer\footnote{Some researchers think  that this means that  the Hartle-Hawking wave function is simply {\it wrong}. A resolution was proposed in \cite{Hartle:2010dq}.  We think that the Hartle-Hawking wave function is an interesting proposal that deserves to be understood better, and that is what we attempt to do in this paper.}.
  We will leave this issue unsettled in this paper. 
  
  The problem of finding a consistent wavefunction for the universe is one that is famously unresolved. In this paper, we have explored the possibility of spacetime manifolds connecting the bra and the ket. We will describe some suggestive configurations in two dimensions, which will not lead to  well defined wavefuntions. We hope that perhaps with a suitable modification, or additional matter content, they might  lead to interesting wavefunctions. Moreover, our discussion will be restricted to two dimensions. It would be interesting to see if similar configurations exist in higher dimensions, a generalization that will likely involve new ideas.

\section{States prepared by Euclidean AdS$_2$ evolution and boundary states}

\subsection{Black holes and boundary states} \la{sec:setup}

\begin{figure}[h]
\begin{center}
\includegraphics[scale=.35]{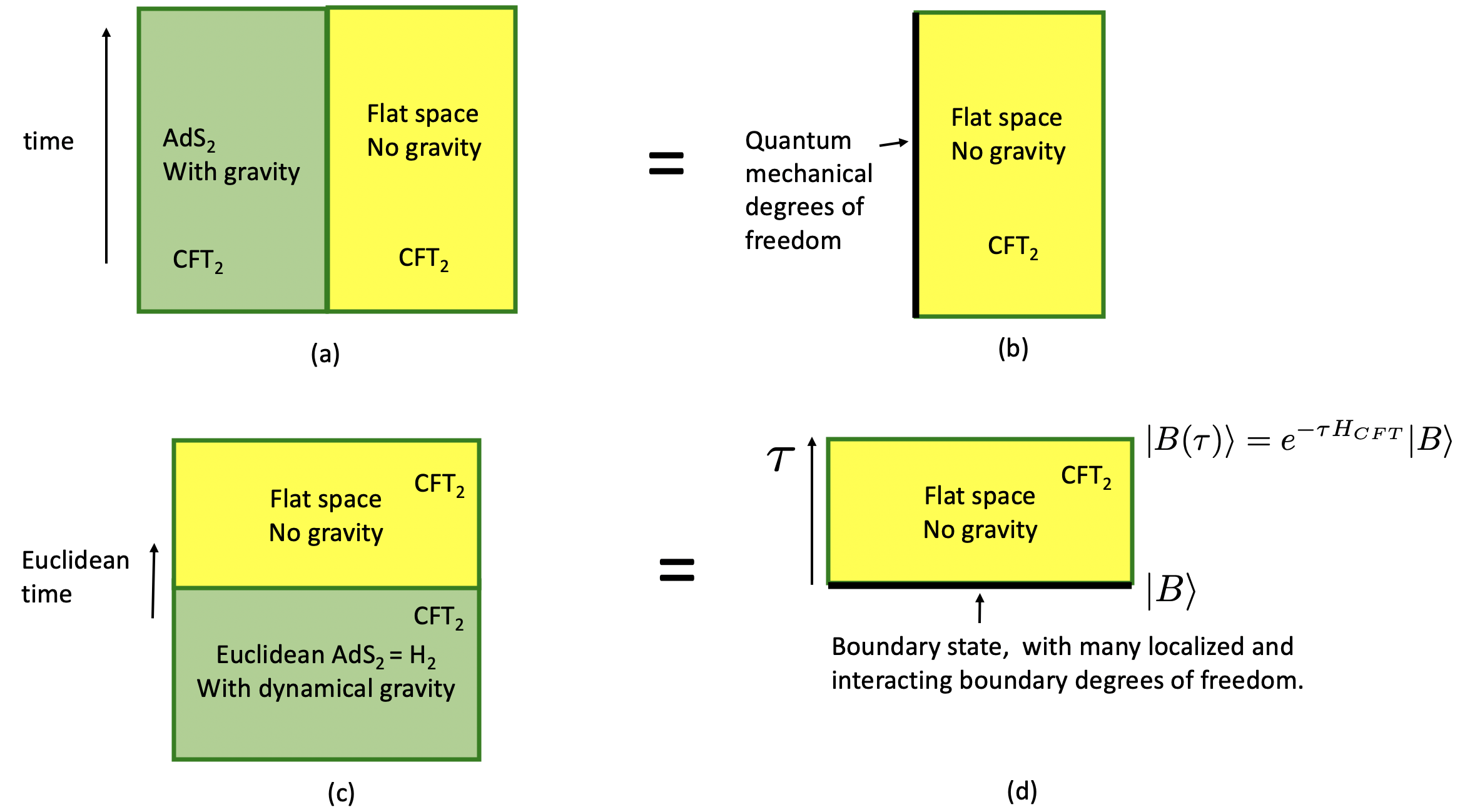}
\caption{ (a) $AdS_2$ with gravity plus flat space. (b) Holographic dual consisting of a flat space CFT interacting with some boundary degrees of freedom. (c) Euclidean rotation. We prepare a state in the CFT by doing Euclidean evolution in the gravity theory. (d) Boundary state interpretation.  }
\label{SetupFig}
\end{center}
\end{figure}

In studies of evaporating black holes, a certain toy model has been studied extensively \cite{Engelsoy:2016xyb,Almheiri:2019psf}. 
This model involves a simple theory of two dimensional gravity, the $AdS_2$  Jackiw Teitelboim theory, coupled to a two dimensional matter CFT. At the $AdS_2$ boundary, this is in turn coupled  to the same CFT but in a flat space with no gravity. We impose transparent boundary conditions for the matter CFT at the interface between the $AdS_2$ region and the flat space region, see figure \ref{SetupFig}.
In other words, we have the two dimensional action
\be \la{action}
I = - { S_0 \over 4 \pi } \left[ \int R + 2 \int K \right] - { 1 \over 4 \pi }\left[  \int \phi (R+2) 
+ 2 \phi_b \int K \right] + I_{CFT} 
\ee 
The term involving $S_0$ gives rise to the extremal entropy. We have boundary conditions that set $\phi = \phi_b$ at the $AdS_2$ boundary and also fix the metric along the boundary.  Finally the CFT action is defined both in $AdS_2$ and in flat space. We can think of this theory as the low energy theory of a possibly  more complicated UV theory\footnote{This theory, as it is,  is not a UV complete gravity theory because there are divergencies that arise when we consider topologies with thin handles. This is the usual tachyon divergence of bosonic string theory.}. 
We  assume that the central charge $c$ of the CFT is large, $c\gg 1$, so that we can neglect the quantum mechanics of the gravitational boundary degrees of freedom \cite{Jensen:2016pah,Engelsoy:2016xyb,Maldacena:2016upp}. Recall that this model does not have propagating gravitational degrees of freedom.  
The quantum fluctuations of the boundary degree of freedom can be neglected because, in the large $c$ limit, it becomes effectively massive (or gapped) and weakly coupled\footnote{We are talking about the boundary mode becoming massive, there are no propagating two dimensional modes.  The massive graviton perspective was emphasized, in a higher dimensional context, in \cite{Geng:2020qvw}.  }, with a coupling going like $1/c$.

\vspace{.5cm}

It is standard to conjecture that the complete gravity theory has a holographic description as a CFT coupled to a quantum mechanical system at the boundary. 
In other words, we could replace the region that has dynamical gravity by a system that lives at the boundary of the flat space region, see figure \ref{SetupFig}(b). 
By going to Euclidean time, and exchanging the Euclidean time direction with the space direction, we can view this system as producing a particular state in the conformal field theory, which we call
 $|B\rangle$. This is a state of the CFT in the flat space theory without gravity. It  can be further evolved in Euclidean time, 
 producing $|B(\tau)\rangle = e^{ - \tau H_{CFT} }|B \rangle$.   The notation suggests that we think of $|B\rangle $ as a boundary state, representing an effective boundary condition for the CFT. Of course, this is generated by the boundary degrees of freedom, which  give rise to a rather non-trivial state. 
States such as $|B(\tau)\rangle$, but with simpler boundary conditions were considered in the past as models for the result of a quantum quench between a gapped theory and a conformal field theory \cite{Calabrese:2005in}. In that context, the state was then evolved in Lorentzian time. After this preparation, we can also evolve it further in Lorentzian time in our case too. In our case, we expect that $|B\rangle$ is a short ranged entangled state because the black hole entropy in figure \ref{SetupFig} (a) is finite, and so the $g$ function is finite (see appendix \ref{App:gfunction} for further discussion). 

Going back to the gravity picture, the state is defined by performing the Euclidean gravitational path integral with fixed metric and dilaton boundary conditions. We also consider the fields of the matter CFT propagating on this geometry with transparent boundary conditions at the boundary. If we consider particular semiclassical solutions, this procedure  prepares a state $|B\rangle_s$ in the conformal field theory. The subscript $s$ indicates that this is the boundary state given by the semiclassical gravity approximation, as opposed to the state $|B\rangle$ which is the one we would get by the full non-perturbative gravity theory.

We will now explore some properties of the state produced in this fashion. In particular, we will analyze the entropies of subregions,  using the gravitational fine grained entropy formula 
\cite{Ryu:2006bv,Hubeny:2007xt,Faulkner:2013ana,Engelhardt:2014gca}. We will find that the entropies computed in this fashion do not obey strong subadditivity, in some particular cases. We will then argue that we should also include bra-ket wormholes, and that they lead to a solution of that problem. Furthermore, these wormholes lead to a very interesting picture for the boundary state. 

Let us also briefly mention that there is a close connection between this boundary state $|B\rangle$ and the S matrix of the Lorentzian problem. If the CFT is a free theory, say of free fermions, then we can define an S-matrix for the Lorentzian problem which tells us what comes out of an extremal black hole after we send in some excitations. This tell us how particles bounce off the boundary, generally creating other particles.  The Euclidean boundary state that we are discussing is an analytic continuation of that $S$ matrix. It can be obtained through a crossing relation\footnote{This relationship was particularly exploited in integrable models \cite{Ghoshal:1993tm}. Of course, in our case we do not have  an integrable theory.}.

\subsection{Subsystem entropies computed using  the naive geometry} \la{sec:naive}

We will start by discussing the simplest Euclidean geometry that contributes to the path integral with these boundary conditions, see figure \ref{SetupFig} (c). The metric and dilaton
\bea \la{metrdil} 
 ds^2 &=& 
 { dz^2 + dx^2 \over z^2 },   ~~~~~~~\phi = { \phi_r \over (-z) }   ~~~~~~{\rm for}~~~{z\leq -\epsilon} \\
 ds^2 & =& 
 { dz^2 + dx^2 \over \epsilon^2 }    ~~~~~~~~~~~~~~~~~~~~~~~~ ~~~{\rm for}~~~ -\epsilon \leq z 
 \eea	
 describe Euclidean $AdS_2$ (or $H_2$) joined to flat space. At the   boundary we set $\phi_b = \phi_r/\epsilon$. We imagine we can take $\epsilon$ to be very small keeping $\phi_r$ constant \cite{Jensen:2016pah,Engelsoy:2016xyb,Maldacena:2016upp}. 
 Notice that the matter CFT lives on a space that is conformally flat. Therefore its state, in the semiclassical description,  is the usual vacuum. In other words, we find that the  semiclassical state is
 $|B\rangle_s = |0\rangle$. 

\begin{figure}[h]
\begin{center}
\includegraphics[scale=.4]{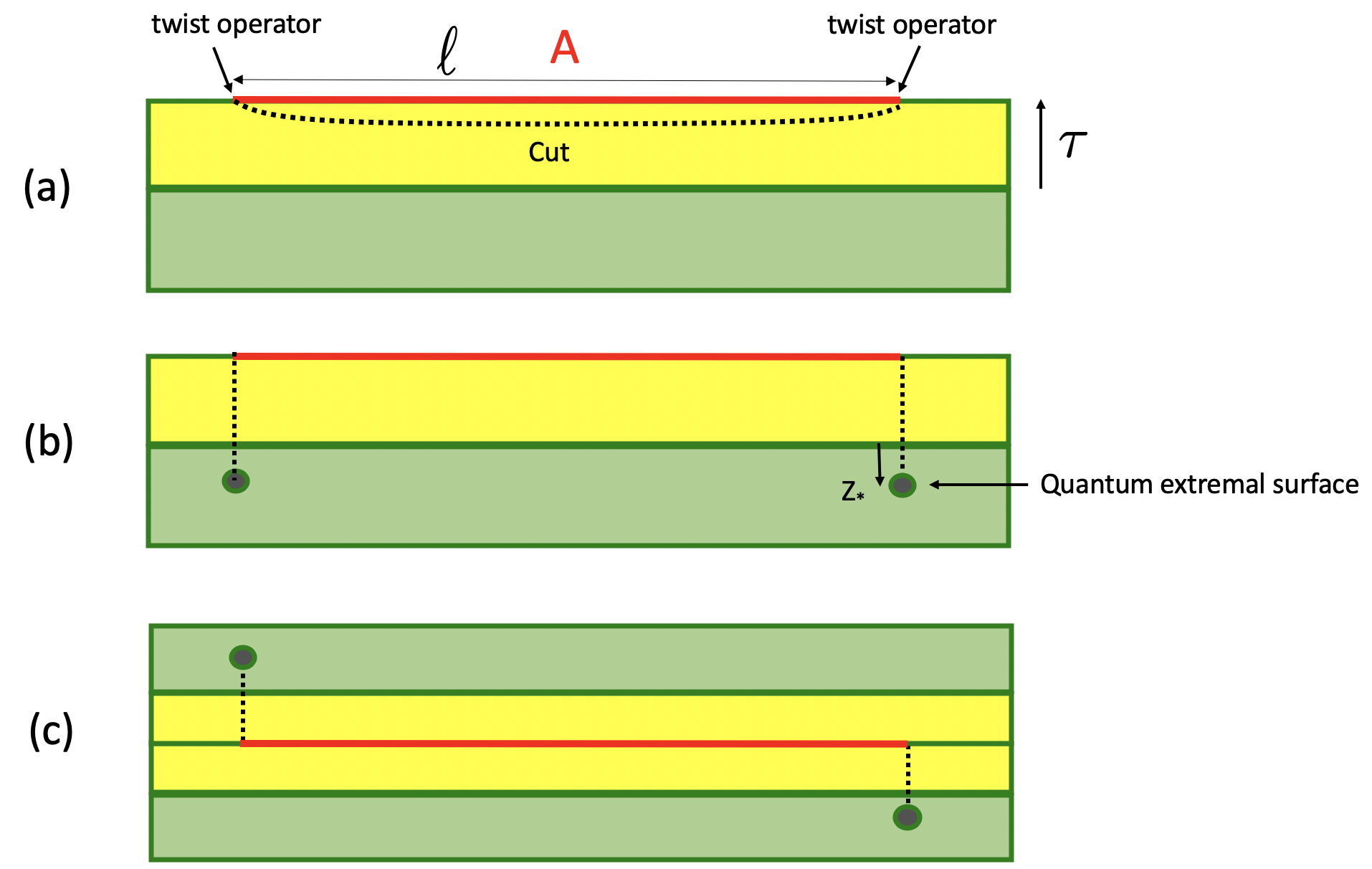}
\caption{ We consider a region $A$ in the state produced by the gravitational (green) plus the field theory (yellow) Euclidean evolution. (a) The naive result with no islands which just gives the field theory entropy of the interval in the vacuum. (b) The entropy when there are non-trivial quantum extremal surfaces. When $\ell$ is large the two intervals, represented by the dotted lines, are very far away. (c) The full computation involves two sides corresponding to the bra and the ket, here represented above and below the red interval. We can put the quantum extremal surface on either side indpendently for each of the two extremal surfaces.    }
\label{IntervalEntropies}
\end{center}
\end{figure}

Interestingly, the gravitational fine grained entropy formula is supposed to give us entropies of subregions of the full state $|B\rangle$ in terms of a formula we can evaluate on the semiclassical spacetime solution. We consider a subregion $A$ consisting of an interval of  length $\Delta x = \ell$ in the $x$ coordinates. We place the interval at 
 $z =\tau $ from the boundary of $AdS_2$, see figure \ref{IntervalEntropies}. 
 The naive answer would be to compute the entropy of such interval in the  usual vacuum  
 \be \la{Snaive}
 S_{\rm no~island} = { c \over 6 } \log \left( {\ell^2 \over \varepsilon_{uv}^2 } \right) 
 \ee
 where $\varepsilon_{uv}$ is a UV cutoff in the definition of the field theory entropy, which should not be confused with $\epsilon$ in \nref{metrdil}. 
 
 We see that \nref{Snaive} becomes large when $\ell$ becomes large. A feature of the fine grained gravitational entropy formula is that we can sometimes decrease the entropy by including certain ``islands''. Computing the entropy using the replica trick, this formula is derived from the existence of replica wormholes \cite{Penington:2019kki,Almheiri:2019qdq}, which in turn can be viewed as the result of inserting extra twist operators in the gravitational region. 
  So we could imagine exploring configurations such as the ones depicted in figure \ref{IntervalEntropies}. These configurations contain two intervals.  We need to add an  extra term involving the dilaton at the two new endpoints. In the regime that $\ell$ is very large,  we assume we can compute their entropy by using an operator product expansion for the underlying twist operators. We will check this assumption a posteriori. To leading order, this gives us just the sum of the entropies of the two intervals. The extremization for each interval is independent of the other. By symmetry, both will be at the same $x$ position. So we only need to consider  
\be \la{Sgen} 
  S_{\rm gen}(z) =  2 \left\{  S_0 + { \phi_r \over - z } + { c \over 6 } 
 \log \left[ {(   \tau -z)^2 \over (-z) \varepsilon_{uv} }  \right] \right\}
 \ee
 and extremize it over $z$,    $\partial_z S_{gen}(z) =0$, to   find the position $z_*$. 
 In the particular case of $\tau=0$ we find 
 \be \la{zstar}
 -z_* =  { 6 \phi_r \over c } ~~~~{\rm for } ~~ \tau=0.
 \ee
 For general $\tau$ we need to solve a quadratic equation, see \nref{quadr} in appendix \ref{App:gfunction}. We should note that $\phi_r/c$ is a distance scale. In the black hole interpretation of figure \ref{SetupFig}(a), it is the evaporation time scale, after we introduce some energy into the black hole \cite{Engelsoy:2016xyb}. This distance scale will appear in many of our formulas.
Inserting (\ref{zstar}) into \nref{Sgen} we get 
 \be \la{SIsland}
 S_{\rm island} = 2 S_0 + { c \over 3} \left[ 1 + \log\left( {6 \phi_r \over c \, \varepsilon_{uv} } \right) \right] .
 \ee
 Notice that the full computation involves two copies of the gravity region, one for the bra and one for the ket.   Therefore, we can also put the quantum extremal surfaces on the other side, 
 see figure \ref{IntervalEntropies}(c). In principle, we should sum over such configurations when we evaluate the Renyi entropies. The effects of the sum  give corrections which are subleading in the $1/c$ expansion.  A more complete discussion of situations with competing saddles can be found in \cite{Marolf:2020vsi,Dong:2020iod}. 
 
  Comparing \nref{Snaive} and \nref{SIsland} we find that   for 
 \be \la{DomDis}
  \ell > { 6 \phi_r \over c } \exp\left( { 6 S_0/ c + 1 } \right)
  \ee 
  the configuration with the non-trivial quantum extremal surface dominates. 
  Note that when the island dominates $\ell \gg |z_*|$, which a posteriori justifies our OPE approximation when computing the entropies of the two intervals using the replica trick.  
  
  In this paper  we work in the regime 
    \be \la{Parrange} 
  c \gg 1 ~,~~~~~~~ { S_0 \over c }= {\rm fixed ~(and~ somewhat~ large) } ~,~~~~~~ { \phi_r \over c } = {\rm fixed } 
  \ee 
  so that the exponential in \nref{DomDis} is not too large. 
     
  Formula \nref{SIsland} gives a  constant result for large $\ell$, which is indeed what we expect for a short range entangled state. This result is simply twice the result for the entropy of a semi-infinite line. In turn, the result for the semi-infinite line is the same as the one obtained in \cite{Almheiri:2019yqk} when considering the entropy of an interval containing the boundary in the interpretation in figure \ref{SetupFig} (a), and it comes from the finiteness of the black hole entropy.   
  
  \subsection{An entropy subadditivity paradox }
   \la{sec:paradox}
  
  We  now consider a problem with the above formulas. We spell out this problem in detail in order to motivate the inclusion of bra-ket wormholes in later sections.  
  
\begin{figure}[h]
\begin{center}
\includegraphics[scale=.4]{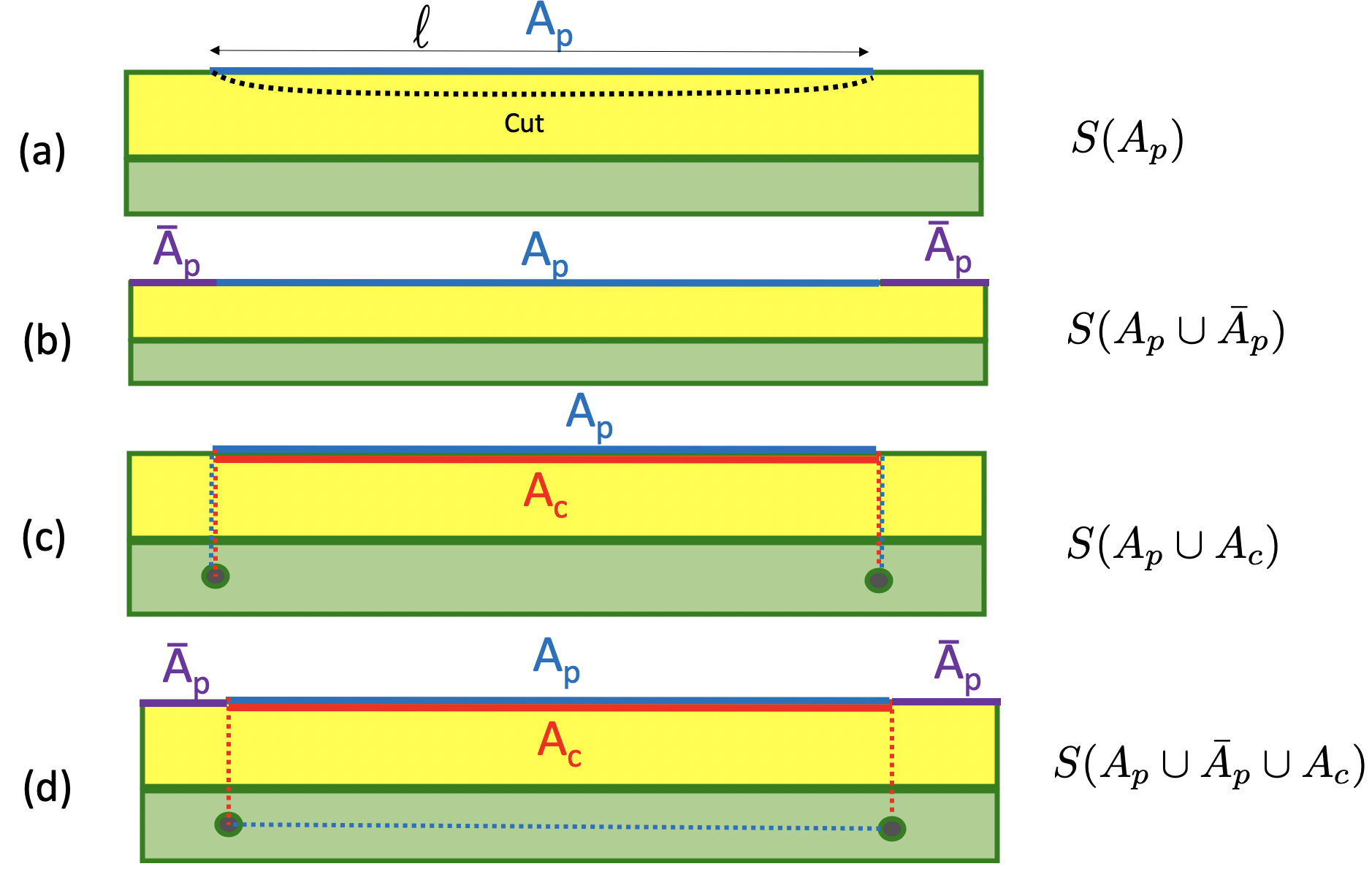}
\caption{ We display the quantum extremal surfaces involved in the computation of the various entropies. Though we are computing them at $\tau=0$ we included a bit of the (yellow) flat space region to emphasize that the subsystems are defined in the region with no gravity.    }
\label{YAMPS}
\end{center}
\end{figure}

   In order to formulate the paradox we assume that the matter CFT consists of two decoupled CFTs. One is a large $c$ CFT$_c$ and the other is a small $c_p$ ``probe'' CFT$_p$.   
  Both exist in the gravity region and the flat space region without gravity. In the semiclassical gravity approximation,  they are only coupled by the boundary graviton, which we are ignoring in our analysis. In the flat space region, they are completely decoupled. Therefore we can define the following subsystems. We pick a long interval lying in the flat space region as before, at $\tau=0$. The first subsystem, called $A_c$   contains the fields of the CFT$_c$ in that given interval. Another subsystem,   called $A_p$,  involves the probe fields of CFT$_p$ on that same interval.  
  Similarly we define $\bar A_p$ to be rest of the space, but containing only the fields of CFT$_p$. 
   We now consider the entropy subaditivity inequality for the three subsystems: $A_c$, $A_p$ and $\bar A_p$, see figure \ref{YAMPS},
  \be
  S( A_c\cup  A_p) +S(A_p \cup \bar A_p)\geq S(A_c \cup A_p \cup \bar A_p)  +S(A_p)\,.
  \label{SSA}
  \ee
  We would like to emphasize that all the subsystems are defined in the non-gravitational region. Consequently, $A_c$, $A_p$ and $\bar A_p$ are inherently distinct quantum degrees of freedom. Nevertheless,   we will see that this inequality fails, and does so by a large amount. 
  
 Specifically, we  choose the following regime in the parameter space:
  \be
\frac{ 6 \phi_r  }{c_p} \exp\left( { 6 S_0/c_p  } \right)\gg  \ell \gg { 6 \phi_r \over c } \exp\left( { 6 S_0/ c + 1 } \right)\,.
  \ee
 This implies that the calculation of the  entropy of $A_c$  is dominated by the island contribution, while the entropy of just the probe fields, of $A_p$,  is given by the no-island expression.  We display the quantum extremal surfaces for each case in figure \ref{YAMPS}. 
  
  Let us calculate all the terms in \eqref{SSA} one by one, using the formulas obtained in the previous section. 
 For  $S(A_p)$ we  get the no island result
   \be
  S(A_p)=\frac{c_p}{3}\log \left( \frac{\ell}{\varepsilon_{uv}} \right)\,.
  \ee
 In the region $A_p \cup \bar A_p$ we   have the vacuum for  the probe fields, a pure state. Then
 \be 
 S(A_p \cup \bar A_p) =0 \,.
 \ee
 For the region $A_c\cup A_p$ we can apply our previous result 
 \eqref{Sgen}  
  \be \la{Ssat}
  S( A_c \cup A_p) =2 S_0+\frac{c+c_p}{3}\left[1+\log\left(\frac{6 \phi_r}{c \varepsilon_{uv}}\right)\right]\,,
  \ee
 where we neglected the change in the position of the island due to addition of the probe fields since this only gives subleading corrections.
For the region $A_c \cup A_p \cup \bar A_p$ we have  emergent twist operators which affects both CFTs, but in the boundary we only have a twist operator for the CFT$_c$. Therefore we get an extra cut affecting the probe fields in the bulk, see figure \ref{YAMPS} (d). It is also important to keep in mind that the emergent twist fields are necessarily   twist fields for both CFTs, since they are just geometrical objects that do not distinguish between the matter species.
We find 
  \be
  S(A_c \cup A_p \cup \bar A_p) =2 S_0+\frac{c}{3}\left[1+\log \left(\frac{6 \phi_r}{c \varepsilon_{uv}}\right)\right]+\frac{c_p}{3}\log \left(\frac{\ell c}{6 \phi_r}\right)\,.
  \ee
 
  Combining these expressions we obtain 
  \be S( A_c\cup  A_p) +S(A_p \cup \bar A_p)- S(A_c \cup A_p \cup \bar A_p)  -S(A_p)
  =\frac{c_p}{3}\left[1-2 \log \left(\frac{\ell c}{6\phi_r}\right)\right]<0\,.
  \ee
We see that strong subadditivity is violated by a quantity of order $c_p \log (\ell c/\phi_r )\gg1$, for $\ell \gg { \phi_r \over c } $.

 Let us now make a few comments.  In spirit, the problem is somewhat similar to the black hole information paradox, as formulated in \cite{Mathur:2009hf,Almheiri:2012rt}. Namely, since the entropy of a large interval is saturated at \eqref{Ssat}, adding probe fields does not increase it by a sufficient amount. This means that $A_p$ and $\bar A_p$ cannot be entangled in the way dictated by the EFT in the bulk. In this sense, $A_p$ and $\bar A_p$ are analogous to modes outside and inside the black hole in \cite{Almheiri:2012rt}. However, our situation is also conceptually different: unlike   the black hole case, both $A_p$ and $\bar A_p$ live in the non-gravitational region and can be simultaneously observed. Thus we need a different resolution of the problem. The key ingredient is introduced in the next section.

\section{Bra-ket wormholes for states prepared by Euclidean AdS$_2$ evolution} 
\la{sec:EuclBK}

\subsection{The bra-ket wormhole }
 \la{sec:BraKet}

\begin{figure}[h]
\begin{center}
\includegraphics[scale=0.27]{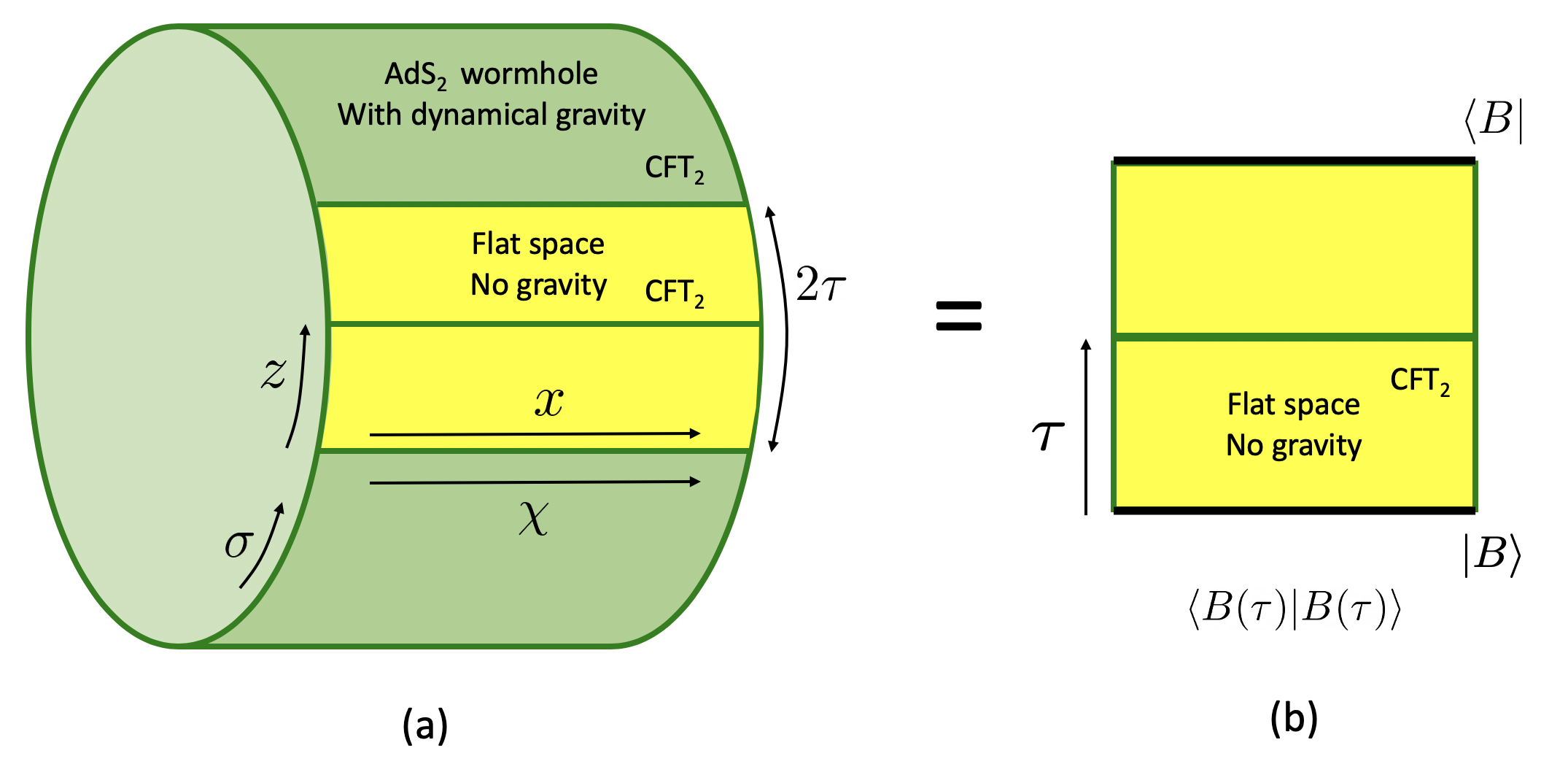}
\caption{(a) We consider a (green) wormhole geometry which connects two asymptotic boundaries, each of which is connected to the edges of a flat (yellow) strip.   (b) In the dual description, we take the bra and the ket of the boundary state and compute the trace after doing some Euclidean evolution. In other words we compute $\langle B(\tau) | B(\tau) \rangle$. We can compute expectation values of operators by inserting them in the flat space region.}
\label{BraketWormhole}
\end{center}
\end{figure}

In this section,  we  consider a different geometry, a wormhole connecting the bra and the ket.
 We will call it the bra-ket wormhole.  It consists of a flat strip with the  metric as in (\ref{metrdil}):
\be \la{metrflat}
 ds^2 ={ dz^2 + dx^2 \over \epsilon^2 } ~,~~~~~~~~~ 0 \leq z \leq 2 \tau
\ee
plus a region with dynamical gravity connecting the two sides of the strip, see figure \ref{BraketWormhole}. 
The strip has width $\Delta z = 2 \tau$. 
The geometry that connects the two sides can be described by the metric
\be \la{metrworm} 
 ds^2  = \frac{d\sigma^2 + d\chi^2}{\sin^2 \sigma}, ~~~~~\quad \sigma_c \leq  \sigma \leq \pi - \sigma_c.
 \ee
 This is to be patched to the two sides of the flat strip. The coordinates $x$ and $\chi$ are related by a rescaling that we determine below. This metric describes Euclidean $AdS_2$ (or $H_2$) in some coordinates that are conformal to a strip. 

This geometry appears naturally when we compute expectation values of local observables in the state $|B\rangle $, or $\langle B |O |B \rangle$. The operator is naturally inserted at $z = \tau$,  which is the point where we glue the bra and the ket to compute the expectation value. The fine grained  entropy is computed as the $n\to 1$ limit of similar expectation values involving $n$ copies of the system, but this is the $n=1$ geometry.


The fact that a connected wormhole topology can appear is familiar when the two boundaries interact \cite{Gao:2016bin,Maldacena:2017axo}. In fact, the flat space strip connecting the two asymptotic regions of the gravity theory can be viewed as adding an interaction between the two sides\footnote{ Other situations where wormhole geometries, even off shell ones,  have been shown to be crucial include \cite{Saad:2018bqo,Saad:2019lba}.}. Actually, the geometry we are discussing is essentially the same as the one discussed in 
\cite{Maldacena:2018lmt} but with euclidean time and space interchanged. Of course,  this interchange is just a matter of how we will interpret the geometry. For this reason, the solution is the same as the one discussed in 
\cite{Maldacena:2018lmt}, which we now proceed to review. 


For simplicity we will first discuss the solution when $\tau=0$.
Since the matter is in a thermal state, we get a non-zero contribution to the stress tensor. 
On a flat cylinder of circumference  $d = \pi$ the stress tensor is 
\be \la{stressflat}
 T_{\sigma\sigma}^{\textrm{flat}} = - T_{\chi\chi}^{\textrm{flat}} = -\frac{\pi c}{6d^2} \quad ~,~~~~~~~~~d = \pi.
\ee
The Weyl factor gives an additional contribution due to the conformal anomaly.
 The anomaly contains two parts. One part is proportional to the metric, whose effect is just to shift $\phi$ by a constant, which we  absorb into the definition of $S_0$. The second part is traceless and given by  
 \be  \la{StressAnomaly}
 T_{\sigma\sigma}^{\textrm{anomaly}} = - T_{\chi\chi}^{\textrm{anomaly}} = \frac{c}{24\pi}.
\ee
Adding (\ref{stressflat}) and (\ref{StressAnomaly}), we get the total stress tensor
\be  \la{stresstensor}
 T_{\sigma\sigma} = - T_{\chi\chi} = -\frac{c}{6\pi} + \frac{c}{24\pi} = - \frac{c}{8\pi}.
\ee
The dilaton solution in the presence of this stress tensor is
\be \la{dilworm}
	\phi = \frac{c}{4} \left( \frac{\frac{\pi}{2} -\sigma}{\tan \sigma} + 1\right) , ~~~~~~{\rm for} ~~~\tau=0
\ee
which asymptotes to $+\infty$ towards both boundaries of the wormhole. 
We now impose the boundary conditions on the metric and dilaton. 
\bea 
\phi &=& { \phi_r \over \epsilon } = { c \pi \over 8 \sigma_c } ~~~\longrightarrow ~~~~ \epsilon = \frac{8\phi_r}{\pi c} \sigma_c
\cr 
{ d x \over \epsilon} &=& { d \chi \over \sigma_c } ~~~~~~~~~~~\longrightarrow ~~~~
\chi = { \sigma_c \over \epsilon } x = { \pi c \over 8 \phi_r } x \la{chix}
\eea

The bra-ket wormhole gravitational action has a suppression of order $e^{-2 S_0}$ relative to the disconnected geometry. This fact will be most clearly seen once we make $x$ (or $\chi$) compact. However, the matter partition function enhances the contribution because  the matter entropy is in a thermal state\footnote{
In \cite{Maldacena:2018lmt}, this was described as negative Casimir energy, since were are exchanging space and euclidean time.}. If the direction $x$ is infinite this matter contribution dominates and this connected geometry is the leading contribution. 


 According to this semiclassical geometry,  conformal field theory state we have in the flat region is a thermal state with temperature\footnote{The subindex in the temperature says how the associated time coordinate is normalized. $\chi$ and $x$ are space coordinates but we imagine that the time coordinate is defined as $\sigma$ in \nref{metrworm} or $z$ in \nref{metrflat} respectively.}
\be \la{Temperature}
 T_\chi  = { 1 \over \pi } ~,~~~~{\rm or }~~~~ T_x = { c \over 8   \phi_r } ~,~~~~ \rho_s = e^{ -  H /T_x}
 \ee 
  Here $H$ is the CFT Hamiltonian that generates translations in the coordinate $z$ in \nref{metrflat}.
 The fact that we get a thermal state is clear because the euclidean time direction is compact in figure \ref{BraketWormhole}. 
 Of course, this temperature, as well as the density matrix $\rho_s$ are  semiclassical concepts. The exact state $|B\rangle$ is supposed to be pure. We will see that the gravitational fine grained entropy formula says that its entropy is zero.   
 
  As a very simple example where something similar happens, consider a free fermion theory in a state produced by a simple Dirichlet boundary condition. In such a state the left and right movers are entangled. However, if we only consider the left movers, they are in a thermal state. If we consider both left and right movers, they are in pure state. With this simple boundary condition it was easy to recognize the operators that show that the state is pure, rather than thermal. For states prepared by the gravity solutions we discussed above, we expect that it would be more difficult to construct such operators.    
 
Even though we discussed entropies, of course,  the existence of the bra-ket wormhole changes   the expectation values of local operators, since we would be computing them in a thermal state rather than in the vacuum. 

\subsection{Entropy of an interval and the disappearance of the paradox}
\la{sec:EntropyWH}

We are now ready to reexamine the  computation of the entropy of an interval.  When we compute  the gravitational fine grained entropy,  we should always use the dominant geometry, which is the wormhole in the current case. 
 We compute the entropy of a region $A$ consisting of an interval of length $\ell$. The no island answer for the entropy is no longer that of an interval in the vacuum,   (\ref{Snaive}). Instead, we get the answer corresponding to an interval in the thermal semiclassical state,    $\rho_s$,  which gives
\be \la{SnaiveWorm} 
	S_{\textrm{no island}} = \frac{c}{3} \log \left[ \frac{\sinh\left(  \Delta \chi\right) }{ \varepsilon_{uv,\chi}}\right]  = { c \over 3 } \log \left[ { 
	\sinh \left( { \pi T_x \ell  } \right) \over  \pi T_x   \varepsilon_{uv} } \right]  ,  
\ee
which grows linearly with the size of the interval.

\begin{figure}[h]
\begin{center}
\includegraphics[scale=0.25]{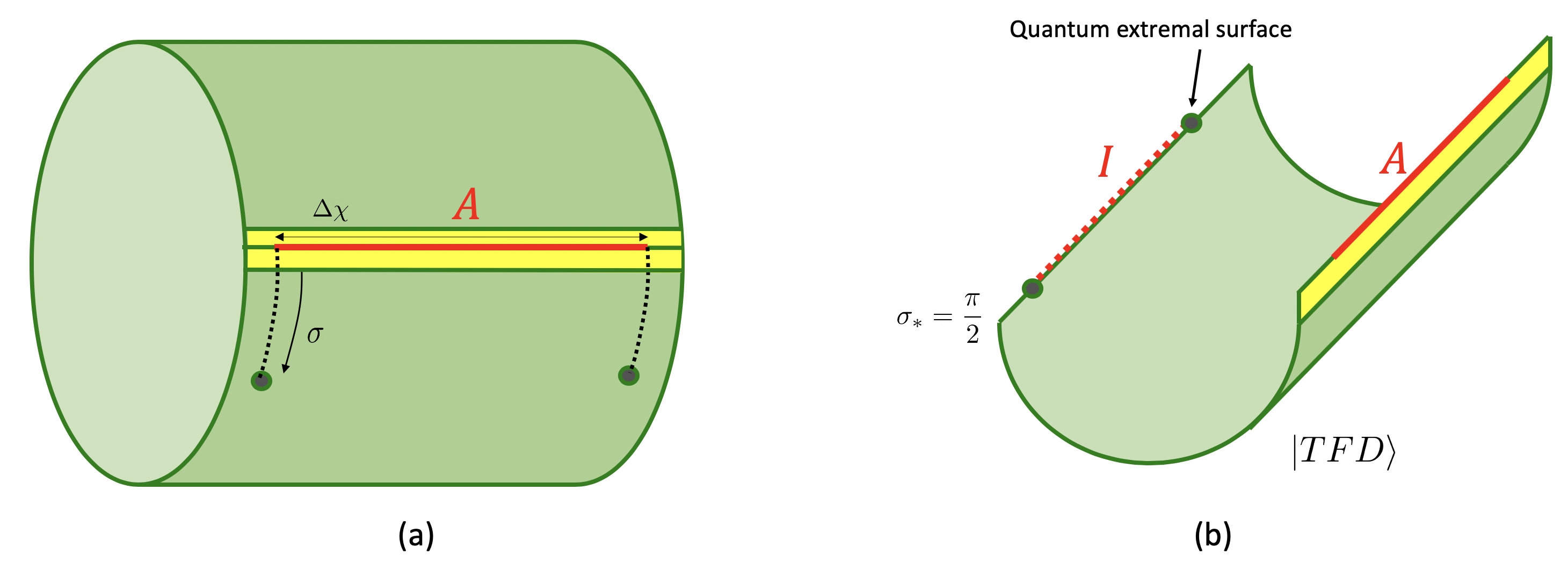}
\caption{(a) We are interested in the entropy of a region $A$ with length $\Delta\chi$. One can look for a configuration with two emergent twist operators in the gravity region, and they pair up with the two boundary twist operators of $A$ in the limit $\Delta\chi \gg 1$. (b) After extremization, the quantum extremal surfaces and the island (denoted by $I$ and the dotted line) are found to be on the throat of the wormhole. The matter entropy that appears in the QES prescription can then be interpreted as the entropy of $A\cup I$ in a $\ket{TFD}$ state.}
\label{fig:WormIsland}
\end{center}
\end{figure}

As before, we look for islands by placing quantum extremal surfaces in the bulk. When $\Delta x = \ell$ is very large, the problem reduces to finding the quantum extremal surface associated to each endpoint separately. 
%
 By symmetry, the emergent twist operator lies at the same $\chi$ position as the original one, see figure  \ref{fig:WormIsland} (a).
The generalized entropy function is given by
\be  \la{SgenWorm}
 S_{\textrm{gen}} (\sigma) = 2 \left\{ S_0 +\frac{c}{4} \left( \frac{\frac{\pi}{2} -\sigma}{\tan \sigma} + 1\right)  + \frac{c}{6} \log \left[ \frac{\sin^2 \sigma }{\sin\sigma\, \varepsilon_{uv,\chi}}\right] \right\}.
\ee
Extremizing over $\sigma$ by setting $\partial_\sigma S_{\textrm{gen}} (\sigma)=0$,  we get 
\be \la{sigmavalue}
 \sigma_* = \frac{\pi}{2},
\ee
which is at the middle of the wormhole, see figure \ref{fig:WormIsland} (b).  Inserting (\ref{sigmavalue}) into (\ref{SgenWorm}) we find
\be \la{SWormInf}
	S_{\textrm{island}} = 2 \left\{ S_0 + \frac{c}{4} + \frac{c}{6} \log\left( \frac{1}{\varepsilon_{uv,\chi}} \right) \right\} .
\ee
As in  (\ref{SIsland}), this is also  IR finite. Comparing (\ref{SWormInf}) with (\ref{SnaiveWorm}), we find that the island  configuration dominates when 
\be 
 \Delta\chi = { \pi c \ell \over 8 \phi_r}  \sim \frac{6S_0}{c},
\ee
which is    shorter than (\ref{DomDis}).   Note that the fact that the entropy saturates at a finite value for very long intervals indicates that the state could be written as a matrix product state, at least at very long distances. We make a few comments on tensor networks in appendix \ref{app:TensorNetworks}.

Despite the similarities in  these results for the entropy, the wormhole geometry has some new features which are absent in the naive geometry. In particular, the wormhole has a $Z_2$ reflection symmetry around $\sigma =\pi/2$, which is precisely where the quantum extremal surfaces lie on in \nref{sigmavalue}. This also means that, when we continue to Lorentzian signature, we can cut the cylinder in half and view the bottom part of the path integral as giving an entangled state consisting of the state in the flat space region and the state at the middle of the wormhole. Of course, from the point of view of the bulk matter theory this is just the thermofield double state,  
see figure\ref{fig:WormIsland}b.
 The matter entropy which appears in the gravitational fine grained entropy formula can then be interpreted as the entanglement entropy of the region $A$ union the island $I$, in the state $\ket{TFD}$.

\subsubsection{Paradox lost}
\la{ParadoxLost}

\begin{figure}[h]
\begin{center}
\includegraphics[scale=0.23]{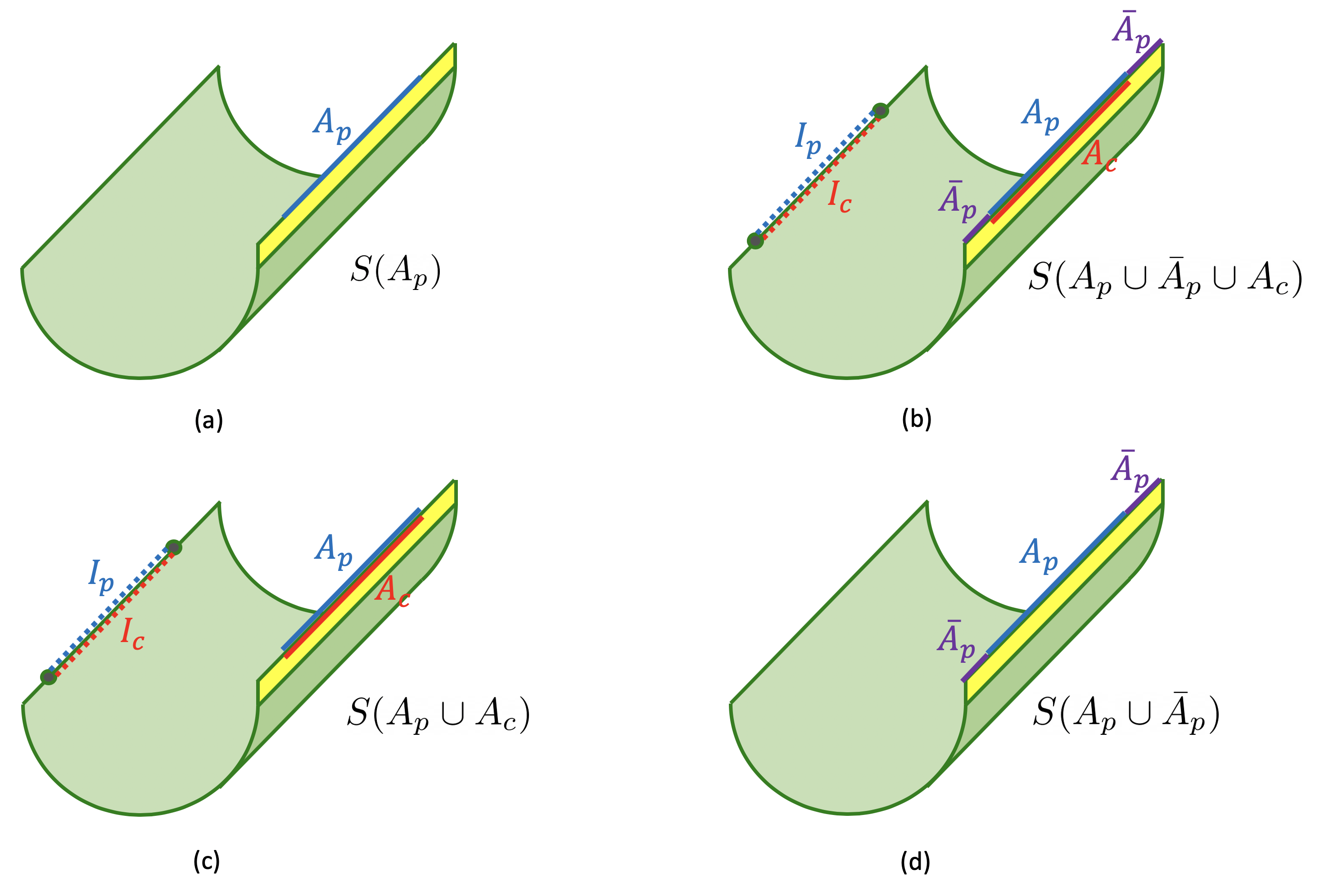}
\caption{We draw the regions involved in the  computation of various entropies. The original regions  in the definition of the entropy are drawn in solid, while the emergent regions (islands)  are drawn in dotted lines. A crucial point is that we can view the euclidean evolution on the depicted half cylinder as preparing a state for the matter theory. This state lives on two lines, the one in the (yellow) non gravitational region and the one in (green) the middle of the wormhole.  }
\label{fig:ParadoxResolution}
\end{center}
\end{figure}

We now check that we do not  run into the strong subaditivity paradox discussed in section \ref{sec:paradox}. 
We consider the same setup with matter given by two CFTs and a distance $\ell$ large enough that we should include an island when we consider $A_c$ but not if we consider $A_p$.  
Since the area terms on both sides of (\ref{SSA}) cancel out, we only need to focus on the matter contribution to the entropy. 
%
%
 The matter region we should consider are shown in figure \ref{fig:ParadoxResolution}. 
 In particular, the cases when we have islands give 
\be \la{resol1}
	S (A_p \cup \bar{A}_p \cup A_c )  =  \textrm{Area terms} + S_{\textrm{bulk}} (A_c \cup I_c ) + S_{\textrm{bulk}} (A_p \cup \bar{A}_p \cup I_p ) ,
\ee
\be \la{resol2}
 S(A_p \cup  A_c) =  \textrm{Area terms} + S_{\textrm{bulk}} (A_c \cup I_c ) + S_{\textrm{bulk}} (A_p \cup I_p ) ,
\ee

For the other two cases we have no islands and we have 
\be\la{resol3}
S (A_p) = S_{\textrm{bulk}} (A_p), 
\ee
\be \la{resol4}
S (A_p \cup \bar{A}_p) = S_{\textrm{bulk}} (A_p \cup \bar{A}_p).
\ee
Note that $S_{\textrm{bulk}}(A_p \cup \bar{A}_p)$ is nonzero because the semiclassical state is thermal, and it is actually infinite in the current case since we have a non-compact spatial direction. For the sake of the argument here, we can imagine the spatial direction being long but finite, which is the case we will study in detail in section \ref{sec:finiteconn}.

Combining (\ref{resol1}) - (\ref{resol4}), we find that the strong subaddivity of the fine grained entropy
\be 
	S(A_p \cup  A) + S (A_p \cup \bar{A}_p)   \geq S (A_p \cup \bar{A}_p \cup A ) + S (A_p)  
\ee
is guaranteed by
\be 
	S_{\textrm{bulk}} (A_p \cup I_p )  + S_{\textrm{bulk}}(A_p \cup \bar{A}_p ) \geq S_{\textrm{bulk}} (A_p \cup \bar{A}_p \cup I_p) + S_{\textrm{bulk}} (A_p),
\ee
which is a strong subadditivity inequality for the probe field in the $\ket{TFD}$ state. Thus we see that the strong subadditivity paradox is resolved in this case. The key point is that the matter computation involves regions which can be viewed as living in the matter state produced by euclidean evolution on the half cylinder. Therefore strong subadditivity should be obeyed since this is a reasonable matter computation. This happens because the quantum extremal surface lies at a $Z_2$ symmetric point, a moment of time reflection symmetry, which makes the analytic continuation to Lorentzian signature straightforward\footnote{One might wonder whether $Z_2$ non-symmetric quantum extremal surfaces exist.  We can check that they do not exist in this model when we consider the cases for general $\tau$ in sec. \ref{sec:checkparadox}.}. 
%
%
This is however not true for the solution in section \ref{sec:naive}, in which the physical interpretation of the bulk entropy term is more obscure. 

\subsection{Compact spatial slice} \la{sec:Compact}

In this section we consider  a compact  spatial direction of length $L$. This subsection can be skipped by a reader in a hurry.   We will also take the amount of Euclidean evolution $\tau$ in the flat space to be general. The focus will be to understand how the two different solutions discussed in section \ref{sec:naive} and \ref{sec:BraKet} generalize and compete. We will restrict our attention to the cases where $L\gg \phi_r/c$ and $\tau\ll L$ where we can construct the solution explicitly.

\subsubsection{The disconnected solution}\la{sec:finitedisc}

\begin{figure}[h]
\begin{center}
\includegraphics[scale=0.23]{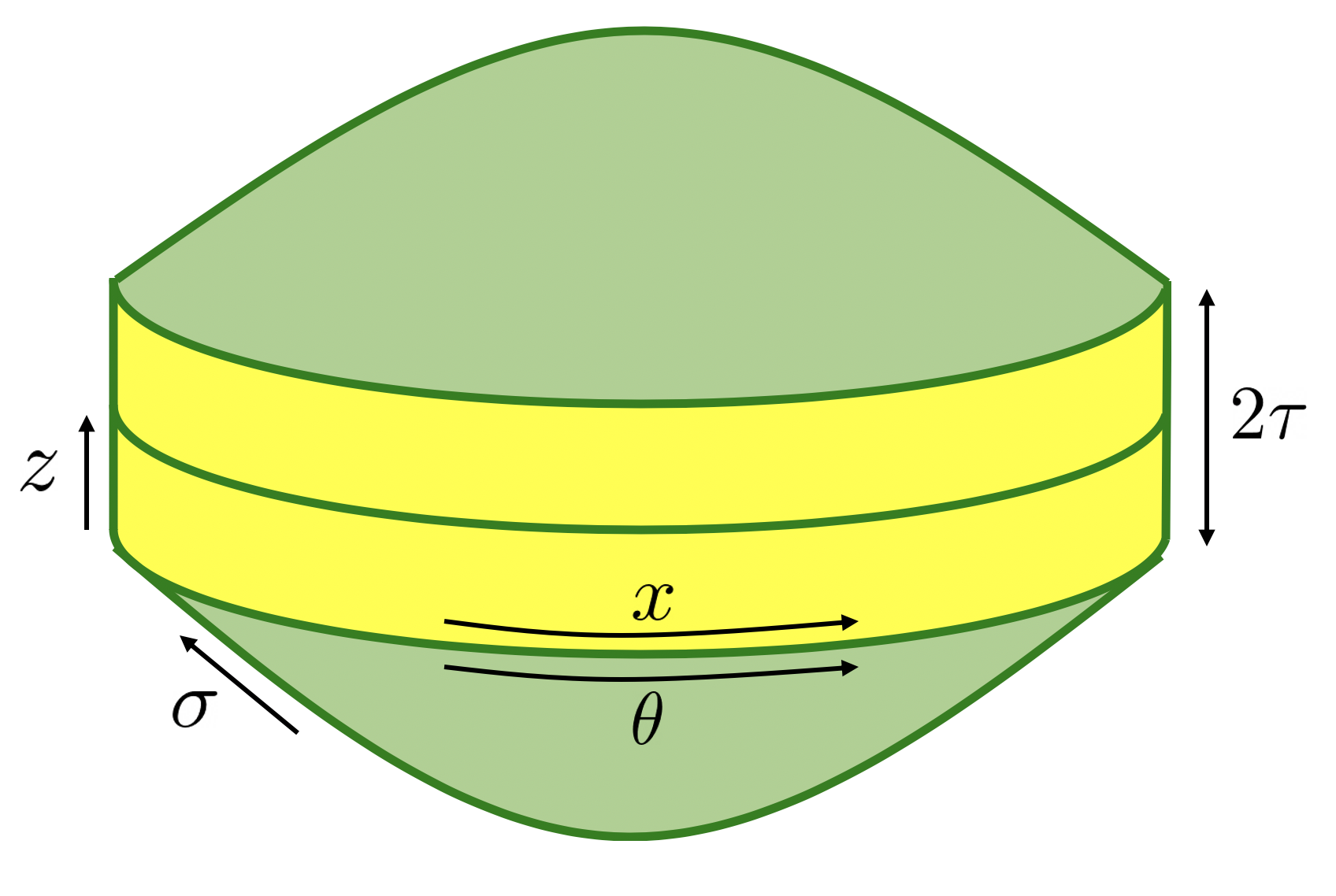}
\caption{We can have a disconnected solution whose gravity region involves two disks. }
\label{fig:FiniteDisconnect}
\end{center}
\end{figure}

 The metric in the flat space region is 
\be \la{MetrCompFlat}
 ds^2 = \frac{dz^2 + dx^2}{\epsilon^2} , ~~~~~~\quad 0 \leq z \leq 2 \tau,~~~~~~  x \sim x+ L.
\ee
We fill in the gravity region with a disconnected geometry involving two disks, each can be described by:
\be \la{MetrCompDisc}
  ds^2 = \frac{ d\sigma^2 + d\theta^2 }{\sinh^2 \sigma} , ~~~~ \, \phi = \frac{2\pi \phi_r}{L} \frac{1}{ \left( - \tanh \sigma\right)}, ~~~~\quad \sigma \leq -\sigma_c,\, ~~~\theta \sim \theta + 2\pi.
\ee
Since $\theta = 2\pi x/L$, the cutoff $\sigma_c$ is related to $\epsilon$ by $2\pi/\sigma_c = L/\epsilon$. In the semiclassical approximation, the matter CFT state at $z=0$ is just  the vacuum state $\ket{0}$ on a cylinder, $|B\rangle_s = |0\rangle$. The length of the cylinder in the flat space region is $\Delta z = 2\tau$. The configuration is shown in figure \ref{fig:FiniteDisconnect}.

We can now compute the partition function of this solution $Z_{\textrm{disc}} = Z_{\rm grav} Z_{\rm CFT}$. We can write the metric of each of the hyperbolic disks as $ds^2 = e^{2\omega}d\hat{s}_{\rm flat}^2$, 
and compute the partition function of matter by using the conformal anomaly 
\be \la{Anomaly}
 Z_{\rm CFT, \textrm{disk}}^{\textrm{anomaly}} = \exp\left[ \frac{c}{24\pi}  \left(\int d^2 x \sqrt{\hat{g}} (\partial \omega)^2 + 2 \int_{\partial} dy \sqrt{\hat{\gamma}} \hat{K} \omega \right) \right].
\ee
But this only gives a (divergent) term  proportional to the boundary length and a term  that can be absorbed into the definition of $S_0$. Apart from the contribution from the two disks, we also have an extra contribution due to the propagation of the vacuum state across the cylinder in the flat space region. The energy of the CFT vacuum state on a cylinder with circumference $L$ is $- \frac{c}{12} \frac{2\pi}{L}$, and the length of the flat space cylinder is $2\tau$. Thus we have
\be 
	Z_{\textrm{CFT},\textrm{flat cylinder} } = \exp \left[ \frac{c}{12} \frac{4\pi \tau}{L}\right].
\ee
Then the full disconnected partition function is
\be \la{pardisconn}
 Z_{\rm disc} =  Z_{\rm grav}	Z_{\textrm{CFT},\textrm{flat cylinder} }  = \exp\left( 2S_0 + \frac{2\pi \phi_r}{L}  + \frac{\pi c \tau }{3L}\right),
\ee
where we also removed a term proportional to the boundary length.

\subsubsection{Connected solution}
\la{sec:finiteconn}

We now look at the connected wormhole geometry, which is described by
\be 
\label{metricWH}
ds^2 = \frac{d\sigma^2 + d\chi^2}{\sin^2 \sigma}, ~~~~\quad \sigma_c \leq \sigma \leq \pi-\sigma_c , ~~~~~~ \, \chi \sim \chi +b,~~~~~ b = { \sigma_c \over \epsilon } L .
\ee
\begin{figure}[h]
\begin{center}
\includegraphics[scale=0.13]{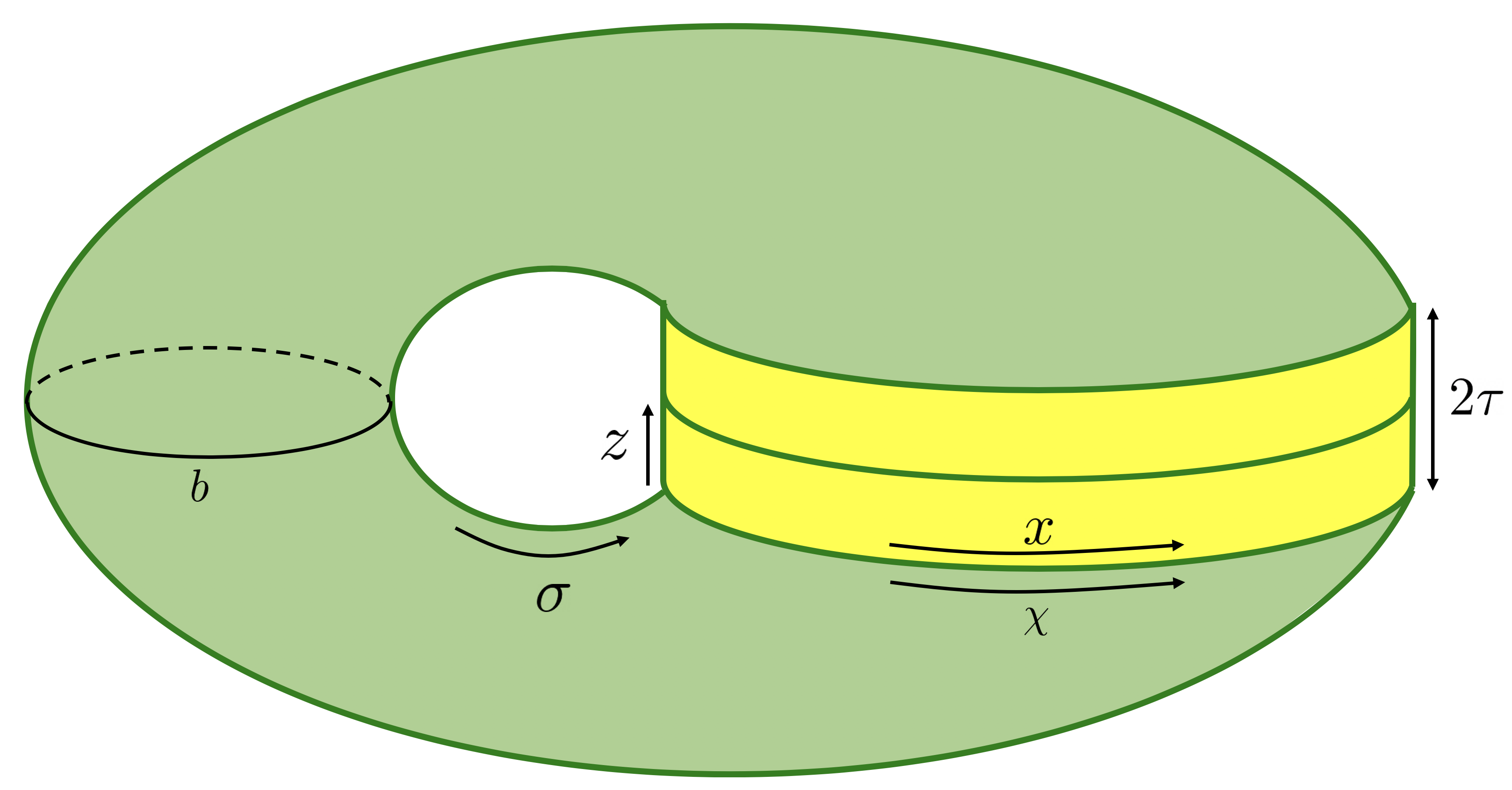}
\caption{An illustration of a bra-ket wormhole in the compact spatial direction case. The minimal geodesic length of the wormhole is given by $b$.}
\label{fig:FiniteWormhole}
\end{center}
\end{figure}
We have a family of geometries labeled by the parameter $b$, which is the length of the throat of the wormhole. The parameter $b$ will be fixed after we solved for the dilaton.  See figure \ref{fig:FiniteWormhole}.

We will first assume that $b\gg 1$, i.e. a very short wormhole, and then justify this assumption a posteriori. The pure gravity configuration also has a zero mode which corresponds to a relative shift, or twist,  between the two sides of the wormhole along its throat. When we have matter, this zero mode gets lifted because the  CFT partition function on a torus decreases when the twist is non-zero. When we have a very long spatial direction compared to the time direction, or in other words, $b\gg 1$, the integral over this mode is dominated by the saddle point where there is no twist, which in the semiclassical description gives us a simple thermal density matrix $\rho_s$ for the matter fields on the geometry. Actually, this integral over the twist enforces the momentum constraint, which does not allow momentum to flow through the wormhole. So when we say that $\rho_s$ is thermal we should always also say that we project onto zero momentum  states.   

\begin{figure}[h]
\begin{center}
\includegraphics[scale=0.25]{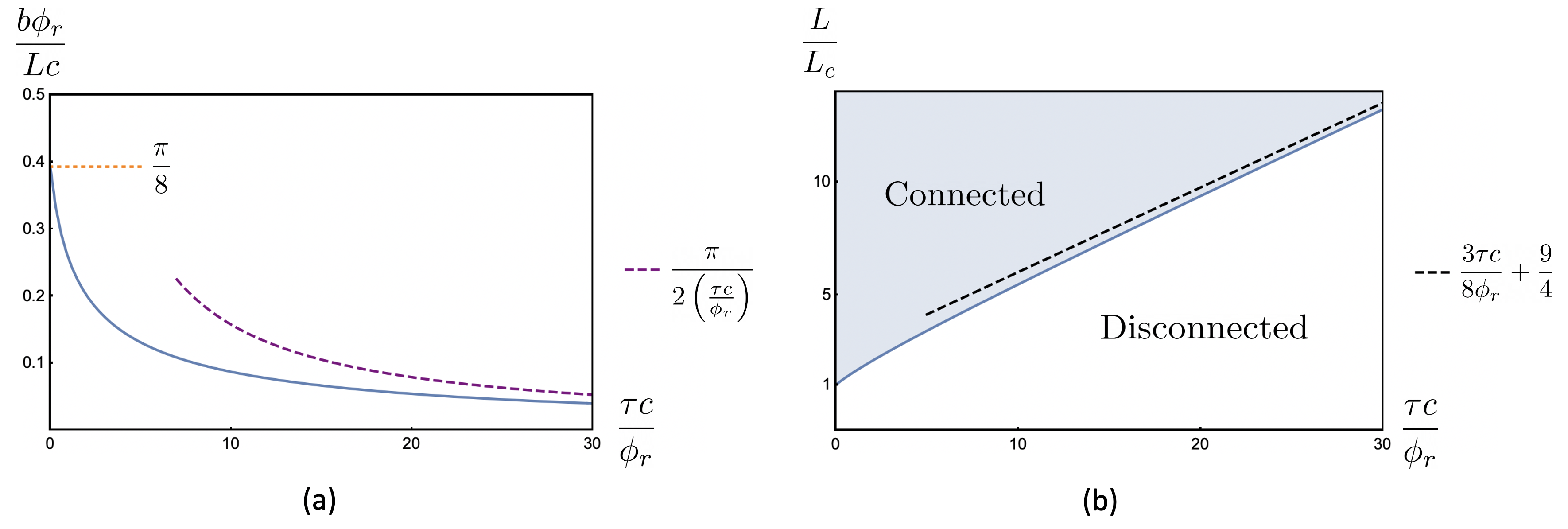}
\caption{(a) We plot the behavior of $\frac{b\phi_r}{L c}$ with respect to $\frac{\tau c}{\phi_r}$ coming from solving (\ref{eqnb}). We also plotted the two approximate expressions in (\ref{solnb1}) and (\ref{solnb2}). (b) We plot the minimal values of $\frac{L}{L_c}$ in order for the wormhole to dominate, as a function of $\frac{\tau c}{\phi_r}$. The blue shaded region is the wormhole phase. We also plotted the approximate expression in \nref{wormsubdom} when $\tau \gg \phi_r/c$. }  
\label{fig:Phases}
\end{center}
\end{figure}

For $b\gg 1$, the thermal stress tensor in the bulk can still be well approximated by that in (\ref{stressflat}), with $d$ replaced by the new length of the wormhole in the $\sigma$ coordinate, which is
\be \la{lengthworm}
	d = \pi  + 2\tau \times\frac{b}{L}. 
\ee 
The anomaly contribution to the stress tensor is the same as in (\ref{StressAnomaly}). Thus the total stress tensor is given by
\be 
	 T_{\sigma\sigma} = - T_{\chi\chi} = -\frac{c}{6\pi} \frac{1}{\left( 1+ \frac{2\tau b}{\pi L}\right)^2} + \frac{c}{24\pi} .
\ee
The dilaton is proportional to the negative energy in the bulk, and is given by 
\be 
\label{dilWH}
 \phi = \left[ \frac{c}{3 \left( 1+ \frac{2\tau b}{\pi L} \right)^2 } - \frac{c}{12}\right]\left( \frac{\frac{\pi}{2} -\sigma}{\tan \sigma} + 1\right).
\ee
Demanding $\phi_b = \phi_r/\epsilon$, and combining with $b/\sigma_c = L/\epsilon$, we find the following equation for $b$:
\be \la{eqnb}
	\frac{\phi_r b}{L} = \frac{\pi}{2} \left[ \frac{c}{3 \left( 1+ \frac{2\tau b}{\pi L} \right)^2 } - \frac{c}{12}\right].
\ee
This equation always has a solution for arbitrary $\tau$ (when we neglect the quantum mechanics of the boundary graviton). It is illuminating to look at the solution in two limits. The first limit is when $\tau$ is small compared to $\phi_r/c$
\be \la{solnb1}
 b \approx \frac{\pi c }{8 \phi_r} L, ~~~~\quad \textrm{for}~~~~~\,\, \tau \ll  \frac{\phi_r}{c}.
\ee
Thus we see that we recovered the rescaling relation derived in (\ref{chix}). The assumption that $b\gg 1$ is a posteriori justified in the regime $L \gg \phi_r/c$. The second limit is when $\tau$ is much greater than $\phi_r/c$
\be \la{solnb2}
 b \approx  \frac{\pi}{2 \tau} L, \quad ~~~~ \textrm{for}~~~~\,\,\tau \gg \frac{\phi_r}{c}.
\ee
Here the assumption $b\gg 1$ is justified when  $\tau \ll L$. Note that 
$b$ decreases as we increase $\tau$, namely the wormhole becomes thin under Euclidean evolution. Alternatively we can say that it becomes longer along the $\sigma $ direction\footnote{ These equations were discussed in a very similar setup in \cite{Maldacena:2018gjk}.}. 
This is consistent with the dual picture in which we are lowering the temperature of the state by performing Euclidean evolution.  We plot the behavior of the dimensionless parameter $\frac{b\phi_r}{L c}$ with respect to $\frac{\tau c}{\phi_r}$ in figure \ref{fig:Phases} (a), where we also superposed the approximate solutions (\ref{solnb1}) and (\ref{solnb2}).  
 
Although the wormhole always exist for any $\tau$, it does not mean that it is the dominant saddle. To compare it with the disconnected saddle, we need to compute the partition function of this solution $Z_{\textrm{con}}= Z_{\rm grav} Z_{\rm CFT}$. Since we have a wormhole geometry, we do not have the $e^{2S_0}$ factor in (\ref{pardisconn}), but we do get a contribution from the boundary term in (\ref{action})
\be \la{pargravWorm}
 	Z_{\rm grav} = \exp \left( -  \frac{\phi_r }{2\pi }\frac{b^2}{L}\right).
\ee
The matter partition function $Z_{\rm CFT}$ can be computed as a product of $Z_{\rm CFT}^{\textrm{flat}}$ and $Z_{\rm CFT}^{\textrm{anomaly}}$. $Z_{\rm CFT}^{\textrm{flat}}$ is the partition function of a flat torus $d\hat{s}^2 = d\sigma^2 + d\chi^2$ with length $d$ given in  (\ref{lengthworm}) and width $b$. Under the assumption $b\gg 1$, the partition function is dominated by the contribution of the ground state propagating along the $b$ direction. Since the ground state energy on a circle with length $d$ is given by $-c\pi/(6d)$, we have
\be \la{parflatworm}
 	Z_{\rm CFT}^{\textrm{flat}} \approx \exp \left(  \frac{\pi c}{6d}b \right) =  \exp \left(  \frac{  c }{6 } \frac{b}{1 + \frac{2\tau b}{\pi L}} \right) 
\ee
where we used (\ref{lengthworm}) in the second step. From \nref{Anomaly} with $e^{ 2 \omega} = 1/\sin^2 \sigma$,
, we compute the anomaly contribution 
	$ \log  Z_{\rm CFT}^{\textrm{anomaly}} = 
	- \frac{cb}{24} 
	$, 
where we dropped a (divergent) term proportional to the length. Of course this is the same term we dropped when we computed the two disks contribution. 
The final answer is then 
\be \la{Zcon}
	Z_{\textrm{con}} = \exp \left(  -  \frac{\phi_r }{2\pi }\frac{b^2}{L} +   \frac{  c }{6 } \frac{b}{1 + \frac{2\tau b}{\pi L}}  - \frac{cb}{24}   \right).
\ee
Extremizing $Z_{\textrm{con}}$ over $b$ gives  (\ref{eqnb}), which is another way to derive that equation. Inserting this value of $b$ in \nref{Zcon} we 
 get the final value for $Z_{\textrm{con}}$, which decreases as $\tau$ increases. 
 In two limits we get, 
 \bea \la{smalltaulim}
  Z_{\textrm{con}} &\approx &  \exp \left[   \frac{ \pi c^2 L }{128 \phi_r} \right], ~~~~{\rm for }~~~~~ \tau \ll { \phi_r \over c } ,
  \\ \la{Paronshell2}
   Z_{\textrm{con}} & \approx &  \exp \left[   \frac{ \pi c L }{48 \tau}  - \frac{ \pi \phi_r L}{8\tau^2} \right], ~~~~~~{\rm for }~~~~~  { \phi_r \over c } \ll \tau . 
 \eea
By comparing \nref{smalltaulim} with (\ref{pardisconn}) we find that the wormhole dominates for 
\be  \la{wormdominate}
  L \gtrsim L_c \equiv \frac{256 \phi_r}{\pi c} \frac{S_0}{c} ~,~~~~~~{\rm and}~ ~~~~~  \tau \ll { \phi_r \over c } 
\ee 
By comparing (\ref{Paronshell2}) with (\ref{pardisconn}), the wormhole dominates for  \be \la{wormsubdom}
 L \gtrsim { 96 \over \pi } \tau { S_0 \over c } + \frac{576\phi_r}{\pi c} \frac{S_0}{c}  =  L_c  \left( { 3 \tau  c \over  8 \phi_r }   + \frac{9}{4}\right) ~,~~~~~~~ { \phi_r \over c } \ll \tau 
 \ee
In general,  we can find that the wormhole always dominates for $L > L_c f(\tau)$, whose form is plotted in figure \ref{fig:Phases} (b).


In appendix \ref{app:Islands} we discuss the entropy of intervals and check that the wormholes starts dominating before the subaditivity paradox can arise.

\subsection{Purity of the state  } \la{sec:purity}

\begin{figure}[h]
\begin{center}
\includegraphics[scale=0.2]{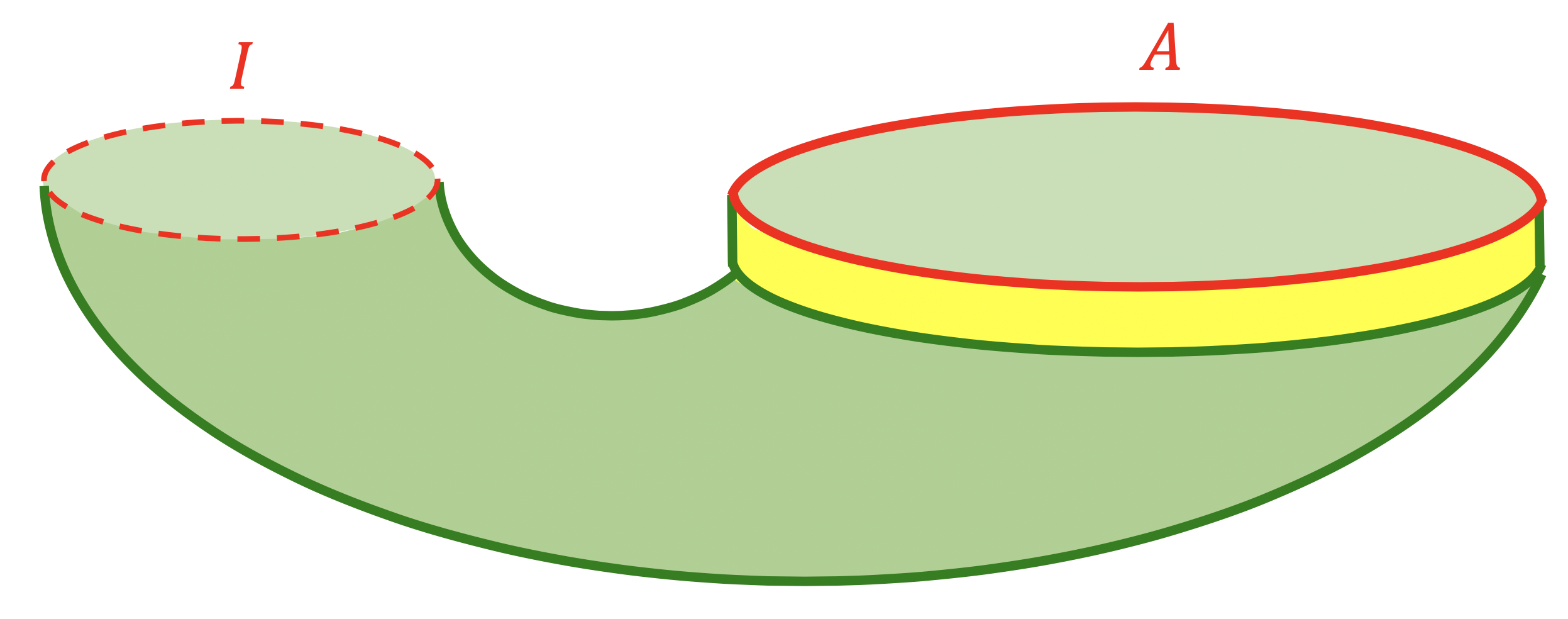}
\caption{When the region $A$ is the entire spatial slice in the flat space region, we have an island that covers the entire throat or the spatial section of the FLRW universe (see section \ref{sec:ClosedUniverse}).}  
\label{fig:Purity}
\end{center}
\end{figure}

In this section we explore the purity of the state prepared by the gravitational path integral. 
First we apply the gravitational fine grained entropy formula to the whole state. 
This is slightly  more clear in the case we have a long compact circle (but is also true in the non-compact case). Now $A$ is the  entire spatial slice   in the flat space region.  Then the island $I$  covers the entire throat of the wormhole,  see figure \ref{fig:Purity}. Since there are no extremal surfaces, the entropy is just the semiclassical entropy of $A \cup I$,
\be 
 S_{\textrm{island}} = S_{\textrm{bulk}} (A\cup I) = 0.
\ee
We get zero since $A$ and $I$ can be viewed  as the two sides of a pure $\ket{TFD}$ state prepared by half of the Euclidean evolution.

We have been proposing that our the gravitational path integral is computing $|B\rangle \langle B| $. In fact, more naively we would say that it computes a  density matrix $\sigma$, which would then give us expectation values of local operators on the strip via 
\be 
 \textrm{Tr} \left[ \sigma (\tau) O\right], \quad \sigma(\tau) \equiv e^{-\tau H_{CFT}} \sigma e^{-\tau H_{CFT}},
\ee
without assuming that $\sigma $ is a pure state. 

\begin{figure}[h]
\begin{center}
\includegraphics[scale=0.25]{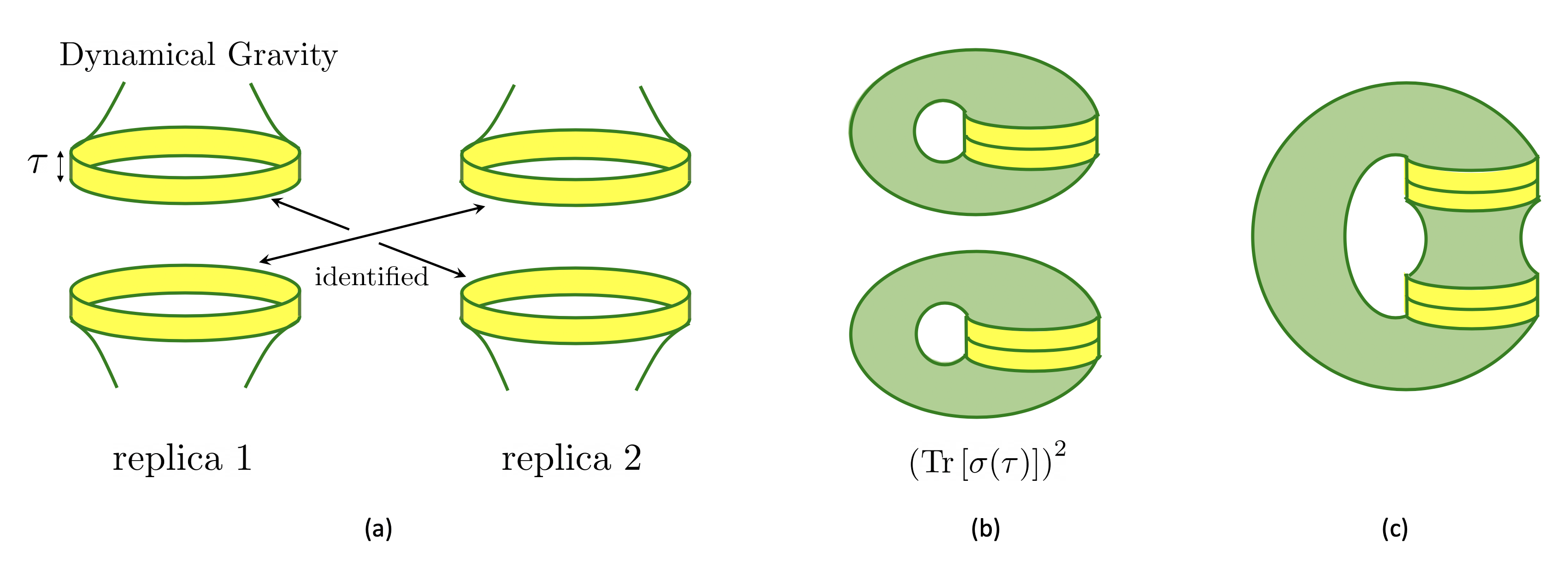}
\caption{(a) The gravity computation of  $\textrm{Tr} \left[ \sigma(\tau)^2 \right] $, where we fix the boundary conditions in the flat space region, while letting the dynamical gravity regions to connect in arbitrary ways. (b) A configuration that is the product of two wormholes. (c) A long wormhole configuration.}  
\label{fig:Purityreplica}
\end{center}
\end{figure}

We can check the purity of the state through a direct replica computation. 
We use gravity to compute the purity $\textrm{Tr} \left[ \sigma(\tau)^2 \right] $ of the (un-normalized) density matrix $\sigma (\tau)$.
In the exact description, the way we compute $\textrm{Tr} \left[ \sigma(\tau)^2 \right] $ would be to take two replicas of the density matrix, and identifying the bra/ket of the first replica to the ket/bra of the second replica, 
 and then letting the dynamical gravity regions to connect in various ways\footnote{For a critique of this procedure see \cite{Giddings:2020yes}.}, 
see figure \ref{fig:Purityreplica} (a).  

A natural way of connecting the dynamical gravity regions is depicted in figure \ref{fig:Purityreplica} (b), which is simply two copies of the wormhole that we discussed in section \ref{sec:finiteconn}. This simply gives us the square of $\textrm{Tr}\left[\sigma (\tau)\right]$\footnote{ When the wormhole is not the dominant saddle, we can replace the wormholes in \ref{fig:Purityreplica} (b) by the disconnected configuration in section. \ref{sec:finitedisc}. We also get $(\textrm{Tr}\left[\sigma (\tau)\right])^2$.}.
On the other hand, there is a different way of connecting the gravity regions as shown in figure \ref{fig:Purityreplica} (c). 
However, this is not an on-shell solution, since the length $d$ of this 
 is greater than $2\pi$ (in the conformally flat coordinate $\sigma$ of \nref{metricWH}), by (\ref{stressflat}) and (\ref{StressAnomaly}) and we see that we no longer have negative energy in the bulk.

Thus at the level of on-shell solutions, we find
\be 
 \textrm{Tr}\left[\sigma (\tau)^2 \right] =  \left(\textrm{Tr}\left[\sigma (\tau) \right] \right)^2
\ee
It remains a question whether off-shell configurations like the one in figure  \ref{fig:Purityreplica} (c) give corrections to the purity, and we make no statement on them here\footnote{ If they contribute off shell, and we have an ensemble,  purity is  restored after using an additional $n\to 0$ replica trick \cite{Engelhardt:2020qpv}.}.

 We comment on the relative entropy between $\sigma$ and the thermal state in appendix \ref{app:relative}.

\section{Lorentzian geometry and a collapsing FLRW cosmology} \la{sec:ClosedUniverse}


\begin{figure}[h]
\begin{center}
\includegraphics[scale=.18]{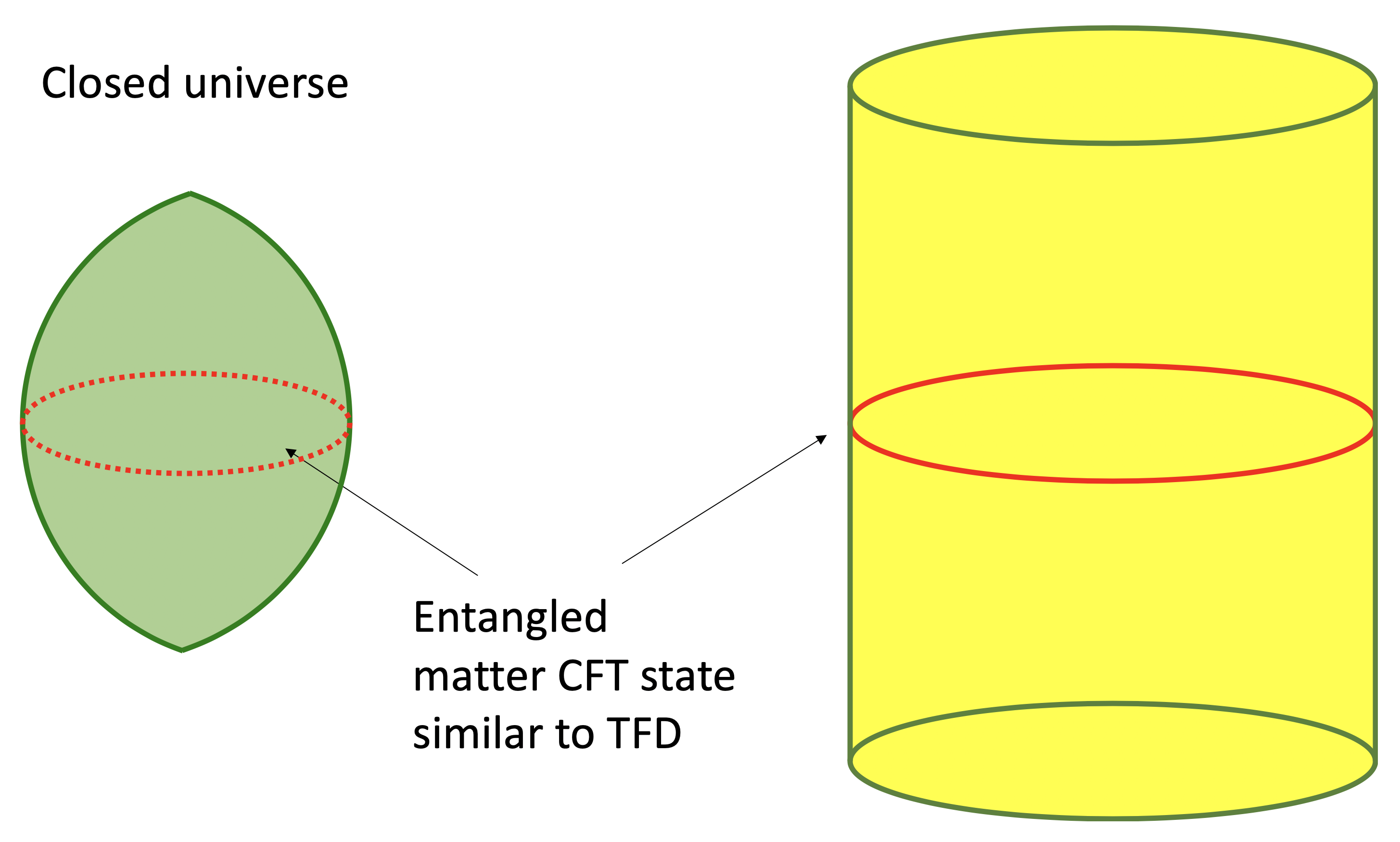}
\caption{Lorentzian continuation of the state prepared by the Euclidean evolution. On the right (yellow) we have a cylinder with no gravity. On the left (green) we have a initially expanding and then collapsing FLRW universe with dynamical gravity. The matter fields are in a thermofield double state, with each side on each of these two  spaces.  }
\label{LorentzianClosed}
\end{center}
\end{figure}

So far, our discussion was purely in Euclidean signature. However,   it is  natural  to consider the Lorentzian time evolution of the state $\ket{B(\tau)}$. We would like to understand how to describe it using gravity. For this purpose,  it is slightly more intuitive to focus on the compact case, assuming $L$ is large enough so that the bra-ket wormhole dominates. 

As we discussed previously, the full wormhole configuration has  a moment of time reflection symmetry which we can view as initial conditions for Lorentzian evolution. In Lorentzian time,  we get two {\rm disconnected} spacetimes, see figure \ref{LorentzianClosed}. We get a flat cylinder  with no gravity and a closed universe with gravity. Semiclassically, we get an initial matter state which is entangled between the two universes. Up to a conformal rescaling,   the matter state is the thermofield double state where one side lives in the cylinder with no gravity and the other in the closed universe. Semiclassically, if we look only at the cylinder,  we get a  thermal state. 
On the other hand, since this state is entangled with a state on the FLRW universe, the gravitational fine grained entropy formula tells us that we have a pure state on the cylinder in the  exact theory, as mentioned  in \cite{Almheiri:2019hni}. Conceptually, this is the same situation that happens when we have a fully evaporated black hole, except that here we have a very explicit and simple construction and we do not encounter the difficulties associated to the  endpoint of black hole evaporation. This also means that the entanglement wedge of the state on the cylinder includes all of the closed universe.

The metric and dilaton on the closed universe, for the case $\tau=0$,  is obtained by analytic continuation from 
\eqref{metricWH} and \eqref{dilWH}. 
\be \la{FLRW}
ds^2=\frac{-d \eta^2+d\chi^2}{\cosh^2\eta}\,,~~~~ \quad \phi=\frac{c}{4}\left(1-\eta \tanh  \eta \right)\, ~,~~~~{\rm for }~~ \tau=0
\ee
In addition, the matter at $\eta=0$ is in a thermal state at temperature $T_\eta = { 1 \over \pi }$. For $\tau> 0$,  we get the same metric, a rescaled dilaton \nref{dilWH} and a different temperature \nref{Temperature}. 
This geometry corresponds to an FLRW  (Friedmann-Lemaitre-Robertson-Walker) universe that emerges from a singularity at $ \eta=-\infty$, expands,   reaches the maximum size at $ \eta=0$, and then recollapses when $ \eta\to\infty$.
We see that the spatial direction shrinks as $\eta \to \pm \infty$, which corresponds to a finite proper time. This produces a geometric singularity if $\chi$ is compact. Even if $\chi$ were not compact, we get that the dilaton is going to $-\infty$, which is a strong coupling region, where fluctuations in the topology are enhanced,  and we cease to trust the semiclassical approximation. 

Since this whole closed universe is in the entanglement wedge of the matter in the cylinder, we could say that we are getting a holographic description for a closed universe cosmology!   Of course, the devil is in the details of entanglement wedge reconstruction. This could be explored   using either the Petz map \cite{Penington:2019kki} or the modular flow method \cite{Chen:2019iro}.

Note that the whole closed universe can be included in the entanglement wedge of an interval, $A$,  that does not cover the full circle around the cylinder.  This happens when the complement of the interval, $\bar A$, is small enough that the dominant contribution to its gravitational fine grained entropy formula comes from the naive semiclassical entropy of the interval. 
 If the cylinder size is very big, then  we can divide the circle into intervals for which their dominant contribution to the entropy will come from islands with  non-trivial quantum extremal surfaces, such as the one discussed in figure \ref{fig:WormIsland}. In this case, we see that the island portion of the closed universe is encoded inside the cylinder interval. Therefore we have an approximate notion of locality for the encoding of the FLRW region of the geometry. 

In \cite{Calabrese:2007mtj}, they  started with a  simple boundary state, after some euclidean time evolution  $|B(\tau)\rangle$, and then studied the Lorentzian time evolution of the entropy of a subregion.   For simplicity,  we   consider just the non-compact case and compute the entropy of a half line. For generic boundary states this is expected to be a finite, but a linearly rising function of time.  In order to compute this, it is useful to first compute the entropy for a situation where we consider the region $A$  located in Euclidean time, at time $\tau'$ with $ 0 < \tau' < \tau$, where $\tau$ is the Euclidean time that we use to construct the state $|B(\tau) \rangle$. Then, after we do the computation, we  set 
$\tau' = \tau \pm i t $ to do the computation at Lorentzian time $t$.   Tracking the quantum extremal surface through this procedure,    we find that it starts at $\eta_*=0$  for $t=0$ and for positive times $t$ it now moves in the negative $\eta$ direction. However, even for large $t$ it does not significantly move in the $\eta$ direction, and furthermore $\eta_*$ remains bounded as $t \to \infty$, for details see appendix \ref{app:IntervalLorentzian}. It is simple to state the basic reason. If we had just the TFD state, and we pick some time $t$ on one of the sides, then on the other side we would want pick the time $\eta =-t$ to minimize the entropy. However, we have two extra effects in the FRLW region:   the conformal factor and the dilaton terms in \nref{FLRW}. Both of them prefer  $\eta =0$, and this preference dominates (at large $\eta$). It is for this reason that $\eta_*$ does not go to large values.  This means that the qualitative form of the entropy is similar to what we would get for the thermofield double when we consider a half space on both sides of the thermofield double (as in \cite{Hartman:2013qma}), giving a linearly rising function of time with precisely the same coefficient that we would find for the thermofield double,   $S \sim  S_0 + \# c +  { c \over 3 } \pi T_t  t $.

\subsection{The doubly holographic setup}
\la{DoublyHolographic}

To gain some further intuition, we can consider the ``doubly holographic" set up introduced in \cite{Almheiri:2019hni}.  Namely the conformal fields are holographic by themselves so that they have a three dimensional dual. In this case, we have a further description of the system as a gravity theory in AdS$_3$. Our flat space region lives on the rigid boundary of the AdS$_3$ space, while the dynamical gravity region of the 2d description corresponds to a dynamical Planck brane extending from the rigid boundary into the bulk of AdS$_3$. 

\begin{figure}[h]
\begin{center}
\includegraphics[scale=.24]{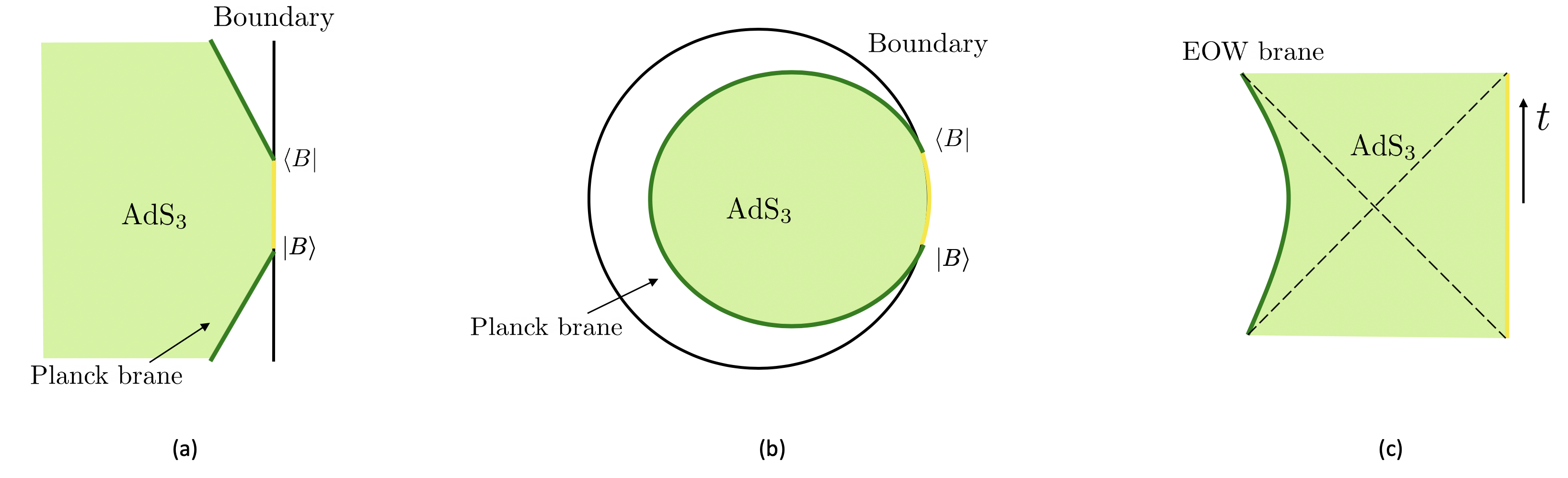}
\caption{We consider a doubly holographic set up, where the flat space region (yellow) is a segment of the AdS$_3$ boundary, while the 2d dynamical gravity regions (green) correspond to Planck branes in the 3d bulk (shaded in light green). The spatial direction is suppressed in the figure, which can be either infinite or compact and is perpendicular to the page. (a) The disconnected configuration for the Planck branes. (b) The connected configuration. (c) We can continue the connected configuration into Lorentzian signature, and the Planck brane becomes an end of the world brane behind the horizon (denoted by the dashed lines).}
\label{fig:DoublyHolo}
\end{center}
\end{figure}

When we consider $\braket{B(\tau)|B(\tau)}$, we have two Planck branes extending from the boundary into the bulk. Naively the two branes are disconnected, see figure \ref{fig:DoublyHolo} (a), and this corresponds to the disconnected saddles we have discussed in sections \ref{sec:naive} and \ref{sec:finitedisc}. However, since the Planck branes are dynamical objects in the 3d theory, we should also allow the possibility that they are connected in the bulk, see figure \ref{fig:DoublyHolo} (b). This would correspond to the connected saddles we studied in sec. \ref{sec:BraKet} and \ref{sec:finiteconn}. We can view the Planck branes as branes created by the boundary conditions of the boundary CFT. The fact that they can be connected was discussed in  \cite{Takayanagi:2011zk,Fujita:2011fp}. The new ingredient in our discussion is the further completely 2d interpretation of such branes. 
The fact that we should allow such connected brane configurations gives us another piece of evidence that  bra-ket wormholes should be included.  
Similar constructions were discussed in  \cite{Cooper:2018cmb,Simidzija:2020ukv,Akal:2020wfl,Antonini:2019qkt}.

Notice that in the Lorentzian geometry of figure \ref{fig:DoublyHolo} (c) the cylinder where the CFT lives is the yellow boundary on the right and the FLRW universe is the green line on the left. The entanglement  is represented as a  geometric connection between these two boundaries. The boundary on the left side is a dynamical gravity configuration and is viewed as part of the state in the CFT. In other words, the state in the CFT is a particular pure state of a black hole (or black brane). 

An advantage of the doubly holographic set up is that,  to  leading order  in $c\gg 1$, the entropy can be computed geometrically in the 3d bulk through the RT/HRT formula. Since the matter fields are holographic, the entropy of an interval $A$ in the semiclassical state can be evaluated by the area of an extremal surface that is homologous to $A$. (This surface is a line in 3d).  The surface can also land on the Planck branes, and when it does, we get an extra area term from the brane that is the value of the dilaton $\phi$ at that point. The prescription gives the same results as our computations in 2d, but it provides a simple geometrical viewpoint of the computation.

\begin{figure}[h]
\begin{center}
\includegraphics[scale=.25]{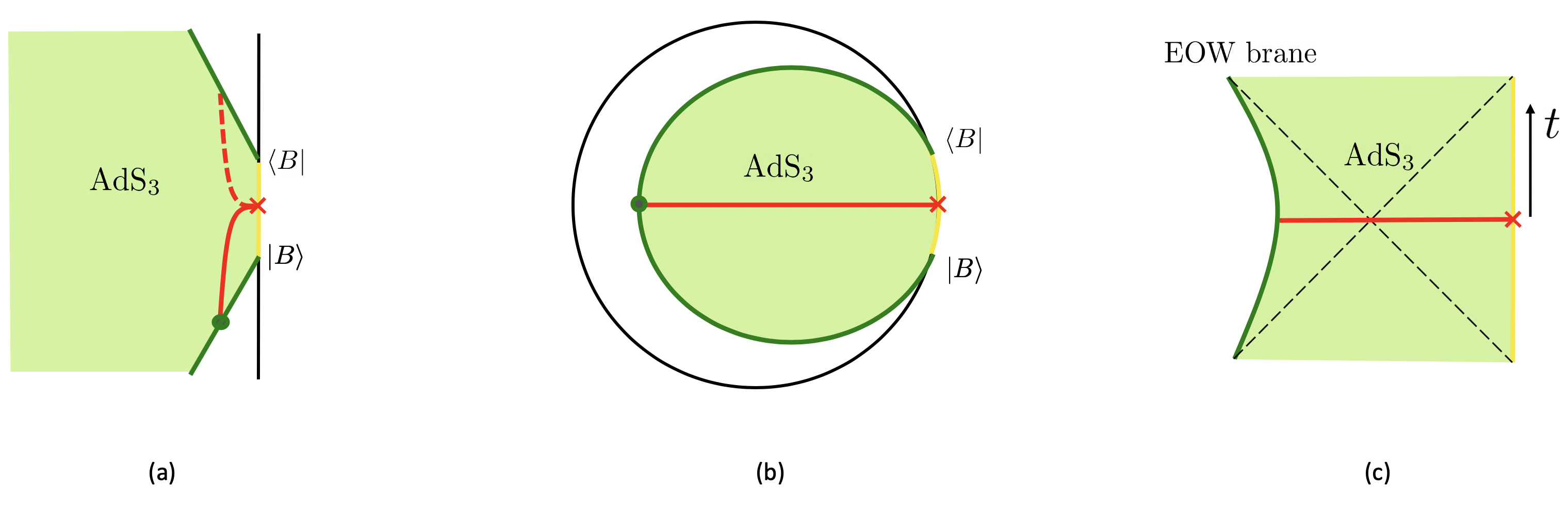}
\caption{(a) In the disconnected geometry, the extremal surface can land on either one of the branes (denoted by the solid and dashed lines), resulting in a $Z_2$ symmetry breaking. (b) In the connected geometry, the extremal surface lands on the symmetric point of the brane and the $Z_2$ symmetry is preserved. (c) When we continue into Lorentzian signature, the entanglement wedge of the boundary includes the entire bulk region to the end of the world brane behind the horizon.
 }
\label{fig:DoublyRT}
\end{center}
\end{figure}

For simplicity, let us consider the case that we have an infinite spatial direction, and the region $A$ is half of the infinite line. Then the only place the extremal surface can land on is on the Planck brane. In the disconnected case, the extremal surface can land on either one of the branes, and the $Z_2$ symmetry is spontaneously broken,  see figure \ref{fig:DoublyRT} (a). In the connected case, the $Z_2$ symmetry is preserved by the solution since the extremal surface lands on the reflection symmetric point of the brane, see figure \ref{fig:DoublyRT} (b). 

In the situation where the Planck branes are connected in the bulk, we can slice the system open around the $Z_2$ symmetric point and continue it into Lorentzian signature, see figure \ref{fig:DoublyHolo} (c). In the Lorentzian picture, the Planck brane is completely detached from the boundary. From the bulk point of view it is an   ``end of the world brane" (EOW brane). 
The entanglement wedge of the boundary includes the whole space up to this end of the world brane. Note that the end of the world brane is behind the horizon, so naively an observer on the boundary can never send a signal to the brane. However, by the idea of entanglement wedge reconstruction, one can in principle communicate with the brane by doing operations that are sufficiently complex.

However, for the disconnected case,   see figure \ref{fig:DoublyRT} (a), if we do the same continuation to Lorentzian time, we get an empty AdS$_3$ space in the bulk. Thus if we compute the entropy of the half infinite line $A$ in the Lorentzian signature, the solution of the extremal surface will be necessarily complex so that it can reach and land on the Planck brane. In particular, this does not lead to a clear interpretation of the entanglement wedge. Fortunately, whenever we have islands,  the connected configuration always dominates over the disconnected one, so we should always use the connected configuration to compute entropy. 

 We comment on the relation between the extra third dimension and the tensor networks in appendix \ref{app:TensorNetworks}.

\section{Euclidean wormholes after projecting on to a typical state} 
\label{sec:Typical}
In section \ref{sec:setup}, we started by saying that the state $\ket{B}$ is defined by performing the Euclidean gravitational path integral with fixed metric and dilaton boundary conditions. The state defined in this way is usually referred to as the Hartle-Hawking state $\ket{HH}$ or no-boundary state, as one sums over all no-boundary manifolds in the gravitational path integral. 
 Later in the discussion, we have argued that when we compute quantities of the form 
 \be \la{overlap}
 \bra{B(\tau)}O\ket{B(\tau)}
 \ee  we should include geometries with different topologies, namely bra-ket wormholes, see figure \ref{fig:FiniteWormhole}. 
From the point of view of holography, 
 this is not too surprising since computing \nref{overlap} is analogous to introducing some interaction between the two quantum systems that describe each of the two sides. In fact, this setup is basically the same as the one considered in  \cite{Maldacena:2018lmt} (based   on \cite{Gao:2016bin}).  
 
 The computation of \nref{overlap} can be viewed as a two step process, first computing the density matrix $\rho = |B\rangle \langle B| $ and then taking the trace $\textrm{Tr}[ \rho O ]$. One could wonder whether the wormholes should be included in the computation of this density matrix, or particular elements of the density matrix. In fact,  \cite{Page:1986vw} gave a prescription for computing the density matrix of the universe which includes such wormholes. As \cite{Anous:2020lka} emphasized, this is different than the Hartle Hawking prescription, which would not   allow such connections. 
 Here we have only argued that bra-ket wormholes should be included if we trace over some of the degrees of freedom.  If we trace over all the field theory degrees of freedom, we have a setup similar to the traversable wormholes of \cite{Maldacena:2018lmt}.  We need to include them to solve the entropy subaditivity paradox\footnote{ Of course, in this discussion we are assuming that the gravitational fine grained entropy formula is correct, since the paradox arose under that assumption.}.  
  
 The discussion of whether one should include wormholes is also related to the factorization puzzle in AdS/CFT \cite{Maldacena:2004rf}. Suppose one  computes the product of two partition functions $Z(\beta_1)\times Z(\beta_2)$ of two holographic CFTs using gravity. If we include wormholes connecting the two asymptotic boundaries,  we violate factorization explicitly, unless the wormholes can be cancelled in some miraculous way. Note, however,  that all of our discussion so far does not violate factorization explicitly, since we  considered quantities like $\bra{B(\tau)}O\ket{B(\tau)}$ , which are not factorized in the dual description. 
In fact, our wormhole appears to be much smarter than just not causing puzzles, it solves the strong subadditivity paradox we found, and also respects the purity of the state, as long as one applies the gravitational fine grained entropy formula.

\begin{figure}[h]
\begin{center}
\includegraphics[scale=0.25]{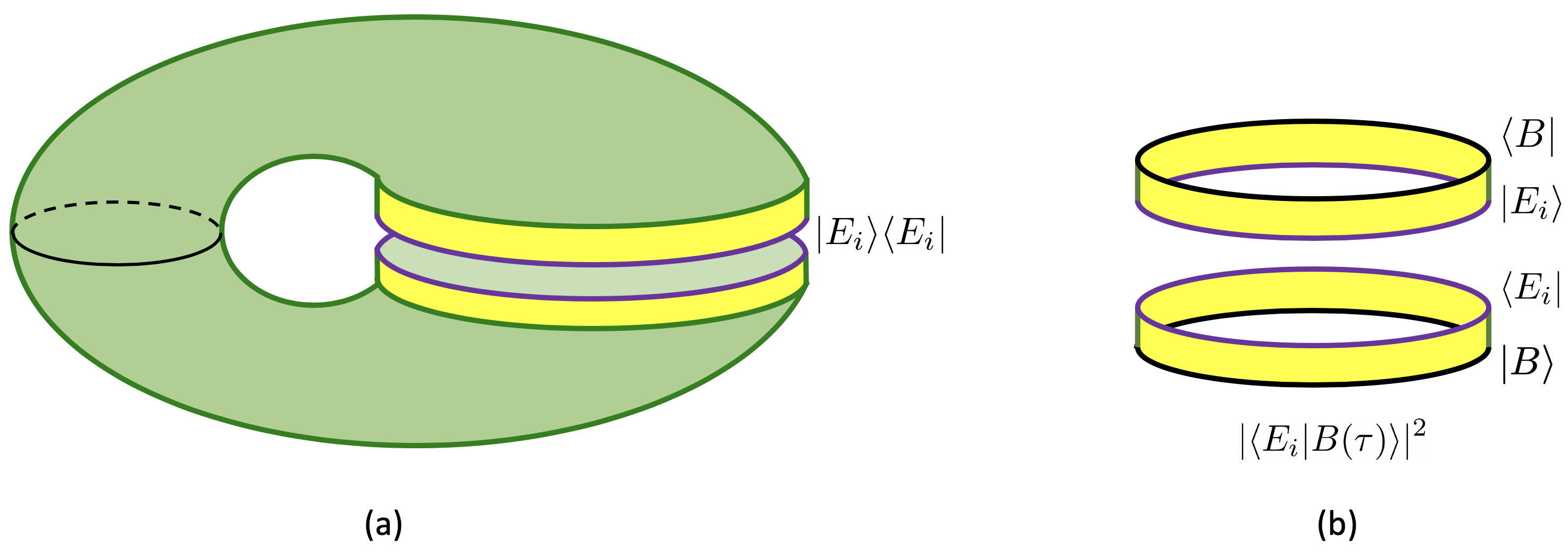}
\caption{Projecting on to a typical energy eigenstate in the (yellow) non-gravitational region. (a) We find the same type of bra-ket wormhole solution. (b) The dual interpretation, we have a boundary state, which we evolve for a time $\tau $ in Euclidean time and then we overlap it with the energy eigenstate $|E_i\rangle$. Independently, we have a similar computation with the bra. }
\label{fig:TypicalState}
\end{center}
\end{figure}
However, we  can change the problem in a way that the factorization puzzle starts to emerge. This discussion was inspired by  \cite{Stanford:2020wkf}.  As we have discussed in section \ref{sec:finiteconn}, at the semiclassical level, the wormhole gives us a thermal state $\rho_s$ on the geometry \nref{Temperature}. In the non-gravitational flat space region, we can now  insert a projection operator onto a typical state $\ket{E_i}$, with the same energy as the one determined by $\rho_s$, which is just the energy of the CFT at temperature $T_x$ \nref{Temperature}. This is an energy and momentum eigenstate (with zero momentum) which implies that  the stress tensor is constant. Therefore we have the same equations and solutions for the wormhole that we had previously, see figure  \ref{fig:TypicalState} (a). 
In the dual description, inserting a projector means we are computing 
 \be \la{ovel}
 |\braket{E_i | B(\tau) }|^2~,
 \ee 
 which factorizes into the product of the amplitude $\braket{E_i | B(\tau)}$ and its complex conjugate, see figure \ref{fig:TypicalState} (b). On the other hand, the gravity computation is not manifestly factorized if we include wormholes as in figure \ref{fig:TypicalState} (a), and thus we do have a factorization puzzle when we consider typical states. 
 As in \cite{Penington:2019kki},  the puzzle becomes more clear if we consider two orthogonal states $|E_i\rangle$ and $|E_j \rangle$, then we cannot use the connected wormhole to compute    $\langle B |E_i \rangle \langle E_j |B\rangle $. On the other hand, we expect that it should be  a random variable with a modulus with an average value given by  \nref{ovel} and also containing a phase  which we do not know how to calculate. 
One important lesson of ER=EPR, and of traversable wormholes,  is that the procedure involved in preparing the state creates a spacetime itself, and that spacetime could be connected with the region we are focusing on. As a simple example, we discuss in appendix \ref{Bmicro} the case where the state which we project on is $|B\rangle $ itself. In that particular example, we find that the factorization puzzle is avoided at the level of on-shell solutions.


 One option is that this computation inherently involves an average over couplings.  This average over couplings has been demonstrated in  pure JT gravity \cite{Saad:2019lba}.\footnote{If we take a single member from the ensemble, then the wormhole only gives a good description for quantities that are self-averaging. The same lesson can be possibly extended to theories that are not a member of the ensemble. When we compute quantities as $\bra{B(\tau)}O\ket{B(\tau)}$, the fact that we sum over all the intermediate states in the middle suggests that the quantity does not depend on the fine detail of the parameters in the theory, which can be an explanation of why they can be computed by a wormhole. }
  A different possibility is that, in the complete theory, one has the freedom to choose whether to include wormholes or not, and different choices lead to the same final result \cite{Marolf:2020xie}.  Notice that in JT gravity with matter the sum over topologies is not well defined due to divergences from very thin wormholes. This suggests that perhaps the full UV theory will be involved in proving factorization when there is a puzzle.

 The energy eigenstates  are highly non-local along the $x$ direction, and therefore $\langle E_n |B\rangle$ does not have a simple dual interpretation as a quantum system evolving along the Euclidean $x$ direction. Nevertheless, it is possible to consider a slightly different setup where such an interpretation is simpler. 
 Let us instead assume that the matter CFT consists of $c$  independent massless scalar fields with a compact circle target space\footnote{This last condition is imposed so that we do not have to worry about complications from the non-compact target space.}. In addition, let us assume that $\tau=0$.  We can now choose a boundary condition for the scalar fields of the form $\varphi |_{\rm Bdy} = \varphi^i_0(x)$, $i=1, \cdots, c$. This computes 
 \be \la{partvc} 
  Z[\varphi_0^i(x)] = \langle B| \varphi^i_0(x) \rangle ~,
 \ee 
  which, by viewing $x$ as Euclidean time,  is expected to give us the partition function of a quantum mechanical system with (Euclidean) time dependent couplings, set by $\varphi_0^i(x)$.  
  Taking the square of \nref{partvc} we expect a factorized answer. Nevertheless, we get a wormhole that connects the two copies as a possible gravity solution, as long as the profile is sufficiently typical that the stress tensor is similar to what we had before. After taking the square of \nref{partvc} and integrating over $\varphi^i_0(x)$ we effectively enact a transparent boundary condition and connect the matter fields through the $AdS$ boundaries. So we see that the connected wormhole solution for the case with connected boundaries is intimately related to the wormhole solution when the boundaries are not connected but have correlated couplings that depend on Euclidean time.  Similar examples of Euclidean wormholes with Euclidean-time dependent couplings   will be discussed in \cite{Marolf:2021kjc}.

\section{A two-dimensional model of inflation and bra-ket wormholes} \la{sec:deSitter}

In this section,  we consider a toy model for inflationary cosmology. It is a two dimensional model with a period of de Sitter evolution followed by a flat space region. Furthermore, we imagine that gravitational effects are important during the de Sitter phase, but can be neglected during the flat space region.  
 
 \begin{figure}[h]
\begin{center}
\includegraphics[scale=.3]{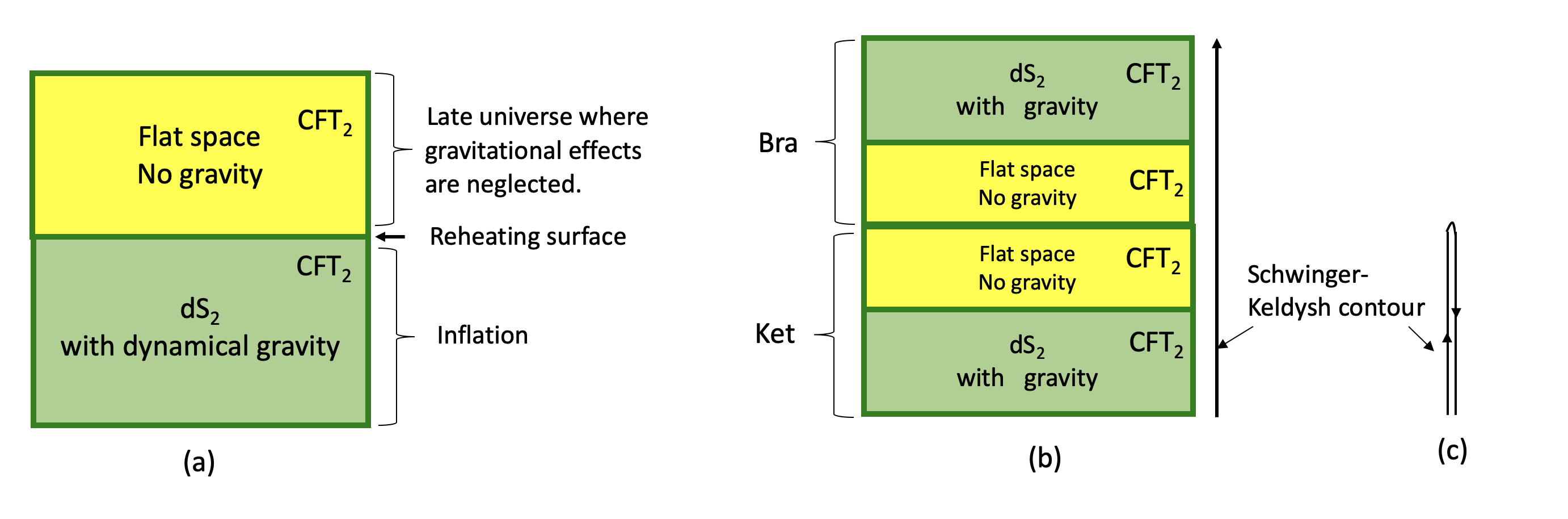}
\caption{ (a) We study a model which consists of de Sitter gravitational evolution followed by an evolution in flat space with no gravity. (b) When we consider expectation values, we join the bra and the ket following the Schwinger-Keldysh contour. In (c) we see a more traditional depiction of the Schwinger-Keldysh evolution. We go forwards in time to prepare the bra, then go backwards in time to prepare the ket.  }
\label{dSSetup}
\end{center}
\end{figure}
More concretely, we study the following model.  We 
consider a $dS_2$ version of JT gravity coupled to  matter that is described by a CFT, see figure \ref{dSSetup}.  At late times we join this to a flat space theory with no gravity. This could be an approximation to a full model of two-dimensional gravity where we interpolate between $dS_2$ JT gravity and flat space JT gravity, but where the dilaton becomes extremely large in the flat space region, so that we can neglect the effects of gravity there. In two dimensions,  we can achieve this by changing the ``potential'' from linear, which gives de Sitter, to constant which gives flat space. If this change happens  at a very large value of $\phi$, we can neglect quantum gravity effects in the flat space region.


The (Lorentzian) action for this model is 
\be \la{actiondS}
I^{(dS)} = { S_0 \over 4 \pi } \left[ \int R + 2 \int K \right] + { 1 \over 4 \pi }\left[  \int \phi (R-2) 
+ 2 \phi_b \int K \right] + I_{CFT} \,.
\ee 
where the action for the CFT is defined both in the de Sitter region as well as in the flat space region. The standard evolution of the conformal fields through the transition between de Sitter and flat space gives us transparent boundary conditions for the conformal fields. In other words the fields evolve continuously from the de Sitter region with gravity to the flat space region with no gravity. The future boundary of de Sitter, which is the location of the ``reheating'' surface, is set where $\phi = \phi_b$, with $\phi_b$ a fixed large value, which we will typically scale as $\phi_b = \phi_r/\epsilon$, keeping $\phi_r$ fixed. Of course, in the real world we do not turn off gravity completely after reheating. Our objective is to define a model that is as simple to analyze as possible, leaving more realistic models for the future\footnote{ In higher dimensions, there are propagating gravitons, so when we talk about turning off gravity we really mean that we select a background spacetime, and we consider gravitons propagating on this fixed background as perturbative fluctuations, including them as one type of ``matter'' field. }.  

Note that we are fixing the geometry of the future, in particular, we are assuming we have a single large connected universe, and not many disconnected ones. If the spatial direction is compact, we will be fixing its length. We can view all these conditions as a certain projection condition on  the most general state resulting from inflation, which could contain disconnected universes. We are not fixing the state of the CFT, this is determined by the gravitational dynamics. We expect to have a well defined state in the CFT, in the flat space region, after all these projections. 

The advantage of turning off gravity in the future is that we can define precise quantum states and we can talk about the von-Neumann entropy of subregions. We can view the whole de Sitter evolution as producing a state $|C\rangle$, which is an initial state for the Lorentzian evolution in the flat space region. We use the letter $C$ for Cosmological.  We will make no assumptions on whether it can be  computed in terms of a quantum system residing at the boundary, which would be the case if some version of dS/CFT was true \cite{Strominger:2001pn,Witten:2001kn,Maldacena:2002vr}. 
 However, the formulas we will find are essentially consistent with that idea. Our main approach is to use  
 the gravitational fine grained entropy formula \cite{Ryu:2006bv,Hubeny:2007xt,Faulkner:2013ana,Engelhardt:2014gca} to calculate subregion  entropies using the semiclassical state   in order to  understand properties of the exact state. 
 
 This setup is very similar to that considered in section \ref{sec:setup}. In both cases we prepare a state in the flat space region. In section \ref{sec:setup} the state was prepared by Euclidean $AdS_2$ evolution. Here it is prepared by Lorentzian $dS_2$ evolution.  This Lorentzian evolution relies on a choice of state and we will discuss some possible choices below.

 \begin{figure}[h]
\begin{center}
\includegraphics[scale=.25]{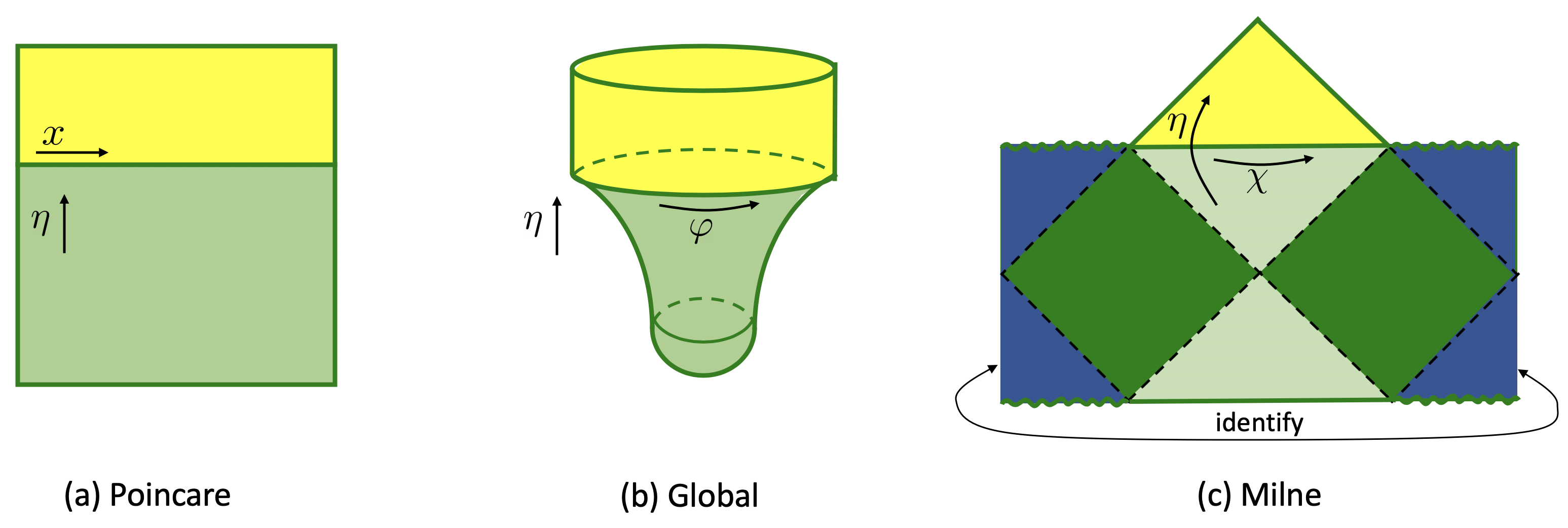}
\caption{We display the various cases we considered. They are distinguished by the isometries preserved by the dilaton. We join a flat space region (yellow) to the future in each case. (a) Poincare case. (b) Global case. (c) Milne case, where the top light green region of de Sitter is the Milne patch covered by the coordinates \nref{Milne}. The two future blue triangles can be described also as a Milne wedge.  This can be viewed as having black holes in $dS_2$, or as a one dimensional universe with two black holes at their ends.  The black hole interiors are shaded in blue, and the dark green regions are the static patches in between the Milne wedge and the black holes.  }
\label{dSCoordinates}
\end{center}
\end{figure}

Before we start, let us mention that there are three different solutions of equations of motion resulting from \eqref{actiondS}, and each of them plays some role in our discussion\footnote{We hope that the use of the same letter for various time coordinates will not cause a confusion.}, see figure \ref{dSCoordinates}, 
 \bea  
 {\rm Global:} ~~ds^2 & = &{ - d\eta^2 +    d\varphi^2 \over \sin^2 \eta }~=~-dt^2+\cosh^2t\, d \varphi^2~,~~~~~~~ \phi= { \tilde \phi_r \over -\tan \eta} ~,      \la{Global}
 \\
 {\rm Poincare:} ~~ds^2 &=& { - d \eta^2 + d x ^2 
 \over \eta^2 } ~=~-dt^2+e^{2t} d x ^2~,~~~~~~~~~~~~~\phi =  { \phi_r \over -\eta } ~,\la{Poincare}
 \\
 {\rm Milne:} ~~ds^2 &=& { - d\eta^2 +  d\chi^2 \over \sinh^2 \eta }
  ~=~-dt ^2 +\sinh^2 t\, d\chi^2~, ~~~~~~~\phi = { \tilde \phi_r \over - \tanh \eta } ~,        \la{Milne}
 \eea
 where we take $\eta< 0$. By taking the central charge of the CFT to be very large, $c\gg 1$, we can ignore the fluctuations of the boundary graviton mode. Then, at a small value of $\eta=  \eta_c $  we glue these metrics to flat Minkowski space  \be \la{metrflatdS}
 ds^2 ={ - dt_M^2 + dx^2 \over \epsilon^2 }\, ~,~~~~~~{\rm and~set}~~~~ \phi = \phi_b = { \phi_r \over \epsilon } ~,~~~~~~~~~~~ \tilde \phi_r = \phi_r { (- \eta_c) \over \epsilon }  
\ee
We will see later how to determine the ratio $(-\eta_c)/\epsilon$, which is a free parameter for now. 
In the Poincare case we set
 $\eta_c = -\epsilon$ so that $\phi_r = \tilde \phi_r $. 

We start by applying the Hartle-Hawking rule to determine the wavefunction. 
Namely, we continue the geometry into the Euclidean time direction so as to produce a reasonable state. 

This is usually done in the global coordinates,  \nref{Global}, whose analytic continuation give us the metric of a sphere and a non-singular imaginary dilaton which breaks the $SU(2)$ isometries to $U(1)$. In this case, the free parameter is fixed in terms of the size of the circle $L$  by demanding that $x \sim x + L $ translates into $\varphi + \varphi + 2 \pi$. This gives a probability 
\be 
|\Psi|^2_{\rm disconnected} \sim e^{ 2 S_0 } \sim e^{S_{dS}}. \la{Discon}
\ee
where $S_0$ is the full entropy including some terms of order $c$ from the matter conformal anomaly, and $ 2S_0$ is the de Sitter entropy. Semiclassiclally, we state that we produce is the vacuum on the cylinder 
$|B\rangle_s = |0 \rangle$. 

When the size of the circle  is very large,  $L \gg \phi_r/c$,
 we can approximate the relevant region of the geometry by the Poincare coordinates, \nref{Poincare}. This is ordinarily done in discussions of inflation applied to our universe. In the Poincare  case, the semiclassical state that is produced is simply the vacuum in the CFT.

We now consider the computation of the entropy of an interval of length $\Delta x =\ell$ that is located in the flat space region. In the semiclassical theory,  the Hartle-Hawking state for the conformal field is just the ordinary vacuum.  When there are no islands,   the entropy of the interval  is identical to \nref{Snaive}, or $S_{\rm no~island} = { c \over 3} \log (\ell/\varepsilon_{uv} ) $. 
As before, we can consider the possibility of islands. In fact, we expect that these dominate for large $\ell$. We can introduce islands and search for the location of quantum extremal surfaces. If $\ell$ is large, we expect that this search boils down to a search for a quantum extremal surface near each end point of the original interval. This is again justified by thinking about an OPE expansion of the twist operators. It should be emphasized that this search for the quantum extremal surface takes place on the two ``leaves'' of the Schwinger Keldysh contour that is used to calculate the state at late times, see figure \ref{dSSetup} (b-c). The quantum extremal surface could be on either side of the contour, the bra or the ket sides. For example, when it is on the ket side we get  a generalized entropy of the form 
\be \la{Slor}
 S_{\rm gen } = 2 \left\{ S_0 + { \phi_r \over - \eta } + { c\over 6 } \log \left[ {(t -\eta)^2 \over (-\eta) \varepsilon_{uv} }\right]  + i { c \pi \over 6 }  \right\} 
 \ee 
 where we have already used that the quantum extremal surface will be at the same spatial position as the endpoint of the interval. The overall factor of two is because we have two endpoints. The imaginary part arises because we have a timelike separation for the underlying twist operators,  and we used the standard prescription for continuing their correlators from spacelike to timelike separations. The extremization of this quantity leads to a result that is essentially the same as \nref{zstar}, or $\eta_* = - { 6 \phi_r \over c }$. Notice that the imaginary term is constant and does not affect the extremization. Furthermore, when we sum over the bra and ket contribution, this imaginary part cancels out. Note that this sum is supposed to be done at the level of the twist operators, in the computation of the Renyi entropies, and could produce terms that are subleading in the large $c$ expansion that we are ignoring. The situation seems similar to the one when there are competing quantum extremal surfaces, as in \cite{Marolf:2020vsi,Dong:2020iod}. 
 
This result is very pleasing because it is saying that the entropy of the interval can never become larger than the de Sitter entropy, $2 S_0 + \mathcal{O}(c)$. However, it must be wrong in detail because we can set up the same paradox that we described in section \ref{sec:paradox}. The general setup is the same, we view the matter CFT as the sum of two CFTs and we select the same choice of regions. 
 In the Euclidean AdS problem, we saw that   this paradox is avoided if we included bra-ket wormholes.\footnote{Note that the presence of this paradox is not unique to the no-boundary state we study. As long as inflation lasts longer than $\sim\log \frac{6 \phi_r}{c\epsilon}$ e-foldings, the paradox will likely occur independently of the initial conditions for inflation.}

Before trying to solve this problem, let us also search for islands in the Milne coordinate system \nref{Milne}.  When we talk about the state produced in the Milne coordinates it is important to realize that we have a physical parameter which is the effective temperature of the state that is produced in the flat region. Alternatively we could think of this parameter as the length of inflation from $t\approx 0$ in \nref{Milne} to the end of inflation.  It turns out that this is set by the free parameter that we had in \nref{metrflatdS}. 
The  free parameter, 
  sets the rescaling between the $\chi$ and 
$x$ coordinates. The ratio between these two coordinates amounts to the physical temperature $T_x$ in the fixed flat space coordinates \nref{metrflatdS} versus the 
temperature $T=1/(2\pi)$ in the $\eta, ~\chi $ coordinates. 
In other words, we define 
\be \la{Tfree}
T_x = { 1 \over 2 \pi } { (-\eta_c) \over \epsilon } = { 1 \over 2 \pi } { d \chi \over d x }  ~,~~~~~~\to ~~~\tilde \phi_r = 2 \pi T_x \phi_r 
\ee
 We will later discuss its value. However, before doing that, we can imagine that we are given this semiclassical geometry and we can compute the entropy of intervals. More precisely, we assume that we have the standard 
Bunch Davies (or Hartle-Hawking) state and we then use the gravitational fine grained entropy formula for computing entropies of intervals in the flat space region.  

Even before we look for islands, we can compute the entropy of some interval of length $\ell$ in the flat space region.  As is well known, for conformal fields,  the usual de Sitter vacuum looks thermal in the Milne coordinates.  
  For an interval of length $\ell$ the no island contribution is just the standard thermal answer at the temperature $T_x$ in \nref{Tfree}, 
 $S_{\rm no~island} = { c \over 3 } \pi T_x \ell $, for $T_x \ell \gg 1$. 
 
\begin{figure}[h]
\begin{center}
\includegraphics[scale=.32]{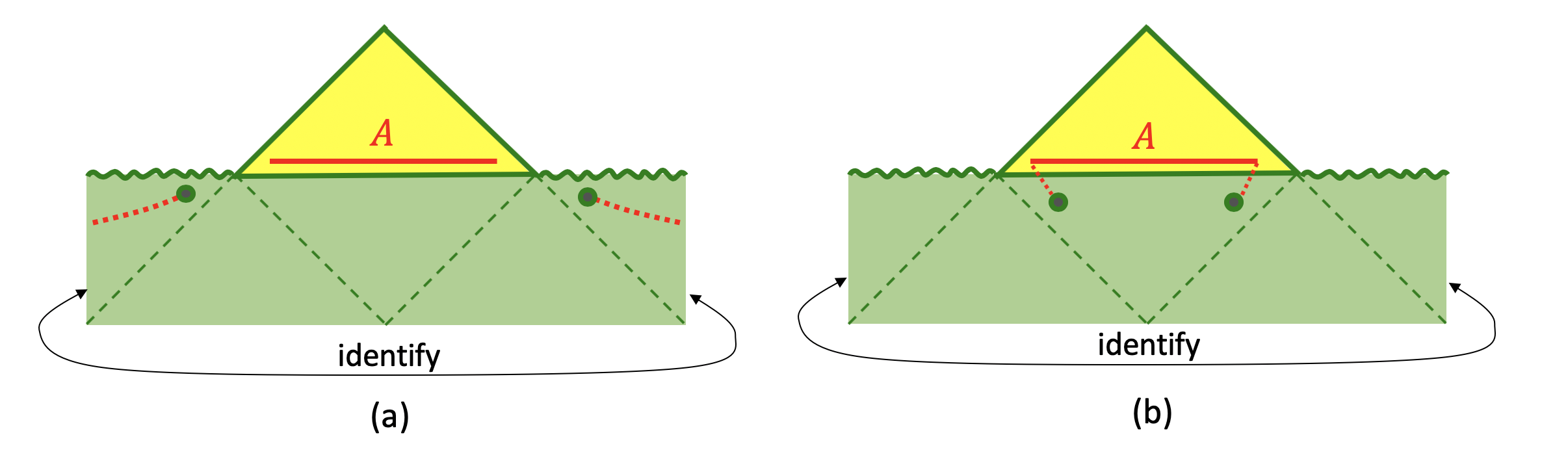}
\caption{We calculate the entropy of the interval $A$ in the Milne case and examine the contribution from islands. We are identifying the left and right parts of the figure because we imagine that the geometry is the usual $dS_2$ spacetime with a compact spatial section. This means that the two black holes can be viewed as entangled with each other.  (a) An island that is in the inside of the black hole region.  (b) Another type of island which can lead to paradoxes.  }
\label{fig:Milneisland}
\end{center}
\end{figure}

 We can now search for islands.
 In these coordinates,  we can think of the solution as an expanding de Sitter region with two black holes, one on the left side and one on the right side. These two black holes are entangled with each other if the full space is de Sitter, which has a compact spatial slice, see figure \ref{dSCoordinates}(c). So this setup is rather similar to the one considered in \cite{Almheiri:2019yqk}, and we therefore expect islands in the interior of these black holes. Indeed, we can make an ansatz and find islands there. For details see appendix \ref{app:islandMilne}.  The final position of the island is shown in figure \ref{fig:Milneisland}(a). 
 These islands do not lead to the subadditivity paradox because they are behind the horizon. 
  
In addition, we have islands that are similar to the ones we found above for the Poincare coordinates. It is clear they should exist for very tiny $T_x$, $T_x \phi_r \ll 1$, in which case we can approximate well a region of this solution by the Poincare coordinates. However, even if we cannot make this approximation, we can still find such islands, see figure \ref{fig:Milneisland}(b). Furthermore, in some regions in parameter space, they dominate over the black hole interior ones, see appendix \ref{app:islandMilne}. Therefore, these lead to paradoxes.

\subsection{Cosmological bra-ket wormholes }
\label{sec:dSbraket}

A crucial point about cosmological observations is that we normally trace over what will continue to happen in the future. This creates a connection between the bra and the ket. In other words, if we think in terms of a Schwinger-Keldysh in-in contour   computation of an observable, the connection arises when we join the ket part of the contour with the bra part of the contour at the time we do the observation.  Lessons from traversable wormholes, or their  euclidean counterparts of section \ref{sec:BraKet}, imply that such direct connections can result in spacetime geometries that are connected between bra and ket in the far past. 
Connections between bra and ket in cosmology were first suggested by the Page prescription for the wavefunction of the universe \cite{Page:1986vw}\footnote{This idea was further analyzed in e.g. \cite{Barvinsky:2006uh}.}. As emphasized in \cite{Anous:2020lka}, this is, in principle, a different prescription than the Hartle-Hawking one. Here we are saying something slightly different. We are saying that, by virtue of tracing out some of the degrees of freedom in the future, we are connecting the bra and the ket computations and therefore there is no reason to forbid connections between bra and ket in the past. In fact, we have seen that we run into trouble if we only consider the naive geometry. Similar bra-ket connections were also suggested by the JT gravity model (without propagating fields) considered in \cite{Penington:2019kki}. 

Once we allow the possibility of such connections, we need to figure out which concrete spacetimes   to consider. Notice that the gravitational action of
 the bra-ket wormholes does not contain the factor of 
$e^{ 2 S_0}$ present in \nref{Discon}.  Thus, in order for them to dominate there should be some enhanced matter contribution. 

\begin{figure}[h]
\begin{center}
\includegraphics[scale=.23]{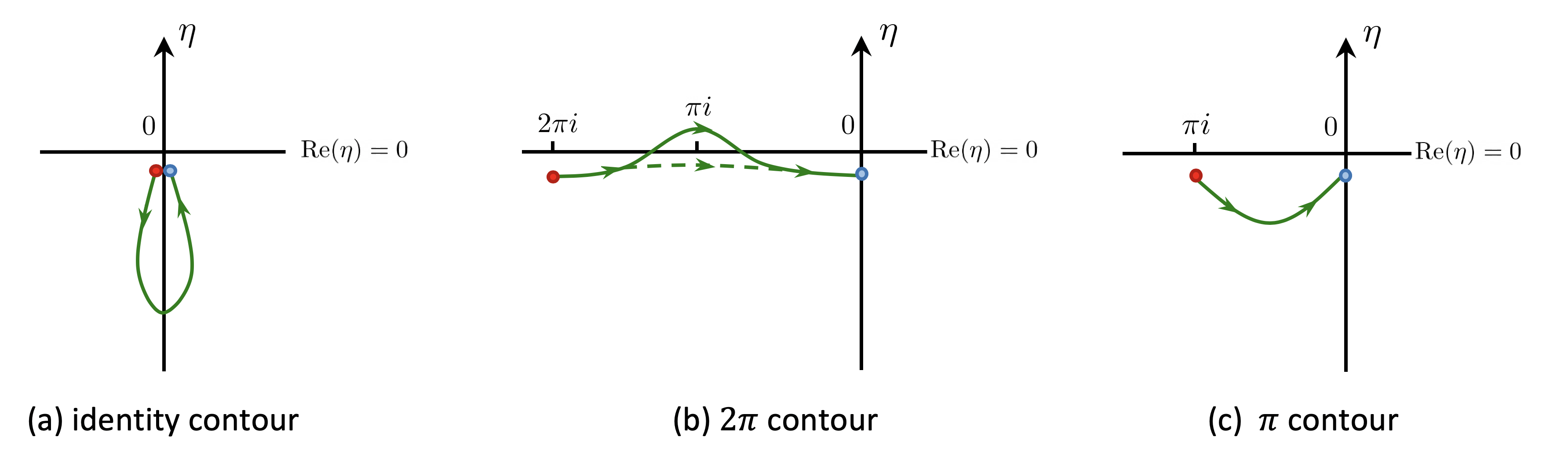}
\caption{Three contours that we consider, the red circle is the bra and the blue circle the ket. We describe them  in the $\eta$ coordinates and real values of $\eta$ correspond to the vertical axis.  (a) The identity contour, we simply go back in time and join the bra and the ket. It does not matter how far back in time we go. (b) The $2\pi$ contour with the Milne metric \nref{Milne}.  Now we have a non-trivial displacement in imaginary time, which will lead to a thermal state for the CFT. Both the solid line and dashed lines contours are equivalent for conformal matter. We suspect that the solid line contour gives the usual Hartle Hawking state also for any massive field. 
(c) The $\pi$ contour, also for the Milne metric \nref{Milne}. It has a shorter displacement in the imaginary direction.  }
\label{fig:contourseta}
\end{center}
\end{figure}

In our case, the search is simplified because we can first impose the dilaton equation of motion which  fixes the metric to be de Sitter. Then the choice of contour boils down to a choice of complex contour in either proper time $t$  or conformal time $\eta$. 
We will consider various possible contours that define distinct bra-ket wormholes. As discussed in detail in \cite{Hertog:2011ky},  in principle, we are allowed to consider any contour in the complex $t$-plane to create a state. Naturally, contours that can be smoothly deformed into one another without crossing a singularity lead to equivalent states.  Presumably, there are some constraints on which contours can and cannot contribute, but at the moment we lack a complete understanding of these constraints. What we can impose with confidence is that the resulting density matrix should be Hermitian and positive-definite. We will consider three candidate contours below, 
depicted in figure 
\ref{fig:contourseta}. When we have a non-trivial displacement along the imaginary $\eta $ direction, it is natural to consider the geometry in the Milne form \nref{Milne}, which is periodic under $\eta \to \eta + i \pi $. 
So in principle we could consider contours where the endpoints are separated by 
$ \eta_{\rm bra} = \pm \eta_{\rm ket} + i n \pi $. We will consider only  three options.\footnote{\textbf{Note added in v4: }  In \cite{Kontsevich:2021dmb,Witten:2021nzp}, Kontsevich, Segal and Witten proposed a criterion to diagnose whether a complex metric is sensible and should be included in the gravitational path integral. The three contours we discuss below do not satisfy the criterion. We check this explicitly in Appendix \ref{app:KSW} and comment on its possible interpretations. }



\begin{figure}[h]
\begin{center}
\includegraphics[scale=.27]{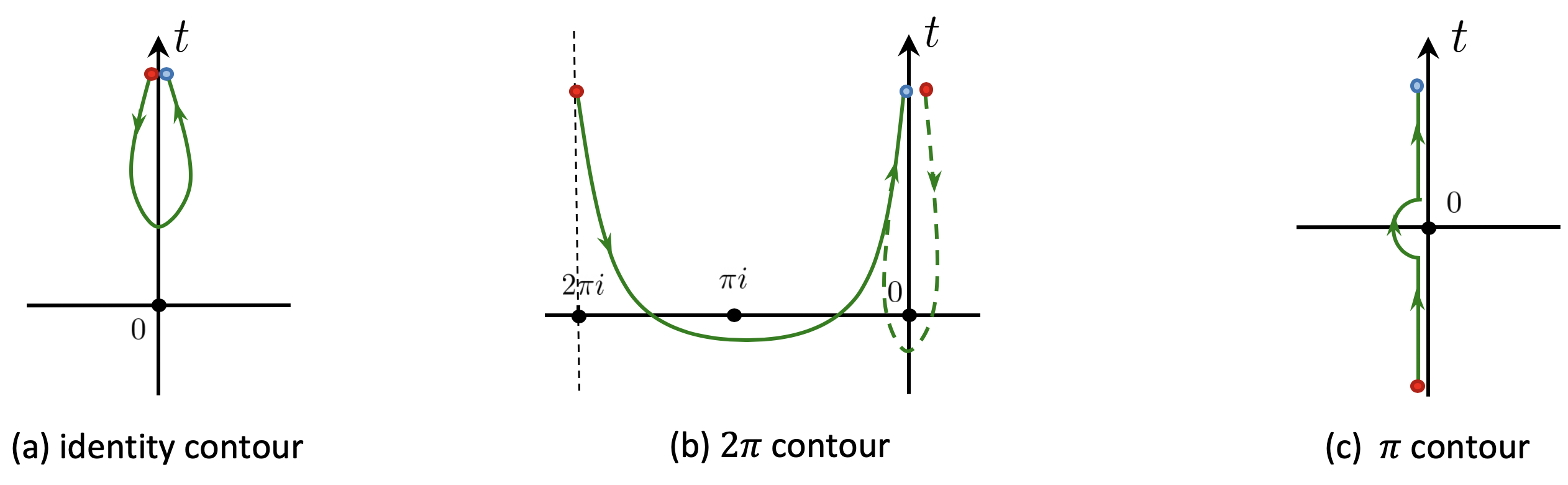}
\caption{We display the same contours as in fig. \ref{fig:contourseta}, but in the $t$ plane. We marked the singularities of the metric (\ref{Milne}) using black dots. The contour can be smoothly deformed as long as it does not cross the singularities.}
\label{fig:contoursT}
\end{center}
\end{figure}

An important property of the contours is the (inverse) temperature which is seen by our CFT matter fields once they are placed on this geometry and evolved along the corresponding contour. This is given by 
\be
{ 1 \over T_\eta(\gamma) } = \beta (\gamma) =i \int_{\gamma}\frac{d t}{a(t)}= i \int_{\gamma} d\eta\,,
\ee
where $\gamma $ is the contour. For the  three contours  we consider, depicted in figures \ref{fig:contourseta}, \ref{fig:contoursT},  the inverse temperatures are 
\be
\beta_{\rm Identity~contour}=0\,,\quad \beta_{2\pi ~{\rm contour} }=2 \pi\,,\quad \beta_{\pi ~\rm contour}= \pi\,.
\ee
We now evaluate the matter stress tensor on these geometries and calculate the backreaction on the dilaton, as in  section \ref{sec:BraKet}. 
We first calculate $\braket{C|C}$,  meaning that we impose periodic boundary conditions on the matter fields along a contour.
We get (dismissing the cosmological constant term that can be absorbed by a shift of $S_0$)
\be
\label{Tandphi}
T_{\eta\eta}=T_{\chi\chi}=\frac{c\pi}{6}\left(\frac{1}{\beta_i^2}-\frac{1}{4 \pi^2}\right)\,, \quad \phi={ \tilde \phi_r \over - \tanh  \eta} -\frac{c \pi^2}{3}\left(\frac{1}{\beta_i^2}-\frac{1}{4 \pi^2}\right)\left( { \eta 
\over \tanh \eta }  -1\right)\,.
\ee
where the second term  in the stress tensor comes from the conformal anomaly.

We can search for a solution as follows. We have already imposed the equation of motion for the dilaton, which fixes the metric to be de Sitter. 
After choosing the contour, we evaluate the matter partition function. At this point, we typically have one free parameter, discussed near \nref{metrflatdS},  which is the length of inflation,  or the effective temperature of the state produced by the wormhole.  We can now do one of two things. We can evaluate the full partition function of gravity plus matter   in terms of this free parameter and then extremize it.  Since the gravity part of the action is purely a boundary term, we see that it cancels in all cases between the bra and the ket contributions. So we only need to extremize the matter part. Alternatively, we can determine the matter stress tensor, solve  for the dilaton and impose its boundary condition. These two procedures give the same answer. Since the gravity theory makes no contribution, we will find that the extremization sets the  traceless part of the  matter stress tensor to zero in the de Sitter region and then the dilaton is unchanged (compared to the one without matter). In practice,  it is a bit more illuminating to attempt to extremize the matter partition function, since sometimes we will not find a solution, and this process will tell us where the free parameter is driven to. 

In computing the matter partition function,  it is important to remember to include the effects of the conformal anomaly, which is non-trivial when the contour includes some Euclidean evolution in the de Sitter region. 
This is easiest to compute in the $\eta$ coordinates, we find only a contribution from the Euclidean parts of the contour since the Lorentzian parts cancel each other. We can evaluate the Euclidean piece of the contour by setting $ \omega = -\log \sinh  (\eta_L + i \eta_E) \sim  (\eta_L + i \eta_E) $ in \nref{Anomaly}, for a very large $\eta_L \ll 0$ and   $\eta_E$ parametrizes the shift into the Euclidean part of the contour. 
Then the anomaly contributes as 
\be \la{dSAnomaly}
\log Z_{\rm anomaly} = { c \over 24 \pi } \int d\chi d\eta_E\, (\partial_{\eta_E}  \omega)^2 \sim -   { c \over 24 } { \Delta \eta_E \over \pi } \Delta \chi
\ee 
 Note that $\Delta \chi$ is the length in the $\chi$ coordinate in \ref{Milne} which we yet need to convert into $\Delta x$, which is the physical coordinate in Minkowski space \nref{metrflatdS}.

\subsubsection{The identity contour} \la{sec:identity}

The simplest possible connection is just an identification between the bra and ket parts of the contour, see figure \ref{fig:contourseta}(a). This is the type of identification we make in the future when we have a Schwinger-Keldysh contour, see figure \ref{dSSetup}(c). So it is fairly natural to assume that we can also have it in the past. We will call this the ``identity'' contour. Unfortunately, this results in an infinite temperature state, independently of the time when we do this identification. In fact, with this identification, the past parts of the bra and ket contours cancel out (as the future parts cancel in the Schwinger-Keldysh contour).  The reason we get this infinite temperature is that we do not have any period of Euclidean evolution, so all states in Hilbert space are equally weighed. 
Though this result looks physically unreasonable, in the sense that it cannot model our universe, it is not clear why it is wrong from the mathematics of the problem. This is related to the fact that if we weigh all present possible  states of the universe equally, the current state of the universe looks extremely unlikely.

Note that in the Euclidean case considered in section \ref{sec:BraKet}, we considered contours of arbitrary length in Euclidean time. The minimum of the action was given by one with a finite amount of evolution in Euclidean time. So the zero length one was disfavored.  

In order to get more insight on this contour, we can regularize its contribution by doing an extra Euclidean evolution by Euclidean time $\tau$ in the flat space region. This decreases the effective temperature to a finite value $T= 1/(2 \tau )$. The Lorentzian dS part of the solution has 
 a free parameter. The matter partition function only depends on the geometry of the flat region outside and it is independent of this free parameter. When $x$ is very long, $x \sim x + L$, it gives 
 \be 
 Z = \exp\left( { \pi c L \over 12 \tau } \right) ~,~~~~~~~ L \gg \tau 
 \ee 
  If we compute the full entropy of the universe, with a compact spatial section, then we find zero if we include the whole lorentzian region as an island. If we compute the entropy of an interval, it depends on the free parameter mentioned above and it is unclear to us how to treat it.

\subsubsection{The $2\pi $ contour } \la{sec:2PiContour}

In order to avoid this problematic contour  we will consider alternative contours that involve some amount of Euclidean evolution. This is already present in the standard Hartle-Hawking prescription (which implies the Feynman's $i\epsilon$ prescription at small scales).  

The simplest contour of this kind that we could think of is the $2\pi $ contour which consists of shifting $\eta $ in  \nref{Milne} by $ 2 \pi $ in the Euclidean direction before joining again to the bra part of the contour
\be 
\eta_{\rm ket} = -2 \pi i + \eta_{ \rm bra} 
\ee
 The contour is depicted in fig. \ref{fig:contourseta} (b). When we have a non-compact spatial direction, it produces a state on the infinite line that has temperature $T_\eta = 1/(2\pi)$ and is exactly the same as the Hartle Hawking state on the Milne patch. It is a thermal state with the usual temperature. This is easy to see by doing a conformal rescaling to the  flat coordinates $ds^2 =-d\eta^2 + d\chi^2 $. In these coordinates, this $2\pi $ contour is just the ordinary flat space thermal Schwinger-Keldysh contour.  
The fact that it is thermal can be viewed also as arising from the fact that the whole future boundary of de Sitter space is covered by the two future Milne patches, see figure \ref{dSCoordinates}(c). When the $\chi$ coordinate is non-compact, this is giving the same as the Hartle-Hawking state in Milne coordinates, see appendix \ref{app:2piHHCFT}. 

However, this contour also allows us to define the state even if we    compactify the direction $\chi$. In this case, we are breaking the de Sitter isometries, but we can continue to define the state using this $2 \pi $ contour.  As we pointed out before,  the gravitational part of the action gives zero. 
The matter part of the action recieves a contribution from the anomaly \nref{dSAnomaly}, with $\Delta \eta_E = 2 \pi $. This precisely cancels the usual torus partition function in the regime of $\Delta \chi \gg 1$, which is $\log Z_{\rm flat} = { c \over 12 } \Delta \chi $.  

This precise cancellation implies that the parameter $T_x$, \nref{Tfree}, is not determined. As we mentioned, this parameter also determines the length of inflation. Notice that the second term in the expression for the dilaton in \nref{Tandphi} vanishes when $\beta = 2\pi$. Maybe by considering fluctuations of the Schwarzian mode, {$T_x$} will be driven either to large or low values.

Another option, is to compactify the $x$ direction, $x \sim x + L$. 
With the metric 
\nref{Milne},  this leads to a singularity at $t=0$, or $\eta = -\infty$. The contour we are choosing   leads to well defined state in the CFT despite this metric singularity, since we avoid it by moving into the complex plane.  Now the matter partition function is enhanced by making $\Delta \chi$ small. Notice that even though $\Delta x = L$ is fixed, $\Delta \chi$ can vary because of the free parameter discussed near \nref{Tfree}. 
  This drives the solution to small  $T_x$ . For such small values  the matter partition function is dominated by the vacuum contribution in the cross channel. This gives a divergent contribution for small $T_x$ that comes from a relatively thin tube joining the bra and the ket. It seems natural to imagine that this divergence is cured by transitioning into the two disconnected disks giving the traditional Hartle Hawking wavefunction. Thus, this contour does not appear to lead to a resolution of the paradox.

 The conclusion is that this contour has a zero mode when $L$ is infinite and it leads to the usual Hartle-Hawking state in Milne coordinates. For $L$ finite,   $T_{x}$ is driven to very small values. 

 \subsubsection{ The $\pi$ contour } \la{sec:PiContour}

 This contour leads to a higher effective temperature, see figure   \ref{fig:contourseta}(c). When we evaluate the matter action including the contribution from the conformal anomaly, we get (for $L\gg 1$), 
 \be \la{matterPi}
 \log Z_{\rm anomaly} + \log Z_{\rm flat} = \left(  - { c \over 24 } + { c \over 6} \right) \Delta \chi = { c \over 8 } 2 \pi T_{x} L 
 \ee 
 Since the gravity part cancels this drives $T_{x} \to \infty$. So, now we are generating a state with very high temperatures. In (\ref{matterPi}) we used that $T_{x}$ governs the rescaling between $x$ and $\chi $ coordinates \nref{Tfree}. 
 
\begin{figure}[h]
\begin{center}
\includegraphics[scale=.38]{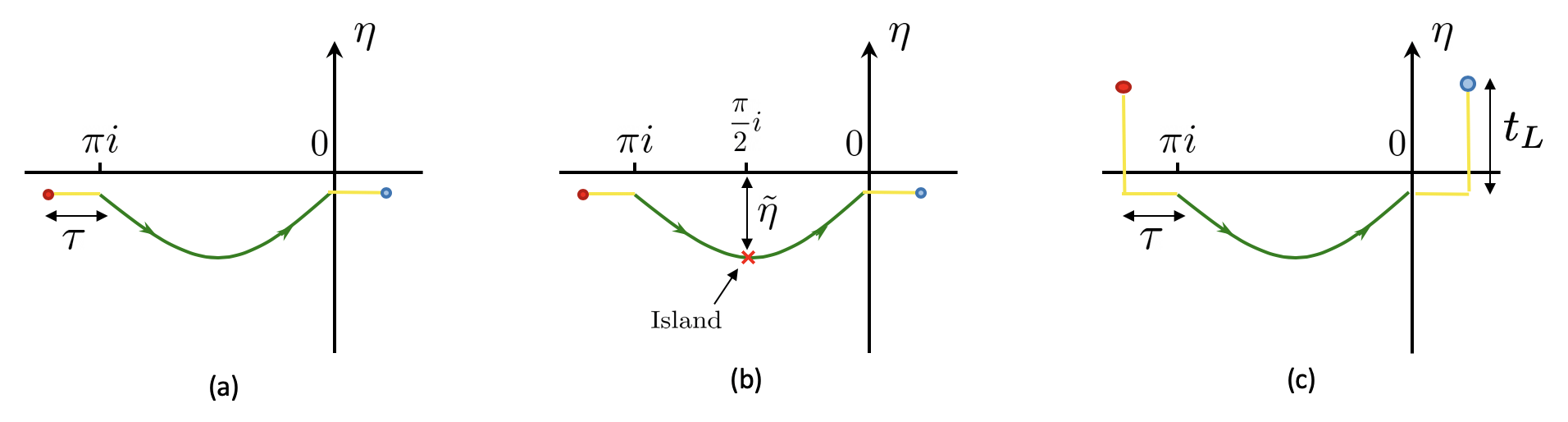}
\caption{Regularized $\pi$ contour. We add further Euclidean time evolution in the flat space region, which is denoted by the yellow lines. (a) Basic contour defining the density matrix. (b) Location of Islands for long intervals. (c) We can further evolve in Lorentzian time, $t_L$, the state prepared in (a). }
\label{fig:dSPiIsland}
\end{center}
\end{figure}

 We can regularize this by adding, in the flat space region, a period of  Euclidean time evolution by an amount  $\tau$, see figure \ref{fig:dSPiIsland}. Now the size of the circle in the $\eta$ coordinates is 
  $ \pi ( 1 + { 4 \tau   \tilde T_{x} })$,  where we defined $\tilde T_x$ to be $T_x$ defined as the rescaling present in \nref{Tfree}, but we have not identified it  with the physical temperature yet.     The matter part of the action then becomes 
  \be 
  \log Z_{\rm anomaly} + \log Z_{\rm flat} = \left(  - { c \over 24 } + { c \over 6( 1 + { 4 \tau   \tilde T_{x} }) } \right) \Delta \chi  =  \left(  - { c \over 24 } + { c \over 6( 1 + { 4 \tau   \tilde T_{x} }) } \right) 2\pi \tilde T_{x} L
  \ee 
  which now has a maximum at $\tilde T_{x} = { 1 \over 4 \tau }$.  Having fixed $\tilde T_x$, we find that $\tilde \phi_r = { \pi \over 2 \tau } \phi_r $. Furthermore, this implies that the total Euclidean time evolution in the $\eta$ coordinates is $\pi ( 1 + 4 \tau \tilde T_x ) = 2 \pi$. This also means that $\tilde T_x = T_x$ is the physical temperature of the state of the CFT prepared by this whole process. In addition, the stress tensor, as well as the correction to the dilaton vanish \nref{Tandphi}. The conclusion is that the state prepared by this procedure agrees, in the de Sitter region,  with the standard Hartle-Hawking state in the Milne coordinates!   
    
So this is the one case where we have managed to fix the free parameter! Furthermore, at this maximum, the result is positive, so that for large enough $L$ it can win over the two separate disk contribution \nref{Discon}. However, it is a bit unphysical because we had to perform an additional Euclidean time evolution after the end of inflation. 
 
For this case, we can now search for islands and we get islands at a value of $\eta$ whose imaginary part lies in the middle of the contour, see figure \ref{fig:dSPiIsland}(b). 
  
The entropy of an interval of length $\ell$, without any island, is given by 
the usual thermal answer 
$S = \frac{c}{3} \pi T_x \ell = \frac{c\pi \ell}{12\tau}$. When the interval is large, this gives us a very large answer. 

As usual, we can also search for an island on the contour. In particular, motivated by the Euclidean discussion, we expect the island to be at half way through the contour, i.e. $\eta=  i \frac{\pi}{2} - \tilde{\eta}$,  see figure \ref{fig:dSPiIsland} (b). Thus we have, see appendix \ref{app:piIslands},   
\be \la{SgenPI} 
 S_{\rm gen} (\tilde{\eta}) = 2 \left\{ S_0 +  {\tilde \phi_r}{ \tanh \tilde \eta} + \frac{c}{6} \log \left( \frac{  \left(2\cosh \frac{\tilde{\eta}}{2}\right)^2 }{ \cosh \tilde{\eta}\,  \varepsilon_{uv,\chi}  }\right) \right\}.
\ee 
Extremizing over $\tilde{\eta}$, we find the following equation
\be \la{tildeEta}
	\cosh \tilde{\eta} \tanh \frac{\tilde{\eta}}{2} = \frac{6\tilde \phi_r}{c},
\ee
which always has solution $\tilde{\eta}>0$. We find that
\be 
 S_{\rm island} = 2 S_0 + \mathcal{O}(c).
\ee
Thus for an interval with size 
$ \ell \gtrsim \frac{24 S_0}{\pi c} \tau $ ,
its entropy is given approximately by the de Sitter entropy $S_{\rm dS}=   2S_0$.

We can also further evolve the final  state in Lorentzian signature, see figure \ref{fig:dSPiIsland}(c). Then  the island  shifts along coordinate $\tilde{\eta}$ and sweeps out the full range of $\tilde \eta$. As expected for a state resulting from a quench \cite{Calabrese:2005in}, for a semi-infinite interval we get an entropy that grows linearly with lorentzian time, see appendix \ref{app:piIslands} for more details. 

 Let us compare this result with what we had in the AdS case. A common feature is that the island is located in the ``middle'' of the Euclidean evolution, hence we can again argue that the CFT is in a $\ket{TFD}$-like
state with some other system. This, in turn, ensures the restoration of strong subadditivity, as long as this solution is the dominant one. Note, however, that even for an interval that is placed at zero Lorentzian time in the flat space region, the island is separated from the slice $\eta=- i \pi/2$.    This is perhaps not surprising since our contour always involves some Lorentzian evolution during the $dS_2 $ region, and when Lorentzian time was added to our boundary state in the $AdS$ case, the island was also shifted in Lorentzian time, as in \eqref{EntrL}. 

 The metric in the region that the islands reside can be written as 
\be \la{MetrSt}
	ds^2 = \frac{d\tilde{\eta}^2 - d\chi^2}{\cosh^2 \tilde{\eta}} = - \left( \frac{-d\tilde{\eta}^2 + d\chi^2}{\cosh^2 \tilde{\eta}} \right).
\ee
If we ignore the overall minus sign,  we find the same FLRW universe as discussed in section \ref{sec:ClosedUniverse}. On the other hand, if we do not ignore it,  for the non-compact case,  this is the metric of the static patch in the Milne solution of
\nref{Milne}, see figure \ref{dSCoordinates}, which lies between the black hole horizon and the cosmological horizon. From this perspective, the island is extended along the timelike direction, $\tilde \eta$, in \nref{MetrSt}. We do not know how this should be interpreted. We note that this foliation of the static patch  is similar to the one that inspires the  ``$dS/dS$'' correspondence \cite{Alishahiha:2004md}.

So far,  we have discussed the case of very large $L$. When $L$ is finite, there is actually a solution. This is obtained as follows. We have seen that for very large $T_x$ we obtain the matter partition function \nref{matterPi}, which diverges. For very small $T_x$ we also expect that it diverges, for reasons similar to the ones we mentioned at the end of section \ref{sec:2PiContour}. This means that there is a minimum of the partition function for some value of $T_x$ of the order $T_x \propto 1/L$. This gives a classical solution. We also get that the stress tensor in the $dS$ region is zero. Unfortunately, this is a {\it maximum} of the action rather than a minimum. It is conceptually similar to the unstable ``hot wormhole''   phase described in \cite{Maldacena:2018lmt,Maldacena:2019ufo}. The matter action at the solution is given by $c$, up to an order one number. Then we expect it to be subdominant to the disconnected solution. In addition, if we regularize the whole problem by adding a bit of euclidean time evolution, $\tau> 0$, then it is  clearly subdominant compared to the bra ket wormhole we discussed above.

 \subsubsection{Bra-ket wormholes and bouncing cosmologies }

   Let us finally make the following observation. The most straightforward interpretation of our bra-ket geometries is that they have a ``Big-bang'' singularity in the past. This is most clear if we compactify the direction $x$ in (\ref{Milne}), $x \sim x + L$, so that $t=0$ is an actual singularity. 
     We nevertheless avoid this singularity by deforming the contour in the Euclidean direction thus providing a non-singular state. The resulting geometry is similar to those that appear in the so-called ``bouncing cosmologies''  where a   circle contracts to zero size and then expands again, see eg. \cite{Khoury:2001bz}.  However,  our interpretation is very different: the contracting part of the geometry is interpreted as the ``bra'' of the wave function of the universe and the expanding one as the ``ket''. It would be interesting to study this type of contour in more realistic models to see whether they could provide for a phenomenologically interesting universe. 
  
\subsection{Summary and discussion of the de Sitter case} \label{summarydS}

To summarize the above discussion, we found that the $dS_2$ JT gravity plus matter CFT theory ``almost'' has classical bra-ket wormhole solutions that bear similarities with the bra-ket wormholes in $AdS_2$ and can resolve the strong subadditivity problem present in the Hartle-Hawking state. It is reasonable to imagine that a proper UV completion of the theory, for example viewing this theory as an approximation to  a higher dimensional theory, will supply the necessary ingredients to stabilize the solution. If this is the case, the length $\pi$ contour appears to be  the most promising one. If the Euclidean action on the resulting solution were still negative, it would dominate over the Hartle-Hawking state and hence provide the leading contribution to the calculation of correlators in our inflationary state $\ket{C}$. Our discussion in section \ref{sec:Typical} on typical states and factorization similarly applies here. 

 \begin{figure}[h]
\begin{center}
\includegraphics[scale=.24]{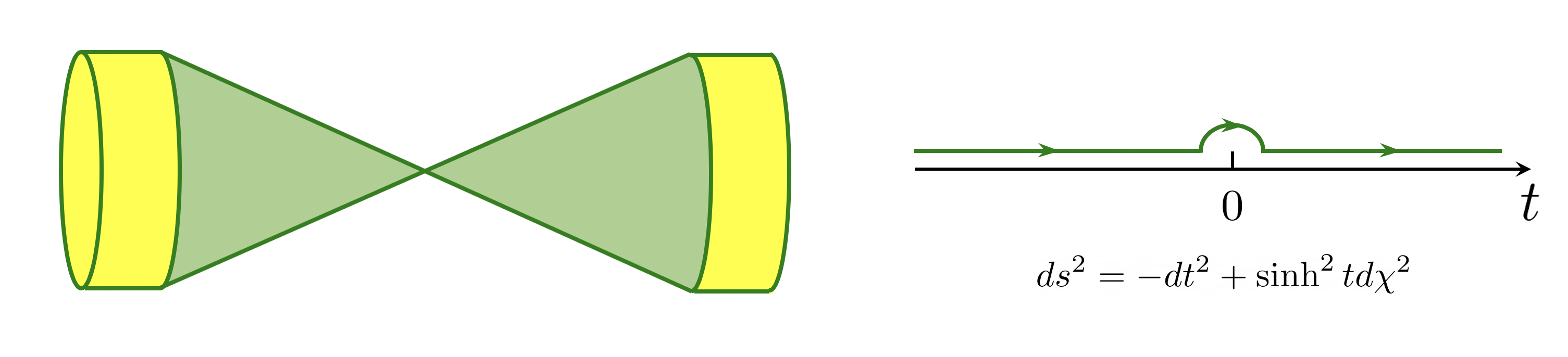}
\caption{The geometry that appear in the $\pi$ contour is very similar to the double cone considered in \cite{Saad:2018bqo}, except for an overall sign in the metric, which  exchanges  space and time and changes the sign of the curvature, taking the locally  $AdS_2$ space to a locally   $dS_2$ one. geometry.   }
\label{fig:DoubleCone}
\end{center}
\end{figure}

We note that the wormhole geometry of our length-$\pi$ contour appears similar to the ``double-trumpet'' solution studied in \cite{Saad:2018bqo} which is responsible for the ``ramp'' in the time dependence of the spectral form factor in SYK-like theories, see figure \ref{fig:DoubleCone}. It is, however, not identical because in their case the bulk theory is $AdS_2$. Also, in \cite{Saad:2018bqo}, the parameter analogous to $T_{x}$  is the black hole temperature and it is natural to fix it (or similarly  fix the energy).  It would be interesting to see if these solutions are related by any analytic continuation. The importance of bra-ket wormholes for de Sitter space was emphasized in \cite{Penington:2019kki}. Their model did not have local matter and the contour considered was most similar to our ``identity'' contour. In the  model in \cite{Penington:2019kki} the actual classical solution also does not exist, however, since the bulk theory was pure JT gravity, the authors were able to calculate the path integral exactly without relying on semiclassical approximation. Other types of connected geometries between bra and ket were discussed in \cite{Dong:2020uxp}\footnote{  We  think that in the  setup of section 3 of \cite{Dong:2020uxp} one will also have to include configurations somewhat similar to the bra-ket wormholes we are discussing, at least in some regions of parameter space, since their geometry can lead to a similar entropy subadditivity paradox. }. 

We should mention that the problem of fixing the parameters of classical solutions in the no boundary approach is something that is unsolved even for the simplest case of inflation. The simplest approach, which does not fix the overall size of the spatial $S^3$ at the reheating surface,  leads to a small amount of inflation, which is analogous to our high temperature problem. There are various approaches to this problem, one involves some effects that focus on the idea that we are unlikely observers \cite{Hartle:2010dq}. Another, more popular one, is to ditch the no boundary proposal all together and search for other measures. What we have done here is to allow for further interesting candidate topologies and we found that they sometimes dominate (the $\pi$ contour) but they still do not lead to a reasonable state. 

In most of our analysis we fixed $L$, the length of the circle in the future. But it is also natural to let it vary. We have not explored this in full detail, since it requires some further understanding of what we are keeping fixed. It seems reasonable to fix boundary conditions on a ``slit'', which would be a finite region of the universe, tracing everything else. A similar idea was recently explored in \cite{Dong:2020uxp}.

Finally, we noted that some of the bra-ket wormholes we considered have some similarities to ``bouncing cosmologies'', but with a different physical interpretation. It would be interesting to study whether this interpretation helps to provide well defined states with promising phenomenology.

\section{Conclusions and discussion}

 In this section we summarize our findings and discuss some open questions. 
 
 In this paper,  we discussed peculiar CFT states  that are prepared by performing a certain gravitational path integral.  We considered two main cases: states produced by Euclidean AdS$_2$ (or $H_2$) evolution and states produced by dS$_2$ evolution.  There is a naive semiclassical geometry giving a first guess for the state.    In order to gain some non-perturbative insight, we used the gravitational fine grained entropy formula. We found that the naive semiclassical geometries lead to an entropy subaditivity paradox. We argued that this paradox is solved by including bra-ket wormholes, which are other semiclassical geometries that connect the bra and ket contours of the gravitational path integral.  Once we include these other geometries this particular paradox disappears. 
 
 In the Euclidean AdS case, we have a fairly traditional setup, very similar to the one modeling  an evaporating blackhole.   The bra-ket wormhole is an explicit semiclassical solution, a version of the ones in \cite{Maldacena:2018lmt}. It dominates the path integral before the paradox can arise and it does not give rise to a paradox by itself. One important feature is that the entanglement entropy of a long interval is never larger than about $2S_0$.  It would be interesting to find examples with definite (non-ensemble) holographic duals. 
 
 Semiclassically, we can interpret the bra-ket wormhole solution as creating a pair of entangled lorentzian states: one living in the CFT with no gravity and the other consisting of a matter CFT in an FLRW expanding and collapsing cosmology. This means that the full entanglement wedge of the CFT state includes the closed universe. Then we expect that this FLRW cosmology is encoded within the state of the boundary CFT. Understanding  in more detail how the cosmological region is encoded within the CFT state would be very interesting. 
 
 Poetically, we can call the FLRW universe a ``puppet universe'', which is controlled via threads of ghostly entanglement to the actual state living on the CFT. In the doubly holographic setup of section \ref{DoublyHolographic} we have an actual geometric connection between the two universes through the third dimension, see figure \ref{fig:DoublyHolo}(c).  The emergence of this apparently detached universe is similar in sprit to a black hole that has completely evaporated, except that the present setup has more symmetries. It would be interesting to find higher dimensional versions of these solutions, specially ones with definite holographic duals, if they exist!

 In section~\ref{sec:Typical},  we pointed out that a minor modification of these  bra-ket wormholes, namely a projection onto typical intermediate states,  leads to Euclidean wormholes that appear to display a factorization problem unless they are viewed as an ensemble average, as in \cite{Saad:2019lba,Coleman:1988cy,Giddings:1988cx,Marolf:2020xie}. This is closely related to  \cite{Stanford:2020wkf}. 
  
Let us now turn to our second model, where we prepare the state via de Sitter evolution. This is a simple model of cosmology. Again we found that the most obvious semiclassical solution violates the strong subaditivity condition for very long intervals. This led us to explore possible bra-ket wormholes. These are constructed by choosing a certain integration contour in the complex time plane. 

The most naive way to join the bra and the ket lead to a very divergent answer. We have not understood why this is excluded from first principles.  In \cite{Penington:2019kki},  this contour did not lead to a problem because there were no propagating fields in their model.  

Next we considered contours with a certain amount of evolution in Euclidean time. The first guess gave a state that is essentially the same as the Hartle Hawking state in the Milne setup. However, this state contains an unfixed zero mode (at leading order), which is the length of inflation. Furthermore, this configuration does not contain the requisite enhancements for it to dominate over the usually discussed de-Sitter no-boundary wavefunction. 

Finally, we studied a contour with half of the previous euclidean time evolution. We found that the effective length of inflation gets driven to zero, giving an infinite temperature state in the CFT. Furthermore,  it dominates over the usual de-Sitter wavefunction. For this last reason,  we do not view the short  inflation problem  as a fatal flaw, rather as a feature of this wavefunction.  We can regulate this by projecting onto low energy states by performing some Euclidean evolution in the flat space region. This gives a well defined solution where we have performed a few computations. This contour has several features in common with the double trumpet that reproduced the ramp in the black hole case \cite{Saad:2018bqo}.  

Of course, one would like to find examples of cosmological wavefunctions, or cosmological measures, which would lead to more physically reasonable states (without extra projections on to lower energy states). This is of course a very well known open problem. We think that solutions where the bra and ket part of the geometries are connected, are very natural in cosmology because we are tracing out over unobserved regions of the universe (see \cite{Dong:2020uxp} for some exploration along these lines). Furthermore, even if some version of dS/CFT was true, the fact that we are not observing the universe in the far future could naturally lead to an average over couplings. In bra-ket wormholes, that sum leads to a connection in the very early universe. 

In the present discussion, we considered a single connected universe. Of course, we could imagine that we create pairs of macroscopic correlated universes, etc. So we should view our computation as projecting the full wavefunction of many universes onto a single macroscopic connected component of length $L$ at the end of inflation. 

With  healthy skepticism, one might say that the above picture is too strange and that bra-ket wormholes should be excluded from the gravitational path integrals. 
 But this leads to a state where the fine grained entropy formula leads to a violation of strong subaditivity.   We view this as {\it excluding} the {\it exclusion} of bra-ket wormholes.
  However, one could doubt the validity of the gravitational fine grained entropy formula  when the quantum extremal surface is timelike separated from the endpoint of the interval. For this reason, it would be nice to derive this rule from first principles.   Of course, we have not proven that  bra-ket wormholes do not lead to other inconsistencies.   

There are some similarities between the bra-ket wormholes we considered and bouncing universes, but the interpretation is different. We view the collapsing region as the bra part of the contour. It looks collapsing because time runs backward along the bra contour.

 The observations in this paper were a formal study of the no-boundary-like probability measures and we did not attempt to find models that could be relevant for the real-world cosmology. Note also that many of the features we discussed are special to two dimensions, and, in particular, we have not found islands for the Hartle-Hawking state in higher dimensional de-Sitter. Nevertheless, it is possible that phenomenologically interesting models in which bra-ket wormholes play an important role do exist and we plan to research this question in the future. An exciting possibility is that, perhaps, we need to go beyond the semiclassical theory to obtain a reasonable measure for the correct computation of probabilities\footnote{ Another popular alternative is that the dynamics of inflation or eternal inflation is the key to the definition of the measure, see, {\it{e.g.}} \cite{Linde:1998gs}.}.

\vspace{1cm}
\textbf{Acknowledgments}

 We are grateful to A. Almheiri, A. Karch, S. Komatsu, A. Levine, H. Lin, R. Mahajan, E. Shaghoulian, E. Silverstein, D. Stanford, E. Witten, Z. Yang, Y. Zhao for discussions.  
 
  Y.C. and V.G. thank the KITP for hospitality where part of this research was done. V.G. is
partially supported by the Simons Foundation Origins of the Universe program (Modern Inflationary
Cosmology collaboration).
J.M. is supported in part by U.S. Department of Energy grant DE-SC0009988, the Simons Foundation grant 385600. J.M. also thanks the National Science Foundation under Grant No. NSF PHY11-25915 which supported his stay at the KITP at UCSB.

\appendix

\section{Discussion on the $g$ function } 
\la{App:gfunction}

\begin{figure}[h]
\begin{center}
\includegraphics[scale=.5]{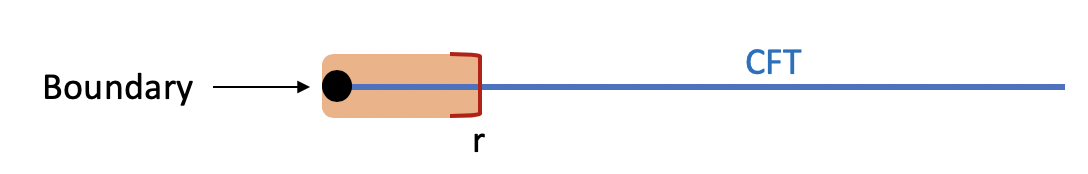}
\caption{ For a CFT with a boundary we define entropic $g$ function at scale $r$ by computing the von Neumann entropy of the shaded region which includes the boundary.  }
\label{gfunction}
\end{center}
\end{figure}

The entropic $g$ function introduced in \cite{Casini:2016fgb} is defined in terms of the entropy of an interval of size $r$ that includes the boundary, see figure \ref{gfunction}:
\be  \la{gfun}
S(r) = { c \over 6 } \log \left( { r \over \varepsilon_{uv} } \right) + c_0 + \log g(r) \,,
\ee
where $c_0$ is a constant that depends on the bulk CFT but not on the boundary conditions. 

We now consider  the usual Lorentzian interpretation for a black hole coupled to a CFT, as in \cite{Almheiri:2019psf}. We think of this system as a CFT with a (non-conformal) boundary. The computation
of the entropy of the interval was done in \cite{Almheiri:2019yqk}. 
It is obtained by using \nref{Sgen}, with $\tau=r$, and setting $\partial_z S_{gen} =0$ to obtain 
\be \la{quadr}
-z_* = { 3 \phi_r \over c } \left[ 1 + w + \sqrt{ 1 + 6 w + w^2 } \right] ~,~~~~~~~w \equiv { r c \over 6 \phi_r   } 
\ee 
We can then insert this answer into the entropy and define the $g$ function as in \nref{gfun}. We get 
\be \la{gfin}
\log g(r) = S_0 - c_0 + { c \over 6 } \left\{ {2  \over 1 + w + w_s} +
\log \left[ {( 1 + 3 w + w_s)^2 \over 2 w (1 + w + w_s) } \right] \right\} ~;~~~~~ w_s \equiv \sqrt{ 1 + 6 w + w^2 } ~~~
\ee
We see that for $r\to \infty$ we get that $\log g$ goes to a constant, in agreement with the idea that in this setup we flow to a conformal invariant boundary condition in the IR. 
For small $r$ it goes as $\log g \sim - { c \over 6 } \log r$ which comes from the fact that $S(r)$ is finite at $r=0$. This is indicating that the boundary has a diverging number of degrees of freedom in the UV. This divergence comes from the matter CFT degrees of freedom and their continuity across $r=0$. 

As expected from the entropic $g$ theorem of \cite{Casini:2016fgb}, we see that \nref{gfin} monotonically decreases as we increase $r$. 

Note that for this problem, we do not have the bra-ket wormholes because the flat space region is infinite in the direction orthogonal to the boundary.

 \section{Computing entropies of intervals for the compact case}

\subsection{Islands for the disconnected configuration} 
\la{app:Islands}

Here we discuss the island solutions in the disconnected geometry with a compact spatial direction and general Euclidean evolution $\tau$. The geometry is given in \nref{MetrCompFlat}, \nref{MetrCompDisc}. In the end, in section \ref{sec:checkparadox},  we will argue that whenever these islands appear, the bra-ket wormhole dominates and changes the answer. 

Let us consider a subregion $A$ with length $\ell$ in the $x$ coordinate, located at $z = \tau$. The angle $\Delta\theta$ that this subregion spans is $\Delta\theta = 2\pi \ell/L$. Thus its entropy in the case without island is given by
\be \la{naivedisconn}
 S_{\textrm{no island}} = \frac{c}{3} \log \left[ \frac{ 2\sin \frac{\Delta\theta}{2}}{\varepsilon_{uv,\theta}}\right]  = \frac{c}{3} \log \left[ \frac{  L \sin \frac{\pi \ell}{L}}{\pi\varepsilon_{uv}} \right]~,~~~~~~~ \varepsilon_{uv,\theta} = { 2 \pi \over L } \varepsilon_{uv}
\ee
We see that the entropy is maximized at $\ell = L/2$ since we have a compact circle. 

\begin{figure}[h]
\begin{center}
\includegraphics[scale=.25]{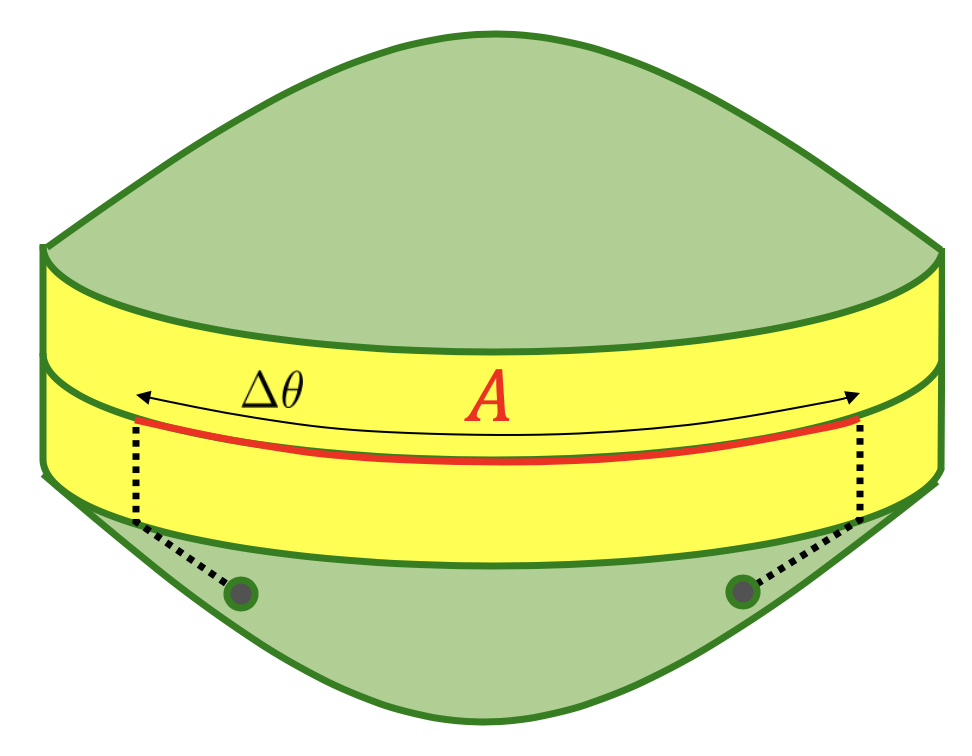}
\caption{We consider a non-trivial configuration of quantum extremal surfaces in the disconnected geometry.}
\label{fig:IslandCompact}
\end{center}
\end{figure}

Now we can also consider a configuration where we have quantum extremal surfaces in the gravity region, as depicted in fig. \ref{fig:IslandCompact}. In the limit that the interval $A$ is large, we will assume that we have a configuration that the computation factorizes into two intervals, each connecting one boundary point of $A$ to an extremal surface in the gravity region with the same $\theta$ coordinate.
The distance between the boundary point and a quantum extremal surface at $\sigma$ is given by $ \Delta\sigma =2\pi \tau/L  - \sigma$. Thus the generalized entropy function is given by
\be  \la{Sgendisconn}
S_{\textrm{gen}} (\sigma) = 2 \left\{  S_0 + \frac{2\pi \phi_r}{L} \frac{1}{(-\tanh \sigma)} + \frac{c}{6} \log \left[\frac{ \left( 2\sinh \frac{ \frac{2\pi \tau}{L}  - \sigma }{2} \right)^2  }{ (-\sinh \sigma) \varepsilon_{uv,\theta} } \right] \right\},
\ee
where the dilaton is given in (\ref{MetrCompDisc}).
Solving $\partial_\sigma S_{\textrm{gen}}(\sigma)=0$ gives us the location $\sigma_*$ of the quantum extremal surface. In the special case of $\tau=0$, we find
\be 
 	\sinh \sigma_{*} =  - \frac{12\pi \phi_r}{c L} ~,~~~~~~~~~~{\rm for} ~~~\tau=0
\ee
As   in section \ref{sec:Compact}, we will be interested in the case that $L\gg \phi_r/c$. Note that since $\sigma_* \ll 1$, the operator product expansion is a valid approximation in getting (\ref{Sgendisconn}). We get
\be \la{Sdisconn1}
 S_{\textrm{island}} = S_{\textrm{gen}} (\sigma_*) \approx 2 S_0 + \frac{c}{3} + \frac{c}{3} \log \frac{6 \phi_r}{c \varepsilon_{uv} }, \quad ~,~~~~~~~\textrm{for} \,\, \tau=0.
\ee

Comparing (\ref{Sdisconn1}) with (\ref{naivedisconn}), we find
\be 
 S_{\textrm{island}}  - S_{\textrm{no island}} = 2S_0 + \frac{c}{3} + \frac{c}{3} \log \frac{6\pi \phi_r}{cL \sin \frac{\pi \ell}{L}}.
\ee
There is a minimal total size $L$ of the spatial direction in order for an island to ever appear, which is given by
\be \la{Lbounddiss1}
	L \geq \frac{6\pi \phi_r}{c} \exp\left( \frac{6S_0 }{c} + 1  \right)
\ee

For general $\tau\neq 0$, extremizing (\ref{Sgendisconn}) gives us a function $\sigma_*(\tau)$. In the limit $\phi_r/c \ll \tau \ll L$, one has
\be 
 \sigma_*(\tau) \approx - \frac{2\pi \tau}{L} + \mathcal{O}\left( 
 \frac{\phi_r}{c L}\right) ,
\ee
and
\be \la{Sdisconn2}
S_{\textrm{island}} = S_{\textrm{gen}} (\sigma_*) \approx 2S_0 + \frac{2\phi_r}{\tau} + \frac{c}{3}\log \frac{8\pi \tau}{L \varepsilon_{uv,\theta}}. 
\ee
Comparing (\ref{Sdisconn2}) with (\ref{naivedisconn}), we find
\be 
 S_{\textrm{island}} - S_{\textrm{no island}} = 2S_0 + \frac{2\phi_r}{\tau} + \frac{c}{3} \log \frac{4\pi \tau}{L \sin \frac{\pi \ell}{L}}.
\ee
Again, for an island to appear, $L$ has a lower bound
\be \la{Lbounddiss2}
	L \geq 4\pi \tau \exp\left( \frac{6S_0}{c} + \frac{6\phi_r}{c \tau}\right).
\ee
(this is bigger than the similar looking distance in \nref{DomDis}.)
As one increases the amount Euclidean evolution $\tau$, one needs a longer total spatial length in order for an island to arise.

Similar to the discussion in sec. \ref{sec:paradox}, when island arises, we will run into a strong subadditivity paradox. As we will check in sec. \ref{sec:checkparadox}, in these cases, we should not use the disconnected geometry to do computation, as the wormhole geometry will be dominating.

\subsection{Islands for the connected configuration} \la{app:Islandsconn}
 
Here we discuss the entropy computations in the connected geometry with finite spatial direction and a general Euclidean evolution $\tau$. The metric and dilaton are given in \nref{metrflat}, \nref{metrworm}, \nref{dilWH}.  We will be considering a region $A$ at $z=\tau$ with length $\ell < L/2$. 

First we consider the entropy computed without introducing any quantum extremal surfaces. Since the region is much smaller than the total spatial direction, we can approximately view the spatial direction as non-compact and use the same formula as (\ref{SnaiveWorm}):
\be \label{SnaiveCompConn}
 S_{\textrm{no island}} = \frac{c}{3} \log  \left[ \frac{ \sinh (\pi T_{\chi}\Delta\chi)}{\pi T_{\chi} \varepsilon_{uv,\chi}} \right] = \frac{c}{3} \log \left[ \frac{\sinh \left(\pi T_x \ell \right)}{\pi T_x \varepsilon_{uv}} \right].
\ee
Here $T_{\chi},T_x$ are the effective temperature of the state measured in $\chi$ and $x$ coordinates:
\be 
 T_{\chi} = \frac{1}{\pi + 2\tau \frac{b}{L}}, \quad T_x = \frac{b}{L} T_{\chi} = \frac{1}{2\tau + \frac{\pi L}{b}},
\ee
where $b$ is determined by (\ref{eqnb}). The entropy grows linearly with the size of the interval.

\begin{figure}[h]
\begin{center}
\includegraphics[scale=.2]{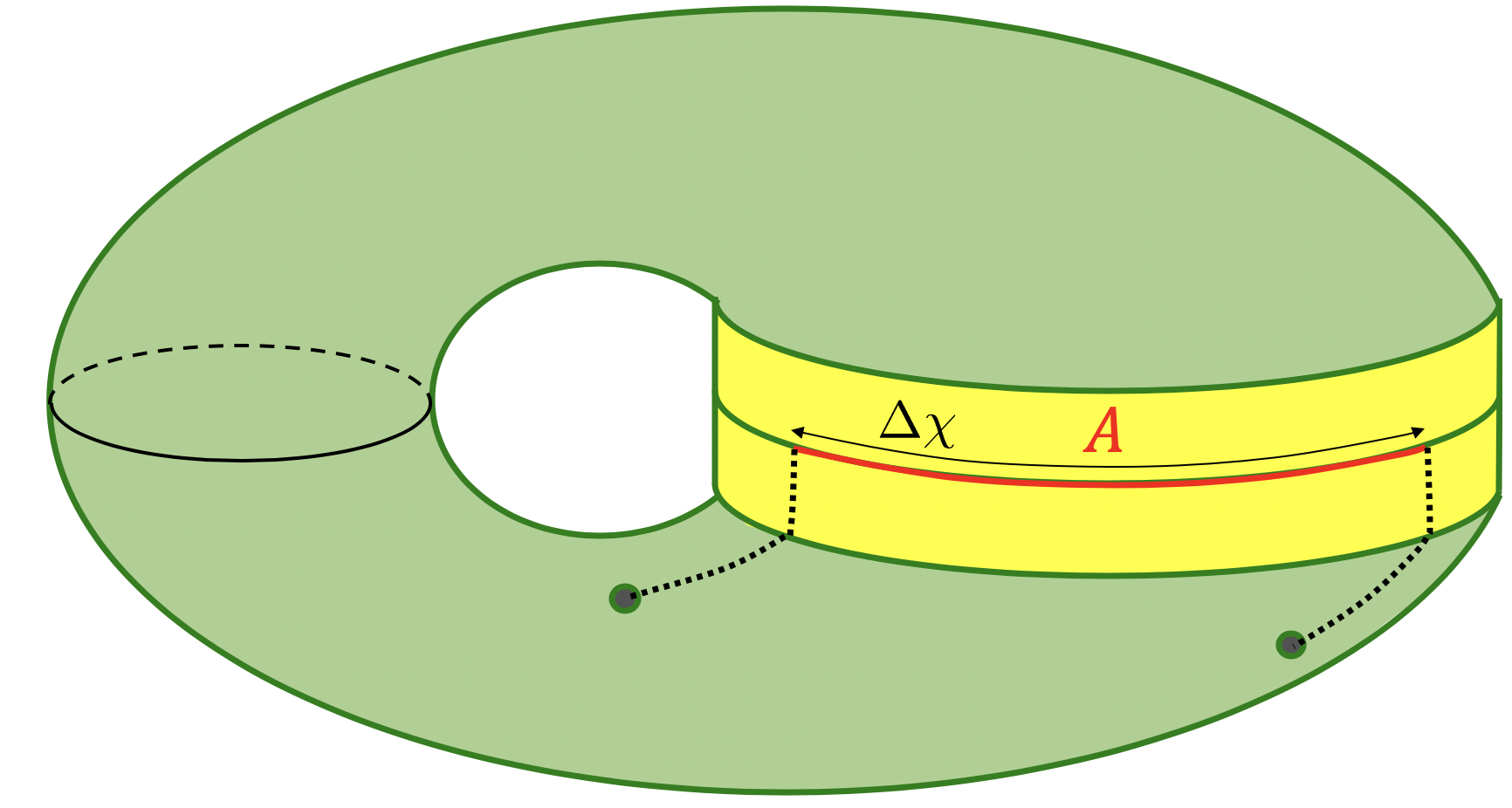}
\caption{We consider a non-trivial configuration of quantum extremal surfaces in the connected geometry.}
\label{fig:IslandCompactConn}
\end{center}
\end{figure}

We now consider the non-trivial configuration where we have quantum extremal surfaces in the dynamical gravity region as in fig. \ref{fig:IslandCompactConn}. When the interval $A$ is large, we are in the limit that we can find the quantum extremal surface associated to each boundary points independently. By symmetry, the quantum extremal surface will be at the same $\chi$ coordinate as the boundary point, and we only need to extremize over their $\sigma$ coordinate.
The dilaton profile has been found in (\ref{dilWH}). For the bulk entropy term in the QES prescription, we need to find the entropy of an interval extending from $z= \tau$ to location $\sigma$ in the bulk. In the $\sigma$ coordinate, the length of the interval is $\sigma + \tau b/L$, while the total length of the wormhole is $1/T_\chi$. When we compute the entropy of an interval, we again approximately treat the spatial direction as non-compact, which is justified as $b\gg 1$. Combining the dilaton and the matter entropy, we have
\be \la{SgenCompConn}
	S_{\rm gen} (\sigma) = 2 \left\{  S_0 +  \left[ \frac{c}{3 \left( 1+ \frac{2\tau b}{\pi L} \right)^2 } - \frac{c}{12}\right]\left( \frac{\frac{\pi}{2} -\sigma}{\tan \sigma} + 1\right) +\frac{c}{6} \log \left[  \frac{ \sin^2  \left( \pi T_{\chi} (\sigma + \frac{\tau b}{L}) \right) }{\left( \pi T_{\chi} \right)^2\sin \sigma\,  \varepsilon_{uv, \chi} }\right]  \right\},
\ee
Solving $\partial_\sigma S_{\rm gen} (\sigma)=0$, we find that we always have $\sigma_* = \pi/2$. Thus the quantum extremal surfaces are always on the throat of the wormhole, and $Z_2$ symmetry breaking does not happen. Taking $\sigma_* = \pi/2$ into (\ref{SgenCompConn}), we find
\be \label{SCompConn}
	S_{\textrm{island}} = 2 \left( S_0  + \frac{c}{3 }  (\pi T_{\chi})^2- \frac{c}{12} + \frac{c}{6} \log \left[ \frac{1 }{(\pi T_{\chi})^2 \varepsilon_{uv,\chi}} \right] \right).
\ee
Comparing (\ref{SCompConn}) and (\ref{SnaiveCompConn}), we find
\be 
S_{\textrm{island}} -S_{\textrm{no island}} = 2S_0 + \frac{2c}{3 } (\pi T_{\chi})^2 - \frac{c}{6} - \frac{c}{3} \log   \left[ \pi T_{\chi} \sinh (\pi T_\chi \Delta\chi)\right] .
\ee
Thus we see that the transition happens when
\be 
 \Delta\chi \geq \frac{1}{\pi T_{\chi}} \left( \frac{6S_0}{c} + \mathcal{O}(1)\right) ,
\ee
or equivalently
\be \la{ellcritical}
	\ell \geq  \ell_c \equiv \frac{1}{\pi T_{x}} \left( \frac{6S_0}{c} + \mathcal{O}(1)\right).
\ee
Since $T_{x}$ decreases as one increases $\tau$, the threshold length for an island to appear increases with $\tau$.

\subsubsection{Checking that the paradox does not appear} \la{sec:checkparadox}

We now proceed to check that the strong subadditivity paradox does not appear. As we checked explicitly in section \ref{app:Islandsconn}, whenever the wormhole is dominating, the quantum extremal surfaces lie on the throat of the wormhole, and we do not run into a paradox based on the argument presented in section \ref{ParadoxLost}. Thus we only need to check that whenever the disconnected saddle is dominating, it will not be convenient to have islands. 

We can first consider the possibility that the wormhole never dominates, namely $L$ is below the threshold given by (\ref{wormdominate}). However, since the right hand side of (\ref{wormdominate}) is always much less than the right hand side of (\ref{Lbounddiss1}) for arbitrary values of $S_0,c$, the island will never appear in the disconnected geometry, and one does not encounter paradoxes.

Another possibility is that a wormhole first dominates when $\tau$ is small, and then stops dominating at some point. We can check this explicitly in the case that $L$ is well above the threshold given by (\ref{wormdominate}), so that the value of $\tau$ where the wormhole becomes subdominant is given by (\ref{wormsubdom}). In other words, we have
\be  \la{Lboundcheck}
 L \lesssim \frac{96 S_0}{\pi c} \tau + \frac{576 \phi_r }{\pi c} \frac{S_0}{c}.
\ee
By comparing (\ref{Lboundcheck}) with (\ref{Lbounddiss2}), we also see that after the wormhole becomes subdominant, the island will not arise in the disconnected geometry. For the intermediate cases where the wormhole becomes subdominant at $\tau \sim \phi_r/c$, we expect the same to be true, since the inequalities used in the above argument are quite loose. This can be explicitly checked with the equation (\ref{eqnb}) for $b$ and the formulas for the partition functions (\ref{pardisconn}) and (\ref{Zcon}).

\subsubsection{Lorentzian time evolution } \la{app:IntervalLorentzian}

To gather further intuition about the emergent closed universe we can manipulate the positions of twist operators that define region $A$ and see how the island moves inside the closed universe. In particular, after a short Euclidean evolution of our state we can evolve it in Lorentzian time by taking $z=\tau+i   t$. Of course if the island includes the entire spacial slice of the closed universe it does not matter at which time it is placed, so we will consider a situation when an island of finite length dominates. 
So from each endpoint of the interval we get 
\be \la{EntrL}
S_0+\frac{c}{4} \left(1- \eta \tanh \eta \right)+\frac{c}{6}\log\left( \frac{\cosh^2(\eta-\tilde t)}{\cosh\eta \,\varepsilon_{uv,\chi}}\right)\
\,, ~~~~~~~~ \tilde t \equiv  { b t \over L } 
\ee 
Extremizing this with respect to $\eta$ we find that for small $t$ the island is located at $\eta\approx- \tilde t$, while taking $t$ large will push the island to some finite, order one, value of $\eta$. So we can move the island a little bit in the Lorentzian time of the closed universe. Importantly, the Euclidean time coordinate of the island does not change. In other words, the islands is at a real physical FLRW time. 
Note that these results are easy to understand. First, if we purely had the TFD state, neglecting the dilaton and the warp factor in \nref{FLRW}, 
the entropy would involve just the term $ { c \over 3 } \log \cosh(\eta - \tilde t) $. 
Then $\eta_* =-\tilde t$ would be the extremum. In other words, if we are free to move the time on the left side, we would want to set it to the case there is no net Lorentzian evolution for the TFD. In the actual problem we also have a dilaton and warp factor contributions. At large $\eta$, $\eta \gg 1$,  the behavior of \nref{EntrL} is roughly $ S \sim - { 5 c \over 12 } |\eta| + { c \over 3 } |\eta - \tilde t| $ which wants to take us back to $\eta$ of order one. Since $\eta_*$ is of order one, the final behavior of the entropy is then 
\be 
S_{\rm half}  = S_0 + \# c +  { c \over 3} \tilde t  = S_0 + \# c + { c \over 3 } \pi T_x t 
\ee 
This is the expected behavior of 
the entropy for a generic boundary state at large times, for a half of an infinite line \cite{Calabrese:2005in}.

  \section{Islands in the dS$_2$ Milne case}

In this appendix, we discuss the island configurations in the Milne case of the de Sitter discussion. The island configurations were drawn in figure \ref{fig:Milneisland}. We will first discuss the island in the Milne wedge (see figure \ref{fig:Milneisland} (b)), and then discuss the island  in the black hole interior (see figure \ref{fig:Milneisland} (a)). 

As we discussed around (\ref{Tfree}), the problem has a free parameter which is the physical temperature $T_x$ in the flat space region. It also corresponds to the coefficient $\tilde{\phi}_r$ in the expression of the dilaton (\ref{Milne}). In this appendix we will fix this parameter, and discuss the islands on a fixed geometry. We take the matter fields on the geometry to be in the Hartle-Hawking vacuum.  

We will also neglect the possible bra-ket wormhole configurations discussed in section. \ref{sec:dSbraket}. However, as we will explain in appendix. \ref{app:2piHHCFT}, the results given by the Hartle-Hawking vacuum for the CFT can be equivalently viewed as the one given by the $2\pi$ contour discussed in section. \ref{sec:2PiContour} for the non-compact case. In other words, here we are only neglecting the identity contour and the $\pi$ contour.

\subsection{An island in the Milne region}\la{app:islandMilne}

\begin{figure}[t!]
\centering  
\begin{tikzpicture}[thick,scale = 1.5]

\draw (-1,0) -- (1,0) -- (0,1) -- (-1,0);
\draw[dashed] (-1,0) -- (-2,-1) -- (0,-3);
\draw[dashed] (-1,0) -- (1,-2);
\draw[dashed] (1,0) -- (2,-1) -- (0,-3);
\draw[dashed] (1,0) -- (-1,-2);
\filldraw[cyan,opacity = 0.05] (0,1) -- (-2,-1) -- (0,-3) -- (2,-1);
\draw[decorate, decoration={snake, segment length=5, amplitude=1}] (1,0) -- (2,0); 
\draw[decorate, decoration={snake, segment length=5, amplitude=1}] (-1,0) -- (-2,0); 

 \draw[->] (-0.3,-0.1)  -- (0.3, -0.1) node[midway,below]{$\chi$};
  \draw[->] (-0.5,-0.35)  arc[radius=1  , start angle=200, end angle=160];
  \draw (-0.7,0.15) node{$\eta$};

\draw[decorate, decoration={snake, segment length=5, amplitude=1}] (1,-2) -- (2,-2); 
\draw[decorate, decoration={snake, segment length=5, amplitude=1}] (-1,-2) -- (-2,-2); 
\draw (-1,-2) -- (1,-2);

\draw[->] (0,-1) -- (0.4,-1+0.4) node[right]{$w^+$};
\draw[->] (0,-1) -- (-0.4,-1+0.4) node[left]{$w^-$};
\end{tikzpicture}
\caption{The conformal field is in the vacuum state of a Minkowski spacetime as shaded in blue, which is conformally equivalent to the original metric. We can parametrize the region with coordinates $w^{\pm}$.}
\label{fig:Milne}
\end{figure}
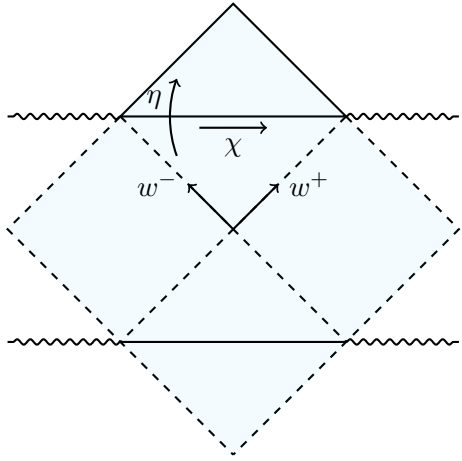

The metric and the dilaton for the Milne patch are
\be 
 ds^2 = \frac{-d\eta^2 + d\chi^2}{\sinh^2 \eta}, \quad ~~~\eta< -\eta_c, ~~~~\chi \in (-\infty,\infty), ~~~ \quad \phi =  \frac{\tilde \phi_r}{-\tanh \eta}
\ee
The Hartle-Hawking vacuum for the CFT is conformally equivalent to the vacuum state on a cylinder, which is also equivalent to the Minkowski vacuum in a Poincare patch on the cylinder. In figure \ref{fig:Milne}, we denoted this Poincare patch in blue. Up to the Weyl factor, the Milne region covers the usual Milne wedge of the Minkowski spacetime. One can parametrize the Minkowski region by coordinates
\be
	w^{\pm} = e^{\eta\pm \chi}, \quad -dw^+ dw^- = e^{2\eta} (-d\eta^2 + d\chi^2).
\ee
The metric of the Milne region and the subsequent flat space region are then given by
\be
	ds^2 = \frac{-dw^+ dw^-}{e^{2\eta}\sinh^2 \eta} \quad (\textrm{Milne region}), \quad \quad 	ds^2 = \frac{-dw^+ dw^-}{e^{2\eta} \epsilon^2 } \quad (\textrm{flat space region}).
\ee

Now we consider a region with size $\Delta \chi$ in the flat space region. For simplicity, we consider the region to be close to the surface where the Milne region is joined to flat space. The ordinary formula for the entropy would be
\be 
   S_{\textrm{no island}} = \frac{c}{6} \log \left[ \frac{|\Delta  w^+ \Delta w^- | }{\varepsilon_{uv}^2}\right] =  \frac{c}{3} \log \left[ \frac{2 \sinh \frac{\Delta \chi}{2}}{\varepsilon_{uv,\chi}}\right].
\ee
The entropy grows linearly with the size of the interval when the region is large, which is in accordance with the fact that we have a thermal state with inverse temperature $2\pi$ in the flat space region.

We can consider the possibility that for each of the boundary points, there is an emergent extremal surface at some $\eta <0$. For each pair of twist operators, the separation in $w^\pm$ coordinate is
\be
 -\Delta w^+ \Delta w^- = -( 1 - e^{\eta} )^2 = -e^{\eta}\left(2\sinh \frac{\eta}{2} \right)^2,
\ee
and thus we have
\be \la{IntMil}
	S_{\textrm{gen}} (\eta) = 2 \left\{ S_0 -  \frac{ \tilde \phi_r}{\tanh \eta} + \frac{c}{6} \log \left[\frac{\left(2\sinh \frac{\eta}{2} \right)^2}{ (-\sinh \eta) \varepsilon_{uv,\chi}} \right] \right\} .
\ee
Extremizing over $\eta$ we get
\be 
  \sinh \eta_* = - \frac{6 \tilde \phi_r}{c}.
\ee
We can have a solution for arbitrary values of $\tilde \phi_r/c$. As a simple case, when $S_0 \gg \tilde \phi_r/c \gg 1$, we have
\be 
   S_{\textrm{island, Milne}} \approx  2 (S_0  + \tilde \phi_r ),
\ee
where we omitted the dependence on the UV cutoff $\varepsilon_{uv,\chi}$. 
Thus the island contribution dominates when 
\be 
	\frac{c}{3} |\Delta \chi| \gtrsim 2(S_0 + \tilde \phi_r).
\ee
The region does not need to be as large as in the Poincare case in order for the transition to happen, which is due to the fact that we have a thermal state in the flat space region, and the naive entropy grows linearly with the size of the region.

\subsection{An island in the black hole interior}\la{app:islandinterior}

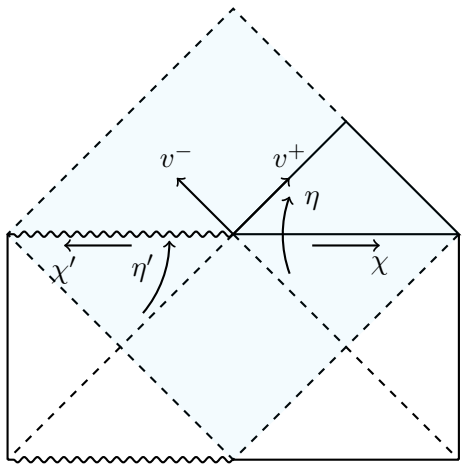
\begin{figure}[t!]
\centering  
\begin{tikzpicture}[thick,scale = 1.5]

\draw (-1,0) -- (1,0) -- (0,1) -- (-1,0);
\draw[dashed] (-1,0) -- (-3,-2);
\draw[dashed] (-1,0) -- (1,-2);
\draw[dashed] (1,0) -- (-1,-2) -- (-3,0) -- (-1,2) -- (0,1);
\filldraw[cyan, opacity = 0.05] (1,0) -- (-1,-2) -- (-3,0) -- (-1,2) -- (1,0);

\draw[decorate, decoration={snake, segment length=5, amplitude=1}] (-1,0) -- (-3,0);
\draw[decorate, decoration={snake, segment length=5, amplitude=1}] (-1,-2) -- (-3,-2); 
\draw (-1,-2) -- (1,-2);

 \draw[->] (-0.3,-0.1)  -- (0.3, -0.1) node[below]{$\chi$};
  \draw[->] (-0.5,-0.35)  arc[radius=1  , start angle=200, end angle=160];
  \draw (-0.3,0.3) node{$\eta$};
 \draw[->] (-1.9,-0.1)  -- (-2.5, -0.1) node[below]{$\chi'$};
 \draw[->] (-1.8,-0.7)  arc[radius=1  , start angle=320, end angle=360];
 \draw (-1.8,-0.3) node{$\eta'$};

\draw [->] (-1,0) -- (-1+0.5,0.5) node[above]{$v^+$};
\draw [->] (-1,0) -- (-1-0.5,0.5) node[above]{$v^-$};

\draw (-3,0) -- (-3,-2);
\draw (1,0) -- (1,-2);
\end{tikzpicture}
\caption{The conformal field is in the vacuum state of a Minkowski spacetime as shaded in blue, which is conformally equivalent to the original metric. We can parametrize the region with coordinates $v^{\pm}$.}
\label{fig:Milneinterior}
\end{figure}

We can also search for an island in the black hole interior. The metric and dilaton in the interior region are  (see figure \ref{fig:Milneinterior}): 
\be 
 ds^2 = \frac{-d\eta'^2 + d\chi'^2}{\sinh^2 \eta'}, \quad \eta' <0,\quad
	\phi =  \frac{\tilde \phi_r}{\tanh \eta'}.
\ee
This can be obtained from \nref{Milne} by setting $\eta = i \pi - \eta'$.
It would be convenient to introduce another set of coordinates $v^{\pm}$, which cover both the flat space region and the black hole interior:
\be \la{nullMilne}
	v^{\pm} = \pm e^{\pm \eta + \chi} ,~~~~ \quad v^{\pm } =\mp e^{\mp \eta' + \chi'} ~,~~~~~~~~ \eta' = i \pi - \eta 
\ee
We have defined $\chi$ and $\chi'$ such that they both increase or decrease when we apply a dilation at $v^\pm =0$. Then the metrics of the interior region and the flat space region are
\be 
 ds^2 = \frac{-dv^+ dv^-}{e^{2\chi'} \sinh^2 \eta'} \quad (\textrm{interior region}), \quad \quad 	ds^2 = \frac{-dv^+ dv^-}{e^{2\chi} \epsilon^2 } \quad (\textrm{flat space region}).
\ee

Since the blue shaded region in figure \ref{fig:Milneinterior} covers another Poincare patch of the cylinder in which the CFT is in the vacuum state, the CFT is in the Minkowski vacuum with respect to the $v^{\pm}$ coordinates. We imagine having two extremal surfaces in the interior region. When the original interval is large, one of the end point will be close to $v^{\pm} = 0$, while the other will be far away. In the interior region, we can also place one of the extremal surfaces to be close to $v^{\pm} =0$, while the other to be far away. Then we are again in an OPE limit, where two twist operators near $v^{\pm}=0$ are close to each other, while the two twist operators far away from $v^{\pm}$ are also close as the spatial direction is identified at $\chi,\chi' \rightarrow \infty$. Let us then focus on one pair of the twist operators, one at $(\eta, \chi) = (0, \chi_0)$, and the other at $(\eta' ,\chi') $. In the $v^{\pm}$ coordinates, their separation is
\be 
 - \Delta v^+ \Delta v^- = -(e^{\chi_0}  + e^{-\eta' + \chi' })(- e^{\chi_0} - e^{\eta' + \chi'}) = 2 e^{\chi_0 + \chi'} (\cosh (\chi_0 - \chi') + \cosh \eta'),
\ee
and thus the contribution to $S_{\textrm{bulk}}$ from one pair of twist operators is
\be 
	 \frac{c}{6} \log \left[\frac{ 2 e^{\chi_0 + \chi'} (\cosh (\chi_0 - \chi') + \cosh \eta') }{  e^{\chi'}(-\sinh \eta' ) e^{\chi_0} \varepsilon_{uv,\chi}   }\right] = \frac{c}{6} \log \left[\frac{ 2  (\cosh (\chi_0 - \chi') + \cosh \eta') }{ (- \sinh \eta' )  \varepsilon_{uv,\chi}  } \right] .
\ee
Since the dilaton has no dependence on $\chi'$, we can extremize over $\chi'$ and get $\chi'_* = \chi_0$. We see that $\chi'_*$ approaches $\pm \infty$ in the same direction as $\chi$, which is consistent with our assumption about the OPE limit. Then the generalized entropy function is given by
\be 
	S_{\textrm{gen}} (\eta')= 2 \left\{ S_0 + \frac{\tilde \phi_r}{\tanh \eta'} +  \frac{c}{6} \log \left[\frac{ 2  (1+ \cosh \eta') }{  (-\sinh \eta')  \varepsilon_{uv,\chi}   } \right]\right\}.
\ee
Extremizing with respect to $\eta'$ we get
\be 
    \sinh \eta'_* = - \frac{6 \tilde \phi_r}{c}.
\ee 
 We also have solutions for arbitrary values of $\tilde \phi_r/c$. As a simple case, let us  also consider the case where $S_0 \gg \tilde \phi_r/c \gg 1$, and 
 \be 
  S_{\textrm{island,interior}} \approx 2 (S_0 - \tilde \phi_r) .
 \ee
We see that the island in the interior leads to a smaller entropy than the island in the Milne wedge and thus dominate when $\tilde \phi_r/c \gg 1$. The transition from the naive answer to $ S_{\textrm{island,interior}}$ occurs when
\be 
   \frac{c}{3} |\Delta \chi | \gtrsim 2 (S_0 -  \tilde \phi_r),\quad ~~~~~ \frac{\tilde \phi_r}{c} = { 2 \pi \phi_r T_x \over c } \gg 1.
\ee
 These interior islands are very similar to the islands discussed in the black hole context \cite{Almheiri:2019psf,Penington:2019npb}, particularly in the setup of \cite{Almheiri:2019yqk}. 
For this reason,   we expect that such islands in the interior also exist for Schwarzschild de Sitter spacetime in any dimensions. They are just the usual black hole islands, except that the black hole is evaporating into an expanding universe.

When $\tilde \phi_r/c$ is small,  we can have a situation where the island in the Milne wedge   dominates over the interior island by winning in the matter entropy term.

\subsection{Equivalence between the $2\pi$ contour and the Hartle-Hawking vacuum for the CFT} \la{app:2piHHCFT}

Here we comment on the connection of the discussion of \ref{app:islandMilne} and \ref{app:islandinterior} to the bra-ket wormhole with the $2\pi$ contour. The observation is that when we consider the reduced density matrix of the Hartle-Hawking vacuum inside the Milne wedge, we get the same thermal state as that from the $2\pi$ contour. 

To see this, we note that in figure \ref{fig:Milneinterior}, the regions covered by $\{\eta,\chi\}$ and $\{\eta',\chi'\}$ are conformal to parts of the two Rindler wedges of the Minkowski spacetime as shaded in blue. Since we are considering conformal fields, we can neglect the non-trivial Weyl factor of the metric. A well-known fact is that the Minkowski vacuum gives rise to a thermal state with inverse temperature $2\pi$ in the Rindler wedge. Thus we know that it is equivalent to the thermal state given by the $2\pi$ contour. 

\begin{figure}[h]
\begin{center}
\includegraphics[scale=0.25]{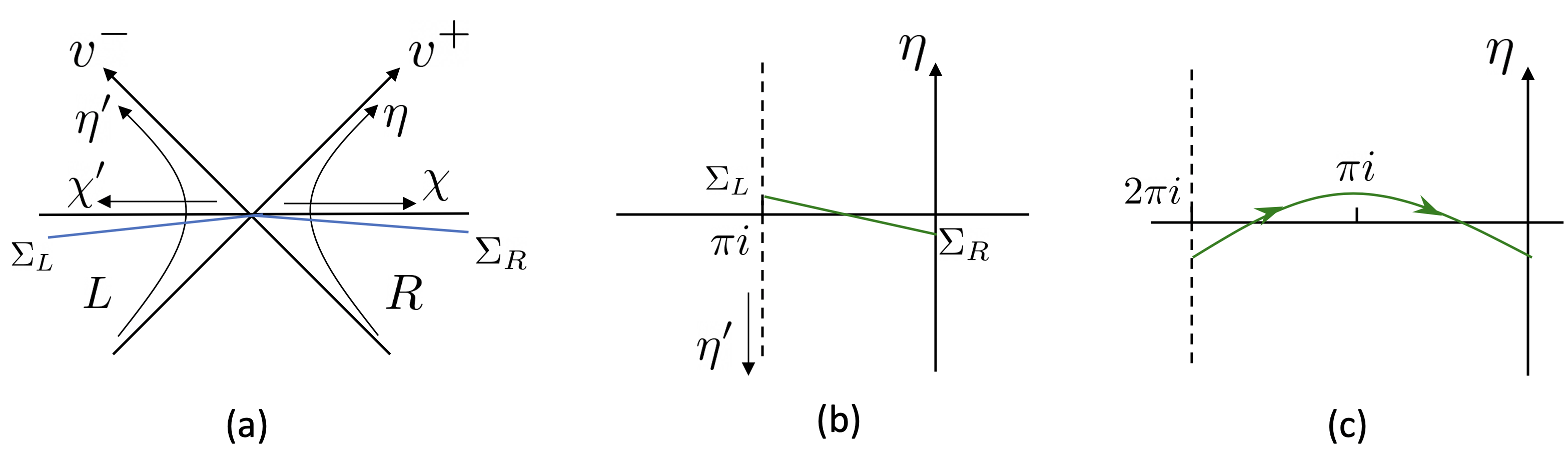}
\caption{(a) We highlight the Rindler structure in figure \ref{fig:Milneinterior}. The black horizontal line denotes the future boundary of the de Sitter spacetime. We consider the Hartle-Hawking vacuum on the slice $\Sigma = \Sigma_L \cup \Sigma_R$ that is close to the future boundary. (b) The contour that prepares the ket state on $\Sigma$. (c) When we combine (b) with the contour for the bra and trace out $\Sigma_L$, we get the $2\pi$ contour for the density matrix.}
\label{fig:Rindler}
\end{center}
\end{figure}

However, we can make the connection between the vacuum and the $2\pi$ contour much more explicit. Let us first consider the entangled state on a slice (the blue line in figure \ref{fig:Rindler} (a)) that is close to the future boundary of the de Sitter spacetime. We would like to find the contour such that we can go smoothly from the right wedge covered by $\eta$ coordinate to the left wedge covered by $\eta'$ coordinate. 
From (\ref{nullMilne}), $\eta$ and $\eta'$ can be expressed in terms of light cone coordinates:
\be 
 \eta = \frac{1}{2} \log v^+ - \frac{1}{2} \log (-v^-),\quad \eta' =   \frac{1}{2} \log v^- - \frac{1}{2} \log (-v^+).
\ee
In the path integral that prepares the ket state of Minkowski vacuum, we demand the boundary condition that the fields are regular at $\textrm{Im} (t)>0$, which selects the negative frequency solutions.  This tells us that in the contour that relates $\eta$ to $\eta'$, we should continue both $v^+$ and $v^-$ through the upper half plane. Thus we find that
\be 
 \eta = - \eta' + i\pi.
\ee
This gives us the contour shown in figure \ref{fig:Rindler} (b). Note that we have $\eta,\eta' <0$ at both ends of the contour. 
Similarly, the contour for the bra satisfies $\eta = -\eta' -i\pi$. To get the contour for the density matrix for $\Sigma_R$, we can simply take one ket contour and one bra contour, and join them by tracing out $\Sigma_L$. This gives us the $2\pi$ contour shown in figure (\ref{fig:contourseta}) (b).

Note that the above argument only relates the Hartle-Hawking state to the $2\pi$ contour with a non-compact spatial direction, while the $2\pi$ contour is also well defined for compact cases. The argument also only applies to conformal fields. However, in \cite{Sasaki:1994yt}, it was shown that for free massive scalar fields in dS$_4$, the Hartle-Hawking vacuum in the Milne wedge (or hyperbolic slicing)   is also equivalent to the state given by a $2\pi$ contour, so we expect there could be more general statement relating the two.

\section{Checking the dS contours against the Kontsevich-Segal-Witten criterion} \label{app:KSW}

\textbf{Note:} This appendix was added in v4. 

In this appendix, we check whether the de Sitter bra-ket wormhole contours studied in sec. \ref{sec:dSbraket} satisfy the allowability criterion proposed by Kontsevich, Segal and Witten in \cite{Kontsevich:2021dmb,Witten:2021nzp}. The conclusion is that they do not satisfy the criterion, as we will check case by case. The criterion is stated as follows. For a complex metric $g$ on a $D$ dimensional manifold, one can pick a real basis in the tangent space such that the metric is diagonal
\begin{equation}
g_{ij} = \lambda_i \delta_{ij}, \, \, i,j = 1,...,D,
\end{equation}
with complex eigenvalues $\{\lambda_i\}$. The metric is only allowable in the gravitational path integral if at every points on the manifold one satisfies the constraint
\begin{equation}\label{constraint}
	\sum_{i = 1}^{D} |\textrm{Arg} (\lambda_i)| < \pi. 
\end{equation}
The allowability criterion originates from demanding the convergence of the path integral of a $p$-form gauge field, for any $p$, see \cite{Kontsevich:2021dmb,Witten:2021nzp} for more details on the derivation. For the case of $D=2$ that we are interested in, the criterion has been proposed earlier by Louko and Sorkin in \cite{Louko:1995jw}. 

\begin{figure}[h]
\begin{center}
\includegraphics[scale=.27]{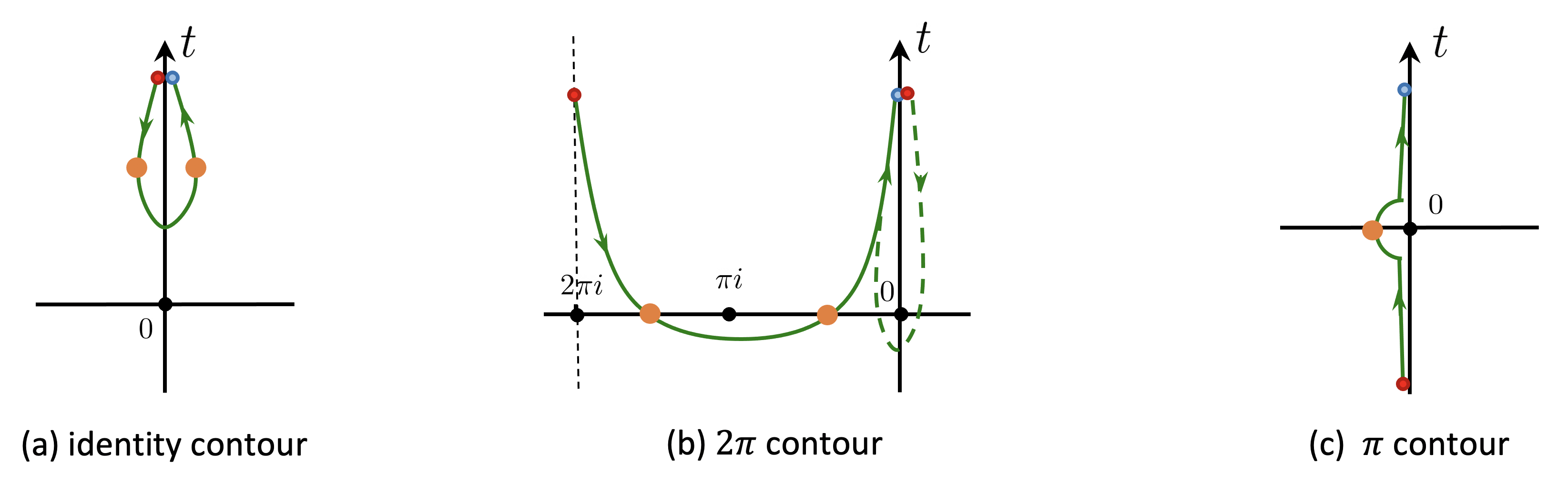}
\caption{The three dS contours studied in sec. \ref{sec:dSbraket}.}
\label{fig:contoursKSW}
\end{center}
\end{figure}

Now we check each of our contours against the criterion. The three contours are drawn in fig. \ref{fig:contoursKSW}.

\paragraph{The identity contour}

We can parametrize the identity using a real parameter $u\in [0,1]$ where $u=0$ corresponds to the red point and $u=1$ corresponds to the blue point in fig. \ref{fig:contoursKSW} (a). The complex metric along the contour can be written as
\begin{equation}\label{identityKSW}
ds^2 = e^{-i\pi} t'(u)^2 du^2 + \sinh^2 t(u) d\chi^2,
\end{equation}
where $t'(0) = -1 + i\epsilon,\, t'(1) = 1+ i\epsilon$ and $\epsilon$ is an infinitesimal positive number. The starting point and the end point lie on the real time axis, i.e. $t(0) , t(1) \in \mathbb{R}$. We write the minus sign in the metric (\ref{Milne}) as $e^{-i\pi}$ such that at the two end points $u=0,1$ the metric satisfies the constraint (\ref{constraint}). Since 
\begin{equation}
\int_0^1 t'(u) du = t(1) - t(0) \in \mathbb{R},
\end{equation}
the imaginary part of $t'(u)$ has to cancel under the integral. Because the imaginary part of $t'(u)$ is positive at both ends $u=0,1$, there must exist $u_* \in (0,1)$ such that $\textrm{Im}[t'(u_*)] = 0$. Therefore at $u_*$, the phase coming from the $du^2$ term in (\ref{identityKSW}) is exactly $\pi$, so it cannot satisfy the constraint (\ref{constraint}). We marked the points with $\textrm{Im}(t'(u_*)) = 0$ in fig. \ref{fig:contoursKSW} (a) as orange dots.

\paragraph{The $2\pi$ contour}

There are two different $2\pi$ contours discussed in sec. \ref{sec:dSbraket}, and we have to check them separately. One can apply the same argument for the identity contour to rule out the dashed line in fig. \ref{fig:contoursKSW} (b). 

For the solid line in fig. \ref{fig:contoursKSW} (b), one applies a different argument. Note that the contour always crosses the line where $t$ is purely imaginary, as it has to go around the point $t=i\pi$ from below. We mark the crossing points in fig. \ref{fig:contoursKSW} (b) as orange dots. At the crossing points $t= i\tau$, the $\sinh^2 t = - \sin^2 \tau$ factor in front of $d\chi^2$ is a negative real number, so the phase from this term is already $\pi$, which violates the constraint (\ref{constraint}).

\paragraph{The $\pi$ contour}

Since the $\pi$ contour starts at positive $t$ and ends at negative $t$, it has to cross the imaginary axis at some $t = i\tau$. We mark the crossing point in fig. \ref{fig:contoursKSW} (c) by an orange dot. At the crossing point, we have the same argument as above, namely $\sinh^2 t = -\sin^2 \tau$ in front the $d\chi^2$ will be a negative real number, which violates the constraint (\ref{constraint}).

~

We've seen that none of the de-Sitter bra-ket wormhole contours we considered satisfy the allowability criterion in \cite{Kontsevich:2021dmb,Witten:2021nzp}. Note that the criterion was derived by coupling the metric to various kinds of matter fields, while in this paper we have only considered the coupling to CFTs. If we only require CFTs to be well-defined on the contour, we will have a milder constraint than (\ref{constraint}). A conformal field theory does not see an overall constant phase in front of the metric, so if a complex metric can be written as a constant phase multiplying an allowable metric, it can still be coupled to a CFT despite being unallowed under (\ref{constraint}). An example is our $\pi$ contour, which, as we discussed in sec. \ref{summarydS}, can be written as an overall minus sign multiplying the ``double-cone" geometry. The ``double-cone" geometry is an allowable geometry as discussed in \cite{Witten:2021nzp}, and so the $\pi$ contour will be allowed if only conformal fields were put on it. Of course, JT gravity coupled to conformal fields is not a UV complete theory by itself, so it is unclear one can justify the validity of the contours by restricting to CFTs only.

Our discussion of the dS bra-ket wormhole contours was motivated partly through an entropy subadditivity paradox, similar to the Euclidean AdS discussion in sec. \ref{sec:paradox}.\footnote{Our Euclidean AdS discussion is unaffected by the allowability criterion, since in that case we have real Euclidean metrics, which are perfectly allowed under (\ref{constraint}).}  Despite the similarity to the Euclidean AdS case, the dS discussion involves time-like separated twist operators. It is plausible that such configurations will also have trouble passing the criterion (\ref{constraint}). However, to check this, it's likely one would have to study the replica geometries with finite replica number $n$, which have not been constructed explicitly.

\section{More on islands for the regularized $\pi$ contour } 
\label{app:piIslands}

\begin{figure}[h]
\begin{center}
\includegraphics[scale=0.27]{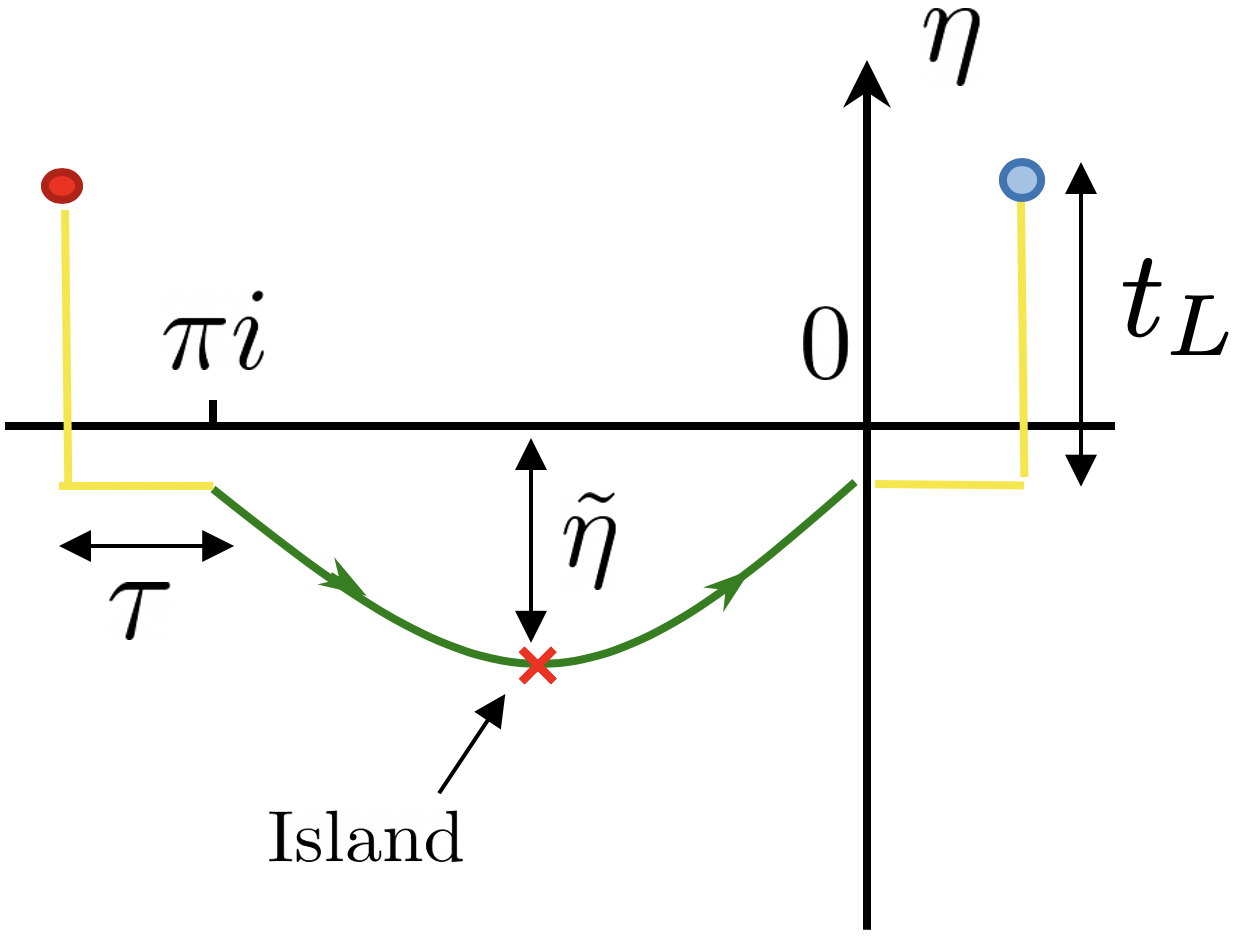}
\caption{We evolve the state described by the $\pi$ contour further in Lorentzian time with amount $t_L$, after an Euclidean regularization with amount $\tau$. The possible location of the island is marked by the red cross. }
\label{fig:LorentzianPiIsland}
\end{center}
\end{figure}

In this appendix, we discuss further details of the islands for the regularized $\pi$ contour in section. \ref{sec:PiContour}. In particular, we imagine evolving the state in Lorentzian signature after an amount $\tau$ of Euclidean regularization, and compute the entropy for an interval at Lorentzian time $t_L$. The contour including the Lorentzian evolution is drawn in figure \ref{fig:LorentzianPiIsland}.

For simplicity, let us consider the case with a non-compact spatial direction and an interval $A$ that is a semi-infinite line. Its naive entropy   is infinite so the island configuration will always dominate. As shown in the figure \ref{fig:LorentzianPiIsland}, we try an ansatz in which the quantum extremal surface is at $\eta= i\pi/2 - \tilde{\eta}$ and then extremize over the value of $\tilde{\eta}$. The distance between the quantum extremal surface and the original twist operator in $\eta$ coordinate is
\be 
 \Delta\eta =   2\pi T_x  (-i \tau  +  t_L )-\left(  i \frac{\pi}{2}  - \tilde{\eta}  \right) =  \tilde \eta + \frac{\pi t_L}{2\tau} - i\pi,
\ee
where we used the value of $T_x = \frac{1}{4\tau}$. Thus the generalized entropy function for the semi-infinite line is given by
\be 
\begin{aligned} 
 S_{\textrm{gen}} (\tilde{\eta} )  & =  S_0 + \frac{\tilde \phi_r}{\coth \tilde \eta}    + \frac{c}{6} \log  \left[ \frac{ \left(  2 \sinh \frac{\Delta\eta}{2}   \right)^2 }{\cosh \tilde\eta \,\varepsilon_{uv,\chi}} \right]   \\
 & =    S_0 + \frac{\tilde \phi_r}{\coth \tilde \eta}    + \frac{c}{6} \log  \left[ \frac{ \left(  2 \cosh \frac{ \tilde \eta + \frac{\pi t_L}{2\tau} }{2}   \right)^2 }{\cosh \tilde\eta \, \varepsilon_{uv,\chi}} \right]   .
\end{aligned}
\ee
Here we have set $\eta = i\pi/2 - \eta$ in the expression for the dilaton and the warp factor appearing in the logarithm,  as compared to \nref{IntMil}. We have also discarded an $i$ inside the logarithm. This $i$ does not affect the extremization procedure. It arises because of a phase appearing in the scale factor of the metric as we go along the contour. For this particular computation the $i$ could be removed by raising the argument of the logarithm to the fourth power and dividing the constant multiplying the logarithms by four. It would be nice to understand how this should be treated in a more proper way. Extremizing over $\tilde\eta$, we get the following equation
\be \la{tildeEta}
 \cosh^2 \tilde\eta \left( \tanh \tilde{\eta} - \tanh \frac{\tilde{\eta} + \frac{\pi t_L}{2\tau}}{2} \right) = \frac{6\tilde{\phi_r}}{c}. 
\ee
This equation has solutions for arbitrary $t_L$. To see the behavior of the entropy under Lorentzian evolution, let us consider a special limit where $t_L \gg \tau$. In this limit, we can write $\tilde{\eta} = \pi t_L/(2\tau) + \delta\tilde{\eta}$ and expand the equation as
\be 
 \frac{e^{2 \left(  \frac{\pi t_L}{2\tau}+ \delta\tilde{\eta} \right) }}{4} \left(  - 2 e^{- 2 \left(    \frac{\pi t_L}{2\tau} + \delta\tilde{\eta}  \right)}  +  2 e^{- 2 \left(    \frac{\pi t_L}{2\tau}  + \frac{\delta\tilde{\eta}}{2}   \right)}  \right) \approx \frac{6\tilde{\phi}_r}{c},
\ee
and find
\be 
 \delta \tilde{\eta} = \log  \left( \frac{12\tilde\phi_r}{c} + 1\right),
\ee
which is an order one number. Thus we have
\be 
S \approx  S_0 +  \tilde{\phi}_r + \frac{c}{6} \log \left[ \frac{4 \cosh \frac{\pi t_L}{2\tau}}{\varepsilon_{uv,\chi}}\right]  = S_0+  \tilde{\phi}_r +  \frac{c}{6} \log\left[ \frac{2}{\varepsilon_{uv,\chi}}\right] + \frac{c}{3} \pi T_{x} t_L, \quad t_L \gg \tau.
\ee
We see that the entropy grows linearly with $t_L$, with a rate that is consistent with the entanglement growth after a quench.

 \section{Relative entropy between the exact state and the thermal state} 
 \la{app:relative}

 We have seen that $\sigma(\tau)$ prepared by gravitational  evolution as discussed in section \ref{sec:purity} is actually pure, despite looking thermal for simple observables. In other words, we have $\sigma (\tau) = \ket{B(\tau)}\bra{B(\tau)}$. In this appendix, we quantify the difference between the two by looking at the relative entropy between $\sigma$ and the precisely thermal one 
$\rho_{\beta}= \frac{1}{Z(\beta)} e^{-\beta H_{CFT}}$  
\be 
\begin{aligned}
 S(\sigma (\tau) | \rho_\beta ) & = \textrm{Tr} \left[ \sigma (\tau) \log \sigma (\tau) - \sigma (\tau) \log \rho_\beta \right] \\
 & =  \beta \langle H_{CFT} \rangle_{\sigma (\tau)} + \log  Z(\beta).
 \end{aligned}
\ee
We can extremize over the value of $\beta$ to find the thermal density matrix that resembles $\sigma(\tau)$ the most. We have
\be \la{bestrel}
 0 = \frac{d}{d\beta}  S(\sigma (\tau) | \rho_\beta )  =   \langle H_{CFT} \rangle_{\sigma (\tau)} -   \langle H_{CFT} \rangle_{\rho_\beta } .
\ee
Of course, this is simply telling us that the best fit of thermal density matrix should have the same energy (or temperature) as the exact $\sigma(\tau)$ in the semiclassical description. When we take the value of $\beta$ that satisfies (\ref{bestrel}), we find
\be
 S (\sigma (\tau) | \rho_\beta )  = S(\rho_\beta),
\ee
namely the relative entropy is given by the thermal entropy of the corresponding temperature. This is a large quantity which scales as the size $L$ of the system. Since the effective temperature decreases as we increase $\tau$, the relative entropy also decreases, which is what we expect as we are projecting into low energy states. 

This discussion is valid for any typical pure state with the same energy as a thermal state.

 \section{Projection into the typical state that is $|B\rangle$ itself} 
 \la{Bmicro} 
 
 Though we do not have a general resolution to the factorization puzzle of section \ref{sec:Typical}, we would like to point out the following calculation where it does not arise, at the level of classical solutions.
 
\begin{figure}[h]
\begin{center}
\includegraphics[scale=0.23]{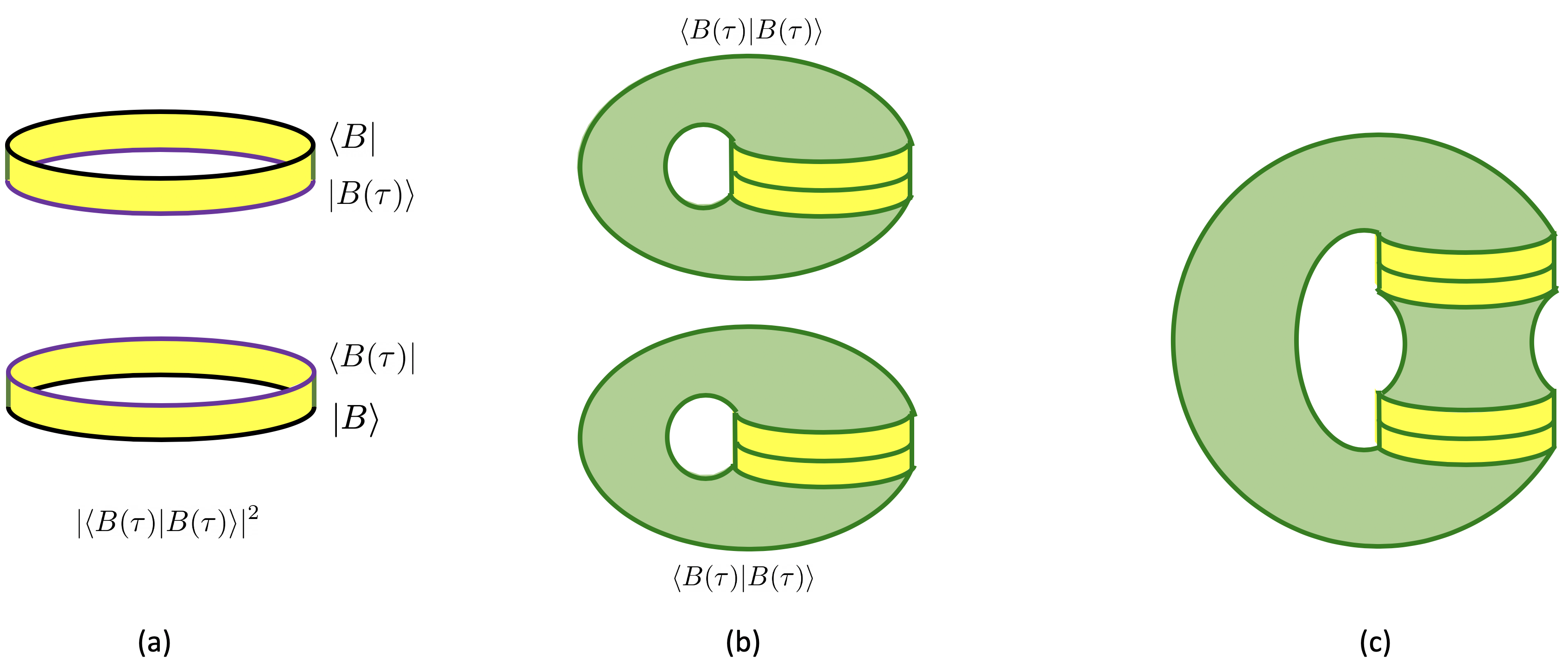}
\caption{We consider the computation of $\langle B(\tau ) | B(\tau ) \rangle \langle B(\tau ) | B(\tau) \rangle$. In (a) we display the holographic dual with boundary states. Starting with the state $|B\rangle$ we evolve by time $\tau$ to get $|B(\tau) \rangle$ and then overlap it with $\langle B(\tau)|$. We do the same with the bra.  In (b) we have the gravity computation. In (c) we have a candidate topology, but there are no solutions with this topology. }
\label{fig:Boverlap}
\end{center}
\end{figure}

   The example involves looking at a special family of typical states, namely the states built on $\ket{B}$ by acting with Euclidean evolution or some other operators. For simplicity, let us consider the state $\ket{B(\tau)}$,  see figure \ref{fig:Boverlap} (a). By construction, this state is a member of the typical states, as it can be approximated by  $\rho_s$, \nref{Temperature}, when computing simple observables. 
Naively we can apply the same argument as in section \ref{sec:Typical}. Namely, we project on to a typical state in the field theory region, except that the typical state we choose is $|B\rangle $ itself, the pure state of the exact theory.  Then one would think that, with this projection,  $|\braket{B(\tau)|B(\tau)}|^2$ is the wormhole in figure \ref{fig:TypicalState} (a). Here we are interpreting the middle $|B(\tau) \rangle \langle B(\tau) |$ as the projection on to a typical state.    However, we have learned that the quantity $\braket{B(\tau)|B(\tau)}$ itself is actually described by a wormhole, and the correct gravity description should be the one in figure \ref{fig:Boverlap} (b), which is the product of two wormholes and gives a factorized  result. The same holds for states that are built on $\ket{B(\tau)}$ by acting with operators. The geometry here is the same as that appearing in the purity discussion of section. \ref{sec:purity}, we are only interpreting it from a different angle.

The major difference from the previous idea is that instead of thinking of the typical state as just a boundary condition where nothing is on the other side, we are imagining that it also comes from an evolution involving gravity, and thus also carries its own dynamical gravity region. When we compute $|\braket{B(\tau)|B(\tau)}|^2$, we let the dynamical gravity decide how the gravity regions connect with each other. A lesson of this simple example is that when we deal with complicated states, we cannot neglect the spacetime involved in producing it. 

As in the purity discussion of section. \ref{sec:purity}, other than the factorizing configuration in figure \ref{fig:Boverlap} (b),  we also get a long wormhole configuration as in figure \ref{fig:Boverlap} (c). As we mentioned in section. \ref{sec:purity}, it is not a solution. It remains a question whether off-shell configurations like it should be included or not, and whether they are consistent with factorization, but what we saw here is an explicit example where factorization is respected at the level of on-shell solutions, while we imposed no restrictions on the type of configurations in the gravitational path integral.

\section{Bra-ket wormholes and tensor networks } 
\la{app:TensorNetworks}

  \begin{figure}[h]
\begin{center}
\includegraphics[scale=.35]{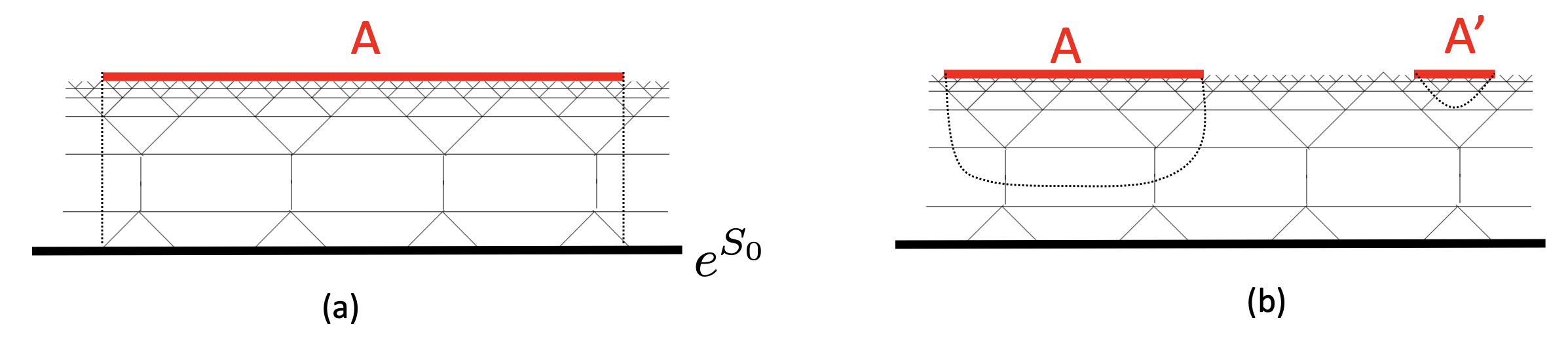}
\caption{ Tensor network representation of the CFT state produced in the setup of section \ref{sec:EuclBK}. We have a tensor network that has two MERA sides joined together, as we expect for a thermal state in a CFT. The top side represent the external legs representing the accessible degrees of freedom in the CFT. At the bottom, we join the legs of the MERA network to a matrix product state that is representing the UV degrees of freedom of the two dimensional gravity theory. These are just internal legs of the final network and not directly accessible.   (a)  A long interval and the RT-like surfaces. The entropy saturates at $2S_0$. (b) $A$ is an intermediate size interval where we see a thermal behavior. $A'$ is a small interval where we see a vacuum like behavior.   }
\label{TensorNetwork}
\end{center}
\end{figure}

It is natural to wonder about the relation between these gravity results  
  and tensor networks. 
Let us consider the state $|B\rangle$ that is produced by Euclidean AdS$_2$ evolution, discussed in section \ref{sec:EuclBK}, taking the spatial line to be infinite,   $L=\infty$. We have seen that the entropy of an interval behaves as follows.  We have the usual vacuum result  for $\ell  \lesssim 1/T_{x}$. It then behaves as the entropy of a thermal state for $ 1/T_{x} \lesssim \ell \lesssim  \ell_c \sim { S_0 \over c } {1 \over T_{x}}$. For larger lengths, $ \ell_c \lesssim \ell$, the entropy saturates at a value of order $2 S_0$. This suggests that, at very long distances,  the state can be represented by a matrix product state. 
These results suggest a picture as indicated in figure   \ref{TensorNetwork}. We have the usual MERA structure near the UV \cite{Vidal:2007hda,Swingle:2009bg}.  We then reflect it to get the structure of a thermofield double. Finally we cutoff   the bottom side    and insert a  particular matrix product state, represented by the thick black line, which takes into account the degrees of freedom of 2d quantum gravity. We expect that there are roughly $e^{S_0}$ indices running along the thick black line. 

In other words, 
the network containing thin lines  creates essentially the TFD state for the matter CFT. The top side in figure \ref{TensorNetwork} represents the flat space spatial line, the bottom side corresponds to a spatial slice at the middle of the wormhole, which is the spatial direction of the FRLW universe.  
 The CFT degrees of freedom on this slice are not directly accessible, and are contracted with a matrix product state that represents the UV degrees of freedom of two dimensional quantum gravity. 
  In principle,  we are allowed to move the thick black line up or down if we also change $S_0 \to S_0 - { c \over 6 } \log ({ 1/\varepsilon_{grav} })$ where $\varepsilon_{grav}$ is the effective cutoff on the CFT in the gravity region. 
  
  In the doubly holographic picture of  section \ref{DoublyHolographic} we can think of this tensor network as giving rise to the extra dimension. In other words, the horizontal red line of figure \ref{fig:DoublyRT} corresponds to the vertical direction in figure  \ref{TensorNetwork}. And the horizontal direction in figure \ref{TensorNetwork} is the dimension orthogonal to the page in figure \ref{fig:DoublyRT}.  The segments given by doted lines in figure \ref{TensorNetwork} are then similar to the RT surfaces of the corresponding intervals.

\small
\bibliographystyle{ourbst}
 \bibliography{EuclideanBraKetWormholes.bib}

\providecommand{\href}[2]{#2}\begingroup\raggedright\begin{thebibliography}{10}

\bibitem{Ryu:2006bv}
S.~Ryu and T.~Takayanagi, {{Holographic derivation of entanglement entropy from
  AdS/CFT}}, \href{http://dx.doi.org/10.1103/PhysRevLett.96.181602}{Phys. Rev.
  Lett. {\bf 96}, 181602, 2006},
  [\href{http://arxiv.org/abs/arXiv:hep-th/0603001}{{arXiv:hep-th/0603001
  [hep-th]}}].

\bibitem{Hubeny:2007xt}
V.~E. Hubeny, M.~Rangamani and T.~Takayanagi, {{A Covariant holographic
  entanglement entropy proposal}},
  \href{http://dx.doi.org/10.1088/1126-6708/2007/07/062}{JHEP {\bf 07}, 062,
  2007}, [\href{http://arxiv.org/abs/arXiv:0705.0016}{{arXiv:0705.0016
  [hep-th]}}].

\bibitem{Faulkner:2013ana}
T.~Faulkner, A.~Lewkowycz and J.~Maldacena, {{Quantum corrections to
  holographic entanglement entropy}},
  \href{http://dx.doi.org/10.1007/JHEP11(2013)074}{JHEP {\bf 11}, 074, 2013},
  [\href{http://arxiv.org/abs/arXiv:1307.2892}{{arXiv:1307.2892 [hep-th]}}].

\bibitem{Engelhardt:2014gca}
N.~Engelhardt and A.~C. Wall, {{Quantum Extremal Surfaces: Holographic
  Entanglement Entropy beyond the Classical Regime}},
  \href{http://dx.doi.org/10.1007/JHEP01(2015)073}{JHEP {\bf 01}, 073, 2015},
  [\href{http://arxiv.org/abs/arXiv:1408.3203}{{arXiv:1408.3203 [hep-th]}}].

\bibitem{Penington:2019npb}
G.~Penington, {{Entanglement Wedge Reconstruction and the Information
  Paradox}},  2019,
  [\href{http://arxiv.org/abs/arXiv:1905.08255}{{arXiv:1905.08255 [hep-th]}}].

\bibitem{Almheiri:2019psf}
A.~Almheiri, N.~Engelhardt, D.~Marolf and H.~Maxfield, {{The entropy of bulk
  quantum fields and the entanglement wedge of an evaporating black hole}},
  2019, [\href{http://arxiv.org/abs/arXiv:1905.08762}{{arXiv:1905.08762
  [hep-th]}}].

\bibitem{Almheiri:2019qdq}
A.~Almheiri, T.~Hartman, J.~Maldacena, E.~Shaghoulian and A.~Tajdini, {{Replica
  Wormholes and the Entropy of Hawking Radiation}},
  \href{http://dx.doi.org/10.1007/JHEP05(2020)013}{JHEP {\bf 05}, 013, 2020},
  [\href{http://arxiv.org/abs/arXiv:1911.12333}{{arXiv:1911.12333 [hep-th]}}].

\bibitem{Penington:2019kki}
G.~Penington, S.~H. Shenker, D.~Stanford and Z.~Yang, {{Replica Wormholes and
  the black hole interior}},  2019,
  [\href{http://arxiv.org/abs/arXiv:1911.11977}{{arXiv:1911.11977 [hep-th]}}].

\bibitem{Hartle:1983ai}
J.~Hartle and S.~Hawking, {{Wave Function of the Universe}},
  \href{http://dx.doi.org/10.1103/PhysRevD.28.2960}{Adv. Ser. Astrophys.
  Cosmol. {\bf 3}, 174--189, 1987}.

\bibitem{Jackiw:1984je}
R.~Jackiw, {{Lower Dimensional Gravity}},
  \href{http://dx.doi.org/10.1016/0550-3213(85)90448-1}{Nucl. Phys. {\bf B252},
  343--356, 1985}.

\bibitem{Teitelboim:1983ux}
C.~Teitelboim, {{Gravitation and Hamiltonian Structure in Two Space-Time
  Dimensions}}, \href{http://dx.doi.org/10.1016/0370-2693(83)90012-6}{Phys.
  Lett. {\bf B126}, 41--45, 1983}.

\bibitem{Cotler:2019nbi}
J.~Cotler, K.~Jensen and A.~Maloney, {{Low-dimensional de Sitter quantum
  gravity}}, \href{http://dx.doi.org/10.1007/JHEP06(2020)048}{JHEP {\bf 06},
  048, 2020}, [\href{http://arxiv.org/abs/arXiv:1905.03780}{{arXiv:1905.03780
  [hep-th]}}].

\bibitem{Maldacena:2019cbz}
J.~Maldacena, G.~J. Turiaci and Z.~Yang, {{Two dimensional Nearly de Sitter
  gravity}},  2019,
  [\href{http://arxiv.org/abs/arXiv:1904.01911}{{arXiv:1904.01911 [hep-th]}}].

\bibitem{Page:1986vw}
D.~N. Page, {{Density Matrix of the Universe}},
  \href{http://dx.doi.org/10.1103/PhysRevD.34.2267}{Phys. Rev. D {\bf 34},
  2267, 1986}.

\bibitem{Maldacena:2018lmt}
J.~Maldacena and X.-L. Qi, {{Eternal traversable wormhole}},  2018,
  [\href{http://arxiv.org/abs/arXiv:1804.00491}{{arXiv:1804.00491 [hep-th]}}].

\bibitem{Stanford:2020wkf}
D.~Stanford, {{More quantum noise from wormholes}},  2020,
  [\href{http://arxiv.org/abs/arXiv:2008.08570}{{arXiv:2008.08570 [hep-th]}}].

\bibitem{Hartle:2010dq}
J.~Hartle, S.~Hawking and T.~Hertog, {{Local Observation in Eternal
  inflation}}, \href{http://dx.doi.org/10.1103/PhysRevLett.106.141302}{Phys.
  Rev. Lett. {\bf 106}, 141302, 2011},
  [\href{http://arxiv.org/abs/arXiv:1009.2525}{{arXiv:1009.2525 [hep-th]}}].

\bibitem{Engelsoy:2016xyb}
J.~Engelsoy, T.~G. Mertens and H.~Verlinde, {{An investigation of AdS$_{2}$
  backreaction and holography}},
  \href{http://dx.doi.org/10.1007/JHEP07(2016)139}{JHEP {\bf 07}, 139, 2016},
  [\href{http://arxiv.org/abs/arXiv:1606.03438}{{arXiv:1606.03438 [hep-th]}}].

\bibitem{Jensen:2016pah}
K.~Jensen, {{Chaos and hydrodynamics near AdS$_2$}},  2016,
  [\href{http://arxiv.org/abs/arXiv:1605.06098}{{arXiv:1605.06098 [hep-th]}}].

\bibitem{Maldacena:2016upp}
J.~Maldacena, D.~Stanford and Z.~Yang, {{Conformal symmetry and its breaking in
  two dimensional Nearly Anti-de-Sitter space}},
  \href{http://dx.doi.org/10.1093/ptep/ptw124}{PTEP {\bf 2016}, 12C104, 2016},
  [\href{http://arxiv.org/abs/arXiv:1606.01857}{{arXiv:1606.01857 [hep-th]}}].

\bibitem{Geng:2020qvw}
H.~Geng and A.~Karch, {{Massive Islands}},  2020,
  [\href{http://arxiv.org/abs/arXiv:2006.02438}{{arXiv:2006.02438 [hep-th]}}].

\bibitem{Calabrese:2005in}
P.~Calabrese and J.~L. Cardy, {{Evolution of entanglement entropy in
  one-dimensional systems}},
  \href{http://dx.doi.org/10.1088/1742-5468/2005/04/P04010}{J. Stat. Mech. {\bf
  0504}, P04010, 2005},
  [\href{http://arxiv.org/abs/arXiv:cond-mat/0503393}{{arXiv:cond-mat/0503393}}].

\bibitem{Ghoshal:1993tm}
S.~Ghoshal and A.~B. Zamolodchikov, {{Boundary S matrix and boundary state in
  two-dimensional integrable quantum field theory}},
  \href{http://dx.doi.org/10.1142/S0217751X94001552}{Int. J. Mod. Phys. A {\bf
  9}, 3841--3886, 1994},
  [\href{http://arxiv.org/abs/arXiv:hep-th/9306002}{{arXiv:hep-th/9306002}}].

\bibitem{Marolf:2020vsi}
D.~Marolf, S.~Wang and Z.~Wang, {{Probing phase transitions of holographic
  entanglement entropy with fixed area states}},  2020,
  [\href{http://arxiv.org/abs/arXiv:2006.10089}{{arXiv:2006.10089 [hep-th]}}].

\bibitem{Dong:2020iod}
X.~Dong and H.~Wang, {{Enhanced corrections near holographic entanglement
  transitions: a chaotic case study}},  2020,
  [\href{http://arxiv.org/abs/arXiv:2006.10051}{{arXiv:2006.10051 [hep-th]}}].

\bibitem{Almheiri:2019yqk}
A.~Almheiri, R.~Mahajan and J.~Maldacena, {{Islands outside the horizon}},
  2019, [\href{http://arxiv.org/abs/arXiv:1910.11077}{{arXiv:1910.11077
  [hep-th]}}].

\bibitem{Mathur:2009hf}
S.~D. Mathur, {{The Information paradox: A Pedagogical introduction}},
  \href{http://dx.doi.org/10.1088/0264-9381/26/22/224001}{Class. Quant. Grav.
  {\bf 26}, 224001, 2009},
  [\href{http://arxiv.org/abs/arXiv:0909.1038}{{arXiv:0909.1038 [hep-th]}}].

\bibitem{Almheiri:2012rt}
A.~Almheiri, D.~Marolf, J.~Polchinski and J.~Sully, {{Black Holes:
  Complementarity or Firewalls?}},
  \href{http://dx.doi.org/10.1007/JHEP02(2013)062}{JHEP {\bf 02}, 062, 2013},
  [\href{http://arxiv.org/abs/arXiv:1207.3123}{{arXiv:1207.3123 [hep-th]}}].

\bibitem{Gao:2016bin}
P.~Gao, D.~L. Jafferis and A.~Wall, {{Traversable Wormholes via a Double Trace
  Deformation}},  2016,
  [\href{http://arxiv.org/abs/arXiv:1608.05687}{{arXiv:1608.05687 [hep-th]}}].

\bibitem{Maldacena:2017axo}
J.~Maldacena, D.~Stanford and Z.~Yang, {{Diving into traversable wormholes}},
  \href{http://dx.doi.org/10.1002/prop.201700034}{Fortsch. Phys. {\bf 65},
  1700034, 2017},
  [\href{http://arxiv.org/abs/arXiv:1704.05333}{{arXiv:1704.05333 [hep-th]}}].

\bibitem{Saad:2018bqo}
P.~Saad, S.~H. Shenker and D.~Stanford, {{A semiclassical ramp in SYK and in
  gravity}},  2018,
  [\href{http://arxiv.org/abs/arXiv:1806.06840}{{arXiv:1806.06840 [hep-th]}}].

\bibitem{Saad:2019lba}
P.~Saad, S.~H. Shenker and D.~Stanford, {{JT gravity as a matrix integral}},
  2019, [\href{http://arxiv.org/abs/arXiv:1903.11115}{{arXiv:1903.11115
  [hep-th]}}].

\bibitem{Maldacena:2018gjk}
J.~Maldacena, A.~Milekhin and F.~Popov, {{Traversable wormholes in four
  dimensions}},  2018,
  [\href{http://arxiv.org/abs/arXiv:1807.04726}{{arXiv:1807.04726 [hep-th]}}].

\bibitem{Giddings:2020yes}
S.~B. Giddings and G.~J. Turiaci, {{Wormhole calculus, replicas, and
  entropies}},  2020,
  [\href{http://arxiv.org/abs/arXiv:2004.02900}{{arXiv:2004.02900 [hep-th]}}].

\bibitem{Engelhardt:2020qpv}
N.~Engelhardt, S.~Fischetti and A.~Maloney, {{Free Energy from Replica
  Wormholes}},  2020,
  [\href{http://arxiv.org/abs/arXiv:2007.07444}{{arXiv:2007.07444 [hep-th]}}].

\bibitem{Almheiri:2019hni}
A.~Almheiri, R.~Mahajan, J.~Maldacena and Y.~Zhao, {{The Page curve of Hawking
  radiation from semiclassical geometry}},  2019,
  [\href{http://arxiv.org/abs/arXiv:1908.10996}{{arXiv:1908.10996 [hep-th]}}].

\bibitem{Chen:2019iro}
Y.~Chen, {{Pulling Out the Island with Modular Flow}},
  \href{http://dx.doi.org/10.1007/JHEP03(2020)033}{JHEP {\bf 03}, 033, 2020},
  [\href{http://arxiv.org/abs/arXiv:1912.02210}{{arXiv:1912.02210 [hep-th]}}].

\bibitem{Calabrese:2007mtj}
P.~Calabrese and J.~Cardy, {{Entanglement and correlation functions following a
  local quench: a conformal field theory approach}},
  \href{http://dx.doi.org/10.1088/1742-5468/2007/10/P10004}{J. Stat. Mech. {\bf
  0710}, P10004, 2007},
  [\href{http://arxiv.org/abs/arXiv:0708.3750}{{arXiv:0708.3750
  [cond-mat.stat-mech]}}].

\bibitem{Hartman:2013qma}
T.~Hartman and J.~Maldacena, {{Time Evolution of Entanglement Entropy from
  Black Hole Interiors}}, \href{http://dx.doi.org/10.1007/JHEP05(2013)014}{JHEP
  {\bf 05}, 014, 2013},
  [\href{http://arxiv.org/abs/arXiv:1303.1080}{{arXiv:1303.1080 [hep-th]}}].

\bibitem{Takayanagi:2011zk}
T.~Takayanagi, {{Holographic Dual of BCFT}},
  \href{http://dx.doi.org/10.1103/PhysRevLett.107.101602}{Phys. Rev. Lett. {\bf
  107}, 101602, 2011},
  [\href{http://arxiv.org/abs/arXiv:1105.5165}{{arXiv:1105.5165 [hep-th]}}].

\bibitem{Fujita:2011fp}
M.~Fujita, T.~Takayanagi and E.~Tonni, {{Aspects of AdS/BCFT}},
  \href{http://dx.doi.org/10.1007/JHEP11(2011)043}{JHEP {\bf 11}, 043, 2011},
  [\href{http://arxiv.org/abs/arXiv:1108.5152}{{arXiv:1108.5152 [hep-th]}}].

\bibitem{Cooper:2018cmb}
S.~Cooper, M.~Rozali, B.~Swingle, M.~Van~Raamsdonk, C.~Waddell and D.~Wakeham,
  {{Black Hole Microstate Cosmology}},
  \href{http://dx.doi.org/10.1007/JHEP07(2019)065}{JHEP {\bf 07}, 065, 2019},
  [\href{http://arxiv.org/abs/arXiv:1810.10601}{{arXiv:1810.10601 [hep-th]}}].

\bibitem{Simidzija:2020ukv}
P.~Simidzija and M.~Van~Raamsdonk, {{Holo-ween}},  2020.
\newblock [\href{http://arxiv.org/abs/arXiv:2006.13943}{{arXiv:2006.13943
  [hep-th]}}].

\bibitem{Akal:2020wfl}
I.~Akal, Y.~Kusuki, T.~Takayanagi and Z.~Wei, {{Codimension two holography for
  wedges}},  2020,
  [\href{http://arxiv.org/abs/arXiv:2007.06800}{{arXiv:2007.06800 [hep-th]}}].

\bibitem{Antonini:2019qkt}
S.~Antonini and B.~Swingle, {{Cosmology at the end of the world}},  2019,
  [\href{http://arxiv.org/abs/arXiv:1907.06667}{{arXiv:1907.06667 [hep-th]}}].

\bibitem{Anous:2020lka}
T.~Anous, J.~Kruthoff and R.~Mahajan, {{Density matrices in quantum gravity}},
  2020, [\href{http://arxiv.org/abs/arXiv:2006.17000}{{arXiv:2006.17000
  [hep-th]}}].

\bibitem{Maldacena:2004rf}
J.~M. Maldacena and L.~Maoz, {{Wormholes in AdS}},
  \href{http://dx.doi.org/10.1088/1126-6708/2004/02/053}{JHEP {\bf 02}, 053,
  2004},
  [\href{http://arxiv.org/abs/arXiv:hep-th/0401024}{{arXiv:hep-th/0401024}}].

\bibitem{Marolf:2020xie}
D.~Marolf and H.~Maxfield, {{Transcending the ensemble: baby universes,
  spacetime wormholes, and the order and disorder of black hole information}},
  2020, [\href{http://arxiv.org/abs/arXiv:2002.08950}{{arXiv:2002.08950
  [hep-th]}}].

\bibitem{Marolf:2021kjc}
D.~Marolf and J.~E. Santos, {{AdS Euclidean wormholes}},  2021,
  [\href{http://arxiv.org/abs/arXiv:2101.08875}{{arXiv:2101.08875 [hep-th]}}].

\bibitem{Strominger:2001pn}
A.~Strominger, {{The dS / CFT correspondence}},
  \href{http://dx.doi.org/10.1088/1126-6708/2001/10/034}{JHEP {\bf 10}, 034,
  2001},
  [\href{http://arxiv.org/abs/arXiv:hep-th/0106113}{{arXiv:hep-th/0106113}}].

\bibitem{Witten:2001kn}
E.~Witten, {{Quantum gravity in de Sitter space}},  in \emph{{Strings 2001:
  International Conference}}, 2001.
\newblock
  [\href{http://arxiv.org/abs/arXiv:hep-th/0106109}{{arXiv:hep-th/0106109}}].

\bibitem{Maldacena:2002vr}
J.~M. Maldacena, {{Non-Gaussian features of primordial fluctuations in single
  field inflationary models}},
  \href{http://dx.doi.org/10.1088/1126-6708/2003/05/013}{JHEP {\bf 05}, 013,
  2003},
  [\href{http://arxiv.org/abs/arXiv:astro-ph/0210603}{{arXiv:astro-ph/0210603}}].

\bibitem{Barvinsky:2006uh}
A.~Barvinsky and A.~Kamenshchik, {{Cosmological landscape from nothing: Some
  like it hot}}, \href{http://dx.doi.org/10.1088/1475-7516/2006/09/014}{JCAP
  {\bf 09}, 014, 2006},
  [\href{http://arxiv.org/abs/arXiv:hep-th/0605132}{{arXiv:hep-th/0605132}}].

\bibitem{Hertog:2011ky}
T.~Hertog and J.~Hartle, {{Holographic No-Boundary Measure}},
  \href{http://dx.doi.org/10.1007/JHEP05(2012)095}{JHEP {\bf 05}, 095, 2012},
  [\href{http://arxiv.org/abs/arXiv:1111.6090}{{arXiv:1111.6090 [hep-th]}}].

\bibitem{Kontsevich:2021dmb}
M.~Kontsevich and G.~Segal, {{Wick Rotation and the Positivity of Energy in
  Quantum Field Theory}}, \href{http://dx.doi.org/10.1093/qmath/haab027}{Quart.
  J. Math. Oxford Ser. {\bf 72}, 673--699, 2021},
  [\href{http://arxiv.org/abs/arXiv:2105.10161}{{arXiv:2105.10161 [hep-th]}}].

\bibitem{Witten:2021nzp}
E.~Witten, {{A Note On Complex Spacetime Metrics}},  2021,
  [\href{http://arxiv.org/abs/arXiv:2111.06514}{{arXiv:2111.06514 [hep-th]}}].

\bibitem{Alishahiha:2004md}
M.~Alishahiha, A.~Karch, E.~Silverstein and D.~Tong, {{The dS/dS
  correspondence}}, \href{http://dx.doi.org/10.1063/1.1848341}{AIP Conf. Proc.
  {\bf 743}, 393--409, 2004},
  [\href{http://arxiv.org/abs/arXiv:hep-th/0407125}{{arXiv:hep-th/0407125}}].

\bibitem{Maldacena:2019ufo}
J.~Maldacena and A.~Milekhin, {{SYK wormhole formation in real time}},  2019,
  [\href{http://arxiv.org/abs/arXiv:1912.03276}{{arXiv:1912.03276 [hep-th]}}].

\bibitem{Khoury:2001bz}
J.~Khoury, B.~A. Ovrut, N.~Seiberg, P.~J. Steinhardt and N.~Turok, {{From big
  crunch to big bang}},
  \href{http://dx.doi.org/10.1103/PhysRevD.65.086007}{Phys. Rev. D {\bf 65},
  086007, 2002},
  [\href{http://arxiv.org/abs/arXiv:hep-th/0108187}{{arXiv:hep-th/0108187}}].

\bibitem{Dong:2020uxp}
X.~Dong, X.-L. Qi, Z.~Shangnan and Z.~Yang, {{Effective entropy of quantum
  fields coupled with gravity}},  2020,
  [\href{http://arxiv.org/abs/arXiv:2007.02987}{{arXiv:2007.02987 [hep-th]}}].

\bibitem{Coleman:1988cy}
S.~R. Coleman, {{Black Holes as Red Herrings: Topological Fluctuations and the
  Loss of Quantum Coherence}},
  \href{http://dx.doi.org/10.1016/0550-3213(88)90110-1}{Nucl. Phys. {\bf B307},
  867--882, 1988}.

\bibitem{Giddings:1988cx}
S.~B. Giddings and A.~Strominger, {{Loss of Incoherence and Determination of
  Coupling Constants in Quantum Gravity}},
  \href{http://dx.doi.org/10.1016/0550-3213(88)90109-5}{Nucl. Phys. {\bf B307},
  854--866, 1988}.

\bibitem{Linde:1998gs}
A.~D. Linde, {{Quantum creation of an open inflationary universe}},
  \href{http://dx.doi.org/10.1103/PhysRevD.58.083514}{Phys. Rev. D {\bf 58},
  083514, 1998},
  [\href{http://arxiv.org/abs/arXiv:gr-qc/9802038}{{arXiv:gr-qc/9802038}}].

\bibitem{Casini:2016fgb}
H.~Casini, I.~Salazar~Landea and G.~Torroba, {{The g-theorem and quantum
  information theory}}, \href{http://dx.doi.org/10.1007/JHEP10(2016)140}{JHEP
  {\bf 10}, 140, 2016},
  [\href{http://arxiv.org/abs/arXiv:1607.00390}{{arXiv:1607.00390 [hep-th]}}].

\bibitem{Sasaki:1994yt}
M.~Sasaki, T.~Tanaka and K.~Yamamoto, {{Euclidean vacuum mode functions for a
  scalar field on open de Sitter space}},
  \href{http://dx.doi.org/10.1103/PhysRevD.51.2979}{Phys. Rev. D {\bf 51},
  2979--2995, 1995},
  [\href{http://arxiv.org/abs/arXiv:gr-qc/9412025}{{arXiv:gr-qc/9412025}}].

\bibitem{Louko:1995jw}
J.~Louko and R.~D. Sorkin, {{Complex actions in two-dimensional topology
  change}}, \href{http://dx.doi.org/10.1088/0264-9381/14/1/018}{Class. Quant.
  Grav. {\bf 14}, 179--204, 1997},
  [\href{http://arxiv.org/abs/arXiv:gr-qc/9511023}{{arXiv:gr-qc/9511023}}].

\bibitem{Vidal:2007hda}
G.~Vidal, {{Entanglement Renormalization}},
  \href{http://dx.doi.org/10.1103/PhysRevLett.99.220405}{Phys. Rev. Lett. {\bf
  99}, 220405, 2007},
  [\href{http://arxiv.org/abs/arXiv:cond-mat/0512165}{{arXiv:cond-mat/0512165}}].

\bibitem{Swingle:2009bg}
B.~Swingle, {{Entanglement Renormalization and Holography}},
  \href{http://dx.doi.org/10.1103/PhysRevD.86.065007}{Phys. Rev. {\bf D86},
  065007, 2012}, [\href{http://arxiv.org/abs/arXiv:0905.1317}{{arXiv:0905.1317
  [cond-mat.str-el]}}].

\end{thebibliography}\endgroup
\end{document}